\documentclass[10pt,aps,showpacs,nofootinbib,prd,aps,epsf,floats,
               amsmath,amssymb,amsfonts,axodraw,floatfix,graphicx,twocolumn]{revtex4-1}
\usepackage{amsmath, amssymb}
\usepackage{paralist}
\newcommand{\mathsym}[1]{{}}

\usepackage{graphicx}
\usepackage{amsmath}
\usepackage{amssymb}
\usepackage{bm}
\newsavebox{\PSLASH}
 \sbox{\PSLASH}{$p$\hspace{-1.8mm}/}
 
\renewcommand{\theequation}{\thesection.\arabic{equation}}
\newcounter{saveeqn}
\newcommand{\add}{\addtocounter{equation}{1}}
\newcommand{\alphaeqn}{\setcounter{saveeqn}{\value{equation}}%
\setcounter{equation}{0}%
\renewcommand{\theequation}{\mbox{\thesection.\arabic{saveeqn}{\alpha{equation}}}}}
\newcommand{\reseteqn}{\setcounter{equation}{\value{saveeqn}}%
\renewcommand{\theequation}{\thesection.\arabic{equation}}}

 \newsavebox{\notrightarrow}
 \sbox{\notrightarrow}{$\to$\hspace{-4mm}/}
 
 \newsavebox{\PARTIALSLASH}
 \sbox{\PARTIALSLASH}{$\partial$\hspace{-1.6mm}/}
 
 \newsavebox{\ASLASH}
 \sbox{\ASLASH}{$A$\hspace{-2.1mm}/}
 
 \newsavebox{\KSLASH}
 \sbox{\KSLASH}{$k$\hspace{-1.8mm}/}
 
 \newsavebox{\LSLASH}
 \sbox{\LSLASH}{$\ell$\hspace{-1.8mm}/}
 
 \newsavebox{\QSLASH}
 \sbox{\QSLASH}{$q$\hspace{-1.8mm}/}
 
 \newsavebox{\DSLASH}
 \sbox{\DSLASH}{$D$\hspace{-2.2mm}/}
 
 \newsavebox{\DbfSLASH}
 \sbox{\DbfSLASH}{${\mathbf D}$\hspace{-2.8mm}/}
 
 \newsavebox{\DELVECRIGHT}
 \sbox{\DELVECRIGHT}{$\stackrel{\rightarrow}{\partial}$}
 
 \newcommand{\blue}{\IfColor{\textCadetBlue}{}}
\newcommand{\black}{\IfColor{\textBlack}{}}
\newcommand{\red}{\IfColor{\textRed}{}}
\newcommand{\green}{\IfColor{\textOliveGreen}{}}
\newcommand{\lil}{\IfColor{\textRedViolet}{}}

\newcommand{\bs}{\boldsymbol}
\begin{document}
\title{Wigner function formalism and\\ the evolution of thermodynamic quantities in an expanding magnetized plasma}
\author{S. M. A. Tabatabaee}\email{tabatabaeemehr\_sma@physics.sharif.ir}
\author{N. Sadooghi}\email{Corresponding author: sadooghi@physics.sharif.ir}
\affiliation{Department of Physics, Sharif University of Technology,
P.O. Box 11155-9161, Tehran, Iran}
\begin{abstract}
By combining the Wigner function formalism of relativistic quantum kinetic theory with fundamental equations of relativistic magnetohydrodynamics (MHD), we present a novel approach to determine the proper time evolution of the temperature and other thermodynamic quantities in a uniformly expanding hot, magnetized, and weakly interacting plasma. The aim is to study the contribution of quantum corrections to this evolution. We first determine the corresponding Wigner function in terms of the solution of the Dirac equation in the presence of a constant magnetic field. Using this function, we then compute the energy-momentum tensor of the above-mentioned plasma, which eventually yields its energy density and pressure. Plugging these quantities in the energy equation of relativistic MHD, we arrive, after choosing an appropriate coordinate system, at a differential equation for the temperature as a function of the proper time. The numerical solution of this equation leads finally to the proper time evolution of the temperature. The latter is then used to determine the evolution of a large number of thermodynamic quantities in this expanding and magnetized plasma. We compare our results with other existing results from relativistic MHD. We also comment on the effect of point to point decaying magnetic fields on the thermodynamic properties of this plasma.
\end{abstract}
\maketitle
\section{Introduction}\label{sec1}
\setcounter{equation}{0}
The primary goal of modern heavy-ion collision (HIC) experiments at the Relativistic Heavy Ion Collider (RHIC) and Large Hadron Collider (LHC) is to produce a plasma of quarks and gluons, and to study its evolution from an early out of equilibrium stage to a late hadronization one. It is widely believed that at $\tau_0\sim 0.2$ fm/c after the collision, the created hot plasma enters a state of local thermal equilibrium. There are strong pieces of evidence that at this stage, the dynamics of the created quark matter is well described by relativistic hydrodynamics (for a recent review see \cite{romatschke2017} and references therein), that, because of extremely large magnetic fields that are also created in noncentral HIC experiments \cite{warringa2007, skokov2009,huang2015,inghirami2019}, is to be extended to relativistic MHD \cite{teller1950}. Assuming firstly that the plasma expands uniformly in the longitudinal direction with respect to the beam direction, and secondly that
the external magnetic field is aligned perpendicular to the plasma velocity, it was recently possible to extend the well-known $1+1$-dimensional Bjorken solution of relativistic hydrodynamics \cite{bjorken, hatsuda-book} to the so-called ``ideal transverse MHD'' \cite{rischke-MHD}, and to determine the proper time evolution of the magnetic field after the onset of hydrodynamics. In \cite{shokri-MHD}, we introduced a more general framework to study nonideal transverse MHD, and determined apart from the evolution of electric and magnetic fields, the spacetime evolution of the temperature in a hot, electromagnetized, and uniformly expanding plasma. In order to explore, in particular, the effect of external electromagnetic fields on the spacetime evolution of the plasma's temperature, the latter is to be compared with the Bjorken $1+1$ dimensional solution in the absence of electromagnetic fields. In a parallel development, we generalized the method introduced in the framework of ``anisotropic hydrodynamics'' \cite{florkowski2010,strickland2010} to anisotropic MHD, and studied the effect of magnetization on the isotropization of a uniformly expanding plasma in and out of equilibrium \cite{tabatabaee2019}. Following the approach introduced in \cite{florkowski2010,strickland2010}, we combined the equations of ideal transverse MHD with the Boltzmann equation of classical kinetic theory in a relaxation time approximation, and determined, in particular, the spacetime evolution of the temperature in this plasma. The latter is in principle affected by the evolution of the anisotropy parameter in this model. A comparison shows, however, no significant difference between this result and the Bjorken solution for the temperature \cite{tabatabaee2019}.
\par
In the present paper, we introduce a novel approach to determine the proper time evolution of the temperature and other thermodynamic quantities in a uniformly expanding, hot, magnetized, and weakly interacting plasma. To do this, we combine the Wigner function formalism of relativistic quantum kinetic theory \cite{wignerfunction} with the energy
equation of relativistic MHD. The aim is to consider the quantum corrections arising from Landau levels, that arise in the presence of constant magnetic fields. In this way, we determine the proper time evolution of  a number of thermodynamic quantities including the temperature $T$, energy density $\epsilon$, transverse and longitudinal pressures $p_{\perp}$, and $p_{\|}$, transverse and longitudinal speeds of sound $c_{s,\perp}$, and $c_{s,\|}$, and magnetic susceptibility $\chi_{m}$, in a uniformly expanding hot QED plasma in the presence of constant and varying magnetic fields $B$.\footnote{Transverse and longitudinal directions are defined with respect to the external magnetic field.} To this purpose, we go through the following steps:
\par
\textit{i}) We first determine the Wigner function of a fermionic hot and magnetized plasma. To do this, we use the solution of the Dirac equation in a constant background magnetic field using the Ritus eigenfunction formalism \cite{ritus1972}.    Our results are comparable with those presented in \cite{wigner-magnetic}.
\par
\textit{ii}) We then use this Wigner function to analytically determine the energy-momentum tensor of a hot plasma in a constant magnetic field. Using this tensor, we determine $\epsilon,p_{\perp}$, and $p_{\|}$ from its diagonal components as functions of $T$. We first show that the resulting expressions (yet in an integral form) are comparable with the corresponding expressions arising from an appropriate thermodynamic potential in the presence of a constant magnetic field. We then consider two different cases of massless ($m=0$) and massive ($m\neq 0$) fermions. In the massless case, we determine these quantities in two different limits of $eB\ll T^{2}$ (high-temperature or weak magnetic field limit) and $eB\gg T^{2}$ (low-temperature or strong magnetic field limit).  In the massive case, we determine these quantities only in the limit  $M_{n}^{2}\ll T^{2}$, where $M_{n}^{2}\equiv m^{2}+2neB$ is the magnetic mass.
\par
\textit{iii}) We then consider the energy equation of relativistic MHD, that is given in terms of $\epsilon, p_{\|}$, and $p_{\perp}$. In a certain Milne coordinate system, this turns out to be simply a first order differential equation in the proper time $\tau$. By plugging the analytical expressions for the above thermodynamic quantities in the two aforementioned limits into this equation, we arrive at a differential equation for $T$ as a function of $\tau$. We solve this equation numerically, and arrive, in particular, at the proper time evolution of the temperature.
\par
\textit{iv}) By plugging the resulting $\tau$ dependence of $T$ into the analytical expressions for $\epsilon, p_{\perp}, p_{\|}, c_{s,\perp}, c_{s,\|}$, and $\chi_{m}$ for $eB\ll T^{2}$ and $eB\gg T^{2}$ in the massless case, and $M_n^{2}\ll T^{2}$ in the massive case, we arrive at the proper time evolution of these quantities in a constant background magnetic field $B$ in these two limits. Finally, in order to consider the more realistic scenario of a decaying magnetic field, we assume that $B$ is constant at each infinitesimal time interval $\delta\tau$, but decays according to a certain profile $B=B(\tau)$ for $B$. Using this $\tau$ dependent magnetic field, it is then possible to determine the $\tau$ dependence of the temperature in a decaying magnetic field. We consider a number of profiles for $B(\tau)$, and determine the proper time evolution of $T$ corresponding to these point to point varying magnetic fields.
\par
It is worthwhile to notice that recently the Wigner function formalism is used to study the kinetic phenomena of chiral plasma \cite{wigner-chiral,rischke2019} and different aspects of QCD phase transition within the Nambu-Jona-Lasinio model \cite{wigner-phase}. The method introduced in the present paper, however, is a novel application of this formalism, that, once combined with the main equations of relativistic (magneto)hydrodynamics yields the evolution of thermodynamic quantities in an expanding plasma.
\par
The organization of this paper is as follows:  In Sec. \ref{sec2}, we present a number of review materials: In Sec. \ref{sec2a}, we review the Ritus eigenfunction method, and present, in particular, an appropriate quantization for the fermionic fields which is used later to determine the Wigner function. In Sec.  \ref{sec2b}, we derive the thermodynamic quantities, $p_{\|},p_{\perp}$, and $\epsilon$, using an appropriate thermodynamic potential. The resulting expressions are later compared with the corresponding expressions arising from the energy-momentum tensor constructed in terms of the Wigner function. In Sec. \ref{sec2c}, we present the main equations of ideal transverse MHD together with the Bjorken solution of $1+1$ dimensional hydrodynamics. The Gubser temperature is also introduced in a slightly different form than the original one in \cite{gubser2010}. Being interested in the evolution of $T$ and other thermodynamic quantities in point to point decaying magnetic fields, we present in Sec. \ref{sec2d}, various solutions for the proper time evolution of $B$.
\par
Our main analytic results are presented in Sec. \ref{sec3}: In Sec. \ref{sec3a}, we derive the Wigner function, the energy-momentum tensor, and the resulting thermodynamic quantities, yet in an integral form. In Sec. \ref{sec3b}, we perform high- and low-temperature approximations, and determine $p_{\|},p_{\perp}$, and $\epsilon$ in these approximations. In Sec. \ref{sec4}, we present the numerical results for the $\tau$ dependence of $T$ in high- and low-temperature approximations. These data are then used to determine the $\tau$ dependence of other thermodynamic quantities. Section \ref{sec5} is devoted to concluding remarks. Comments on Gubser temperature as well as useful formulae in high- and low-temperature approximations are presented in Appendices \ref{appA} and \ref{appB}.

\section{Review material}\label{sec2}
\setcounter{equation}{0}
To be self-content, we present in what follows a number of review materials. We start by introducing the solutions of the Dirac equation in the presence of a constant magnetic field in the Ritus eigenfunction formalism (Sec. \ref{sec2}). In particular, we present the quantization of fermions with a positive charge. In Sec. \ref{sec3a}, we use this quantization to determine the energy-momentum tensor of a magnetized plasma via the Wigner function formalism. In Sec. \ref{sec2b}, we use the thermodynamic potential of a hot plasma in a constant background magnetic field, and present the corresponding analytical expressions for $p_{\|}, p_{\perp}$, and $\epsilon$. In Sec. \ref{sec2c}, we then briefly introduce the ideal relativistic transverse MHD. In particular, assuming the plasma to expand uniformly, we present the energy and Euler equation of ideal transverse MHD in the Milne coordinate system, as well as the Bjorken and Gubser solutions for the temperature. Finally, in Sec. \ref{sec2d}, we present a number of phenomenological and theoretical solutions for the proper time dependence of the magnetic field. These expressions are then used in Sec. \ref{sec4}, to determine the proper time dependence of the temperature for point to point decaying magnetic fields.
\subsection{Fermions in a constant magnetic field}\label{sec2a}
One of the mostly used methods to solve the Dirac equation of charged fermions in the presence of a constant magnetic field is the Ritus eigenfunction method \cite{ritus1972}. We outline it in this section. A complete derivation of this method in a $3+1$  dimensional spacetime in a multiflavor system is previously presented in \cite{sadooghi2016}.  In what follows, we present, in particular, an appropriate quantization for a positively charged fermion. We use this quantization in Sec. \ref{sec3a} to determine the Wigner function and the energy-momentum tensor of a QED plasma.
\par
Let us start with the Dirac equation for a charged fermion with mass $m_q$,
\begin{eqnarray}\label{N1}
\left(\gamma\cdot\Pi^{(q)}-m_{q}\right)\psi^{(q)}=0.
\end{eqnarray}
Here, $\Pi_{\mu}^{(q)}=i\partial_{\mu}+eq A_{\mu}^{\text{ext}}$, with $e>0$ and $q$ the charge of this fermion. By fixing the gauge field  $A_{\mu}^{\text{ext}}$ as $A_{\mu}^{\text{ext}}=(0,0,Bx,0)$ with $B>0$, we arrive at a magnetic field $\boldsymbol{B}$ aligned in the third spatial direction $\boldsymbol{B}=B\boldsymbol{e}_{z}$. In order to solve \eqref{N1}, we use the ansatz $\psi_{+}^{(q)}=\mathbb{E}_{n,+}^{(q)}u(\tilde{p}_{q}^{(\kappa)})$, as the positive frequency solution and $\psi_{-}^{(q)}={\mathbb{E}}_{n,-}^{(q)}v(\tilde{p}_{q}^{(\kappa)})$, as the negative frequency solution. Here,  $\mathbb{E}_{n,\kappa}^{(q)}$ with $\kappa=\pm 1$ satisfies the Ritus eigenfunction relation
\begin{eqnarray}\label{N2}
\left(\gamma\cdot \Pi^{(q)}\right)\mathbb{E}_{n,\kappa}^{(q)}=\kappa\mathbb{E}_{n,\kappa}^{(q)}\left(\gamma\cdot \tilde{p}_{q}^{(\kappa)}\right),
\end{eqnarray}
with $\kappa=+1$ ($\kappa=-1$) for positive (negative) frequency solution. Using the above expressions for $\psi_{\kappa}^{(q)}$ and the relation \eqref{N2}, it is possible to show that the spinors $u$ as well as $v$ are the free Dirac spinors satisfying $$\left(\gamma\cdot \tilde{p}_{q}^{(+)}-m_{q}\right)u(\tilde{p}_{q}^{(+)})=0,$$
as well as
$$\left(\gamma\cdot \tilde{p}_{q}^{(-)}+m_{q}\right)v(\tilde{p}_{q}^{(-)})=0.$$
Moreover, as it is shown in \cite{taghinavaz-ritus, fayazbakhsh-ritus, fayazbakhsh2014}, the Ritus momentum for a particle with charge $q$ is given by
\begin{eqnarray}\label{N3}
\tilde{p}^{\mu}_{q,\kappa}=(E_{n}^{(q)}, 0,-\kappa s_{q}\sqrt{2n|qeB|},p_{z}),
\end{eqnarray}
with $\kappa=\pm 1$ for positive and negative frequency solutions, $n$ labeling the Landau levels in the external magnetic field, and $s_{q}\equiv \text{sgn}(qeB)$. In this paper, we assume $eB>0$. Following the method presented in \cite{taghinavaz-ritus, fayazbakhsh-ritus,fayazbakhsh2014}, the Ritus eigenfunctions $\mathbb{E}_{n,\kappa}^{(q)}, \kappa=\pm 1$ are given by
\begin{eqnarray}\label{N4}
\mathbb{E}_{n,\kappa}^{(q)}(\xi_{\kappa}^{s_q})=e^{-i\kappa\bar{p}\cdot\bar{x}}\mathbb{P}_{n,\kappa}^{(q)}(\xi_{\kappa}^{s_q}),
\end{eqnarray}
with $\xi^{s_{q}}_{\kappa}\equiv (x-\kappa s_{q}p_{y}\ell_{B}^{2})/\ell_{b}$ and $\ell_{b}\equiv |qeB|^{-1/2}$ as well as $\bar{p}^{\mu}\equiv \left(p_{0},0,p_y,p_z\right)$ as well as $\bar{x}^\mu\equiv \left(t,0,y,z\right)$. The physical, on the mass-shell fermion possesses the following energy dispersion relation:
\begin{eqnarray}\label{N5}
E_n^{(q)}=\left(2n|qeB|+p_z^2+m^2\right)^{1/2}.
\end{eqnarray}
Here, $E_{n}^{(q)}\equiv p_0$.
In \eqref{N4}, $\mathbb{P}_{n,\kappa}^{(q)}$ is defined by
\begin{eqnarray}\label{N6}
\hspace{-0.8cm}\mathbb{P}_{n}^{(q)}(\xi_{\kappa}^{s_q})=P_{+}^{(q)}f_{n}^{+s_q}(\xi^{s_q}_{\kappa})+\Pi_{n}P_{-}^{(q)}f_{n}^{-s_q}(\xi_{\kappa}^{s_q}),
\end{eqnarray}
with $\Pi_{n}\equiv 1-\delta_{n,0}$, and the projectors $P_{\pm}^{(q)}$ given by
\begin{eqnarray}\label{N7}
P_{\pm}^{(q)}\equiv \frac{1\pm is_q \gamma_1\gamma_2}{2}.
\end{eqnarray}
As it turns out, $P_{\pm}^{(+)}=P_{\pm}$ and $P_{\pm}^{(-)}=P_{\mp}$ with $P_{\pm}\equiv \frac{1}{2}\left(1\pm i\gamma_1\gamma_2\right)$. The functions $f_{n}^{\pm s_q}(\xi_{\kappa}^{s_q})$, appearing in \eqref{N6} are given by
\begin{eqnarray}\label{N8}
\begin{array}{rclcrcl}
f_n^{+s_q}&\equiv&\Phi_{n}(\xi_{\kappa}^{s_q}),&&n&=&0,1,2,\cdots,\\
f_{n}^{-s_q}&\equiv& \Phi_{n-1}(\xi_{\kappa}^{s_q}),&&n&=&1,2,\cdots,
\end{array}
\end{eqnarray}
with $\Phi_{n}$ given in terms of Hermite polynomials $H_n(z)$
\begin{eqnarray}\label{N9}
\Phi_{n}(\xi_{\kappa}^{s_q})\equiv a_n\exp\left(-\frac{(\xi_{\kappa}^{s_q})^{2}}{2}\right)H_n(\xi_{\kappa}^{s_q}).
\end{eqnarray}
Here, $a_n\equiv \left(2^{n}n!\sqrt{\pi}\ell_{b}\right)^{-1/2}$.
\par
As in \cite{sadooghi2016}, we introduce at this stage the following quantization for a positively charged fermion ($q=+1$) with mass $m\equiv m_{+}$ in the presence of a constant magnetic field:
\begin{widetext}
\begin{eqnarray}\label{N10}
\psi_{\alpha}(x)&=&\frac{1}{V^{1/2}}\sum_{n=0}^{\infty}\sum_{s=1,2}\int\frac{dp_y dp_z}{(2\pi)^2}\frac{1}{\sqrt{2E_n}}\bigg\{[\mathbb{P}_{n}^{(+)}(\xi^{+})]_{\alpha\sigma}u_{s,\sigma}(\tilde{p}_{+})a_{\bs{\bar{p}}}^{n,s}e^{-i\bar{p}\cdot \bar{x}}+[\mathbb{P}_{n}^{(+)}({\xi}^{-})]_{\alpha\sigma}v_{s,\sigma}(\tilde{p}_{-})b^{\dagger n,s}_{\bs{\bar{p}}}e^{+i\bar{p}\cdot \bar{x}}\bigg\},\nonumber\\
\bar{\psi}_{\alpha}(x)&=&\frac{1}{V^{1/2}}\sum_{n=0}^{\infty}\sum_{s=1,2}\int\frac{dp_y dp_z}{(2\pi)^2}\frac{1}{\sqrt{2E_n}}\bigg\{a_{\bs{\bar{p}}}^{\dagger n,s}\bar{u}_{s,\rho}(\tilde{p}_{+})
[\mathbb{P}_{n}^{(+)}(\xi^{+})]_{\rho\alpha}
e^{+i\bar{p}\cdot \bar{x}}+b^{n,s}_{\bs{\bar{p}}}\bar{v}_{s,\rho}(\tilde{p}_{-})[\mathbb{P}_{n}^{(+)}({\xi}^{-})]_{\rho\alpha}e^{-i\bar{p}\cdot \bar{x}}
\bigg\},\nonumber\\
\end{eqnarray}
\end{widetext}
with $\bar{\bs{p}}=(0,p_y,p_z)$, $E_n\equiv E_n^{(+)}$ defined in \eqref{N5}, $\tilde{p}_{\kappa}\equiv \tilde{p}_{+,\kappa}$ and  $\xi^{\kappa}\equiv \xi^{+}_{\kappa}$ for $\kappa=\pm 1$. Here, $a_{\bs{\bar{p}}}^{n,s}, a_{\bs{\bar{p}}}^{\dagger n,s}$ and $b_{\bs{\bar{p}}}^{n,s},b_{\bs{\bar{p}}}^{\dagger n,s}$ are two sets of creation and annihilation operators satisfying the commutation relations
\begin{eqnarray}\label{N11}
\hspace{-0.3cm}\{a_{\bs{\bar{p}}}^{n,s}, a_{\bs{\bar{q}}}^{\dagger m,s^{\prime}}\}=(2\pi)^{2}V\delta^{2}\left(\bar{\bs{p}}-\bar{\bs{q}}\right)
\delta_{s,s^{\prime}}\delta_{n,m},\nonumber\\
\hspace{-0.3cm}\{b_{\bs{\bar{p}}}^{n,s}, b_{\bs{\bar{q}}}^{\dagger m,s^{\prime}}\}=(2\pi)^{2}V\delta^{2}\left(\bar{\bs{p}}-\bar{\bs{q}}\right)
\delta_{s,s^{\prime}}\delta_{n,m},
\end{eqnarray}
and the spinors $u_{s,\alpha}, \bar{u}_{s,\alpha}$ as well as $v_{s,\alpha}, \bar{v}_{s,\alpha}$ satisfying
\begin{eqnarray}\label{N12}
\sum_{s}u_{s,\alpha}(\tilde{p}_{+})\bar{u}_{s,\beta}(\tilde{p}_{+})&=&\left(\gamma\cdot\tilde{p}_{+}+m\right)_{\alpha\beta},\nonumber\\
\sum_{s}v_{s,\alpha}(\tilde{p}_{-})\bar{v}_{s,\beta}(\tilde{p}_{-})&=&\left(\gamma\cdot\tilde{p}_{-}-m\right)_{\alpha\beta}.
\end{eqnarray}
In \cite{sadooghi2016}, we used the above quantization relations \eqref{N10} to derive the propagator of a positively charged fermions. In the present paper, however, we use it to determine the Wigner function of a hot and magnetized fermionic plasma.
\subsection{Thermodynamic quantities arising from the thermodynamic potential of a magnetized QED plasma}\label{sec2b}
As aforementioned, in Sec. \ref{sec3a} we use the above quantization \eqref{N10}, and determine the thermodynamic  quantities $\epsilon, p_{\|}, p_{\perp}, \cdots$ using the Wigner function formalism. We then compare the resulting expressions in the integral form with the expressions arising from the effective potential of a magnetized QED plasma at finite temperature $T$ and zero chemical potential $\mu$. It is given by \cite{fayazbakhsh-ritus,fayazbakhsh2014}
\begin{eqnarray}\label{N13}
\Omega_{\text{eff}}=-\frac{eB}{2\pi}\sum_{n=0}^{\infty}\alpha_{n}\int\frac{dp_{z}}{2\pi}\bigg\{E_{n}+2T\ln\left(1+e^{-\beta E_{n}}\right)\bigg\},\nonumber\\
\end{eqnarray}
with $\beta\equiv T^{-1}$ and  the spin degeneracy factor $\alpha_{n}\equiv 2-\delta_{n,0}$. It arises from the free fermion determinant $\Omega_{\text{eff}}=-\frac{1}{\beta V}\ln Z$ with $V$ a three-dimensional volume and $Z$ the fermionic partition function \cite{kapusta-book}. We use the standard definitions
\begin{eqnarray}\label{N14}
\hspace{-0.5cm}p_{\|}=-\Omega_{\text{eff}},\quad s=-\frac{\partial \Omega_{\text{eff}}}{\partial T},\quad M=-\frac{\partial \Omega_{\text{eff}}}{\partial B},
\end{eqnarray}
to determine the parallel pressure $p_{\|}$, the entropy density $s$, and the magnetization $M$. We arrive at
\begin{eqnarray}\label{N15}
p_{\|}&=&\frac{eB}{2\pi^2}\sum_{n=0}^{\infty}\alpha_{n}\int_{0}^{\infty}dk_z E_{n}\nonumber\\
&&+\frac{eB}{\pi^{2}}\sum_{n=0}^{\infty}\alpha_{n}\int_{0}^{\infty} \frac{dk_z}{E_n} k_{z}^{2}f_{\text{F}}(E_n)\nonumber\\
Ts&=&\frac{eB}{\pi^{2}}\sum_{n=0}^{\infty}\alpha_{n}\int_{0}^{\infty}\frac{dk_z}{E_n}(E_{n}^{2}+k_{z}^{2})f_{\text{F}}(E_n),\nonumber\\
BM&=&\frac{eB}{2\pi^{2}}\sum_{n=0}^{\infty}\alpha_n\int_{0}^{\infty}\frac{dk_z}{E_n}(E_n^2+neB)\nonumber\\
&&+\frac{eB}{\pi^{2}}\sum_{n=0}^{\infty}\alpha_{n}\int_{0}^{\infty}\frac{dk_z}{E_n}(k_z^{2}-neB)f_{\text{F}}(E_n),\nonumber\\
\end{eqnarray}
with the Fermi-Dirac distribution function
\begin{eqnarray}\label{N16}
f_{\text{F}}(E_n,k_z)\equiv\frac{1}{e^{\beta E_n}+1},
\end{eqnarray}
and  the energy $E_n$ arising from \eqref{N3} with $q=+1$
\begin{eqnarray}\label{N17}
E_{n}=\left(m^{2}+2neB+k_{z}^{2}\right)^{1/2}.
\end{eqnarray}
For later convenience,  we have used an appropriate partial integration (PI) in \eqref{N5} to reformulate the following integral:
\begin{eqnarray}\label{N18}
T\int_{-\infty}^{+\infty}\frac{dk_z}{2\pi}\ln\left(1+e^{-\beta E_n}\right)\stackrel{\text{PI}}{=}\frac{1}{\pi}\int_{0}^{+\infty}\frac{dk_z}{E_n}\frac{k_z^{2}}{e^{\beta E_n}+1}. \nonumber\\
\end{eqnarray}
Combining the expressions for $p_{\|}$, $s$, and $BM$, the energy density $\epsilon$ and the transverse pressure $p_{\perp}$ are given by
\begin{eqnarray}\label{N19}
\hspace{-1cm}\epsilon&=&-p_{\|}+Ts
\nonumber\\
\hspace{-1cm}&=&-\frac{eB}{2\pi^{2}}\sum_{n=0}^{\infty}\alpha_{n}\int_{0}^{\infty}dk_z E_{n}\left(1-2f_{\text{F}}(E_n,k_z)\right),
\end{eqnarray}
and
\begin{eqnarray}\label{N20}
p_{\perp}&=&p_{\|}-BM\nonumber\\
&=&-\frac{(eB)^{2}}{2\pi^{2}}\sum_{n=0}^{\infty}n\alpha_{n}\int_{0}^{\infty}\frac{dk_z}{E_n}\left(1-2f_{\text{F}}(E_n,k_z)\right). \nonumber\\
\end{eqnarray}
In Sec. \ref{sec4}, we show that the matter ($T$-dependent) part of the above  expressions for $\epsilon, p_{\|}$, and $p_{\perp}$ in the massless case coincides with the results arising from the energy-momentum
tensor determined in the Wigner function approach.
\subsection{Bjorken flow and the ideal transverse MHD}\label{sec2c}
The above expressions for the thermodynamic quantities are, in particular, functions of the temperature $T$ and the magnetic field $eB$. Assuming thermodynamic equilibrium with a fixed temperature, the $T$ dependence of the corresponding integrals to $\epsilon, p_{\|},p_{\perp},\cdots$ can be numerically determined for a  fixed $eB$ and  fermion mass $m$. The goal of the present paper is, however, to determine the proper time dependence of these thermodynamic quantities in a \textit{uniformly expanding ideal plasma}, where, in particular, the temperature depends on the space and time. In this case, because of their dependence on $T$, all the other thermodynamic quantities become also spacetime dependent. Hence, to determine their dynamics, it is enough to determine the evolution of $T$ for a fixed value of $eB$.
\par
To do this, we use the ideal relativistic MHD. This turns out to be a useful tool to describe the dynamics of an ideal and locally equilibrated fluid in the presence of electromagnetic fields. Let us thus assume our magnetized plasma to be in local thermal equilibrium. Assuming furthermore its electric conductivity to be infinitely large, we can neglect the electric field, and concentrate only on the presence of a magnetic field. As it is described in \cite{rischke-MHD, shokri-MHD,tabatabaee2019}, relativistic MHD is described by a set of equations consisting of the energy-momentum conservation,
\begin{eqnarray}\label{N21}
\partial_{\mu}T^{\mu\nu}=0,
\end{eqnarray}
the homogeneous and inhomogeneous Maxwell equations,
\begin{eqnarray}\label{N22}
\partial_{\mu}\tilde{F}^{\mu\nu}=0, \qquad \partial_{\mu}F^{\mu\nu}=J^{\nu},
\end{eqnarray}
with the current
\begin{eqnarray}\label{N23}
J^{\nu}=\rho_{e}u^{\nu}+\partial_{\mu}M^{\mu\nu}.
\end{eqnarray}
The total energy-momentum $T^{\mu\nu}$ in \eqref{N21} consists of two parts
\begin{eqnarray}\label{N24}
T^{\mu\nu}=T_{f}^{\mu\nu}+T_{\text{em}}^{\mu\nu}.
\end{eqnarray}
In an infinitely conducting fluid with vanishing electric field, the fluid part of $T^{\mu\nu}$  is given by \cite{shokri-MHD,tabatabaee2019}
\begin{eqnarray}\label{N25}
T_{f}^{\mu\nu}=\epsilon u^{\mu}u^{\nu}-p_{\perp}\Xi_{B}^{\mu\nu}+p_{\|}b^{\mu}b^{\nu}.
\end{eqnarray}
Here,  $u^{\mu}=\gamma(1,\bs{v})$ is the fluid four-velocity arising from $u^{\mu}=dx^{\mu}/d\tau$ with the proper time $\tau\equiv\sqrt{ t^{2}-\bs{x}^{2}}$ and $\gamma$ the Lorentz factor. For the metric $g^{\mu\nu}=\text{diag}\left(1,-1,-1,-1\right)$, it satisfies $u_{\mu}u^{\mu}=1$. The transverse projector $\Xi_{B}^{\mu\nu}$ in \eqref{N25} is given by $\Xi_{B}^{\mu\nu}\equiv g^{\mu\nu}-u^{\mu}u^{\nu}+b^{\mu}b^{\nu}$ with $b^{\mu}\equiv B^{\mu}/B$ satisfying $b_{\mu}b^{\mu}=-1$. The latter arises from the definition of $B^{\mu}$ in terms of the field strength tensor $F^{\mu\nu}$, $B^{\mu}\equiv \frac{1}{2}\epsilon^{\mu\nu\alpha\beta}F_{\nu\alpha}u_{\beta}$, leading to $B^{\mu}=(0,\bs{B})$ in the local rest frame of the fluid $u^{\mu}=(1,\bs{0})$. As concerns the electromagnetic part of $T^{\mu\nu}$, it is given by the standard Maxwell tensor \cite{rischke-MHD,shokri-MHD},
\begin{eqnarray}\label{N26}
T_{\text{em}}^{\mu\nu}=F^{\mu\lambda}F_{\lambda}^{~\nu}+\frac{1}{4}g^{\mu\nu}F^{\rho\sigma}F_{\rho\sigma}.
\end{eqnarray}
Using $F^{\mu\nu}=-Bb^{\mu\nu}$ with $b^{\mu\nu}\equiv \epsilon^{\mu\nu\alpha\beta}b_{\alpha}u_{\beta}$, $T^{\mu\nu}_{\text{em}}$ is equivalently given by
\begin{eqnarray}\label{N27}
T^{\mu\nu}_{\text{em}}=\frac{1}{2}B^{2}\left(u^{\mu}u^{\nu}-\Xi_{B}^{\mu\nu}-b^{\mu}b^{\nu}\right).
\end{eqnarray}
In \eqref{N22}, the dual field strength tensor is given by
\begin{eqnarray}\label{N28}
\tilde{F}^{\mu\nu}=B^{\mu}u^{\nu}-B^{\nu}u^{\mu}.
\end{eqnarray}
Moreover, the magnetization tensor $M^{\mu\nu}$ in \eqref{N23} is given by $M^{\mu\nu}\equiv -Mb^{\mu\nu}$.
Contracting \eqref{E21} with $u_{\nu}$ and $\Delta_{\alpha\nu}=g_{\alpha\nu}-u_{\alpha}u_{\nu}$, and assuming that the fluid velocity $\bs{v}$ is perpendicular to the direction of the magnetic field $\bs{B}$, we arrive at the energy and Euler equations in transverse ideal MHD, respectively,
\begin{eqnarray}\label{N29}
D\epsilon+\theta(\epsilon+p_{\perp})=0,
\end{eqnarray}
and
\begin{eqnarray}\label{N30}
\hspace{-0.5cm}\left(\epsilon+p_{\perp}+B^{2}\right)Du_{\rho}-\nabla_{\rho}\left(p_{\perp}+\frac{1}{2}B^{2}\right)=0.
\end{eqnarray}
Here, with $p_{\perp}=p_{\|}-BM$, $D\equiv u^{\mu}\partial_{\mu}$, $\theta\equiv \partial_{\rho}u^{\rho}$, and $\nabla^{\mu}\equiv \Delta^{\mu\nu}\partial_{\nu}$ with $\Delta^{\mu\nu}\equiv g^{\mu\nu}-u^{\mu}u^{\nu}$.
For the plasma under consideration, we make a number of other assumptions. First, we assume that the magnetic field $\bs{B}$ is aligned in the $z$-direction, and that the  fluid velocity $\bs{v}$, being perpendicular to the direction of the magnetic field, is directed in the $y$ direction. Then, assuming
\begin{enumerate}
\item translational invariance in the $x$-$z$ plane,
\item a uniform expansion of the fluid in the longitudinal $y$ direction, and
\item boost invariance along the $y$ direction,
\item boost invariance of $\epsilon, p$ and $B$,
\end{enumerate}
and replacing $v_{y}$ in $u^{\mu}=\gamma(1,0,v_y,0)$ with $v_y=y/t$, we arrive at the Bjorken flow \cite{bjorken,hatsuda-book}
\begin{eqnarray}\label{N31}
u^{\mu}=\left(\cosh\eta,0,\sinh\eta,0\right).
\end{eqnarray}
Here $\eta\equiv \frac{1}{2}\ln\frac{t+y}{t-y}$ is the boost variable in the Milne coordinates. In these coordinates, and for a plasma expanding in the $y$-direction, the proper time reads $\tau=\sqrt{t^{2}-y^{2}}$. Using this parametrization, we also obtain
\begin{eqnarray}\label{N32}
\frac{\partial}{\partial t}&=&+\cosh\eta\frac{\partial}{\partial\tau}-\frac{1}{\tau}\sinh\eta\frac{\partial}{\partial\eta}, \nonumber\\
\frac{\partial}{\partial y}&=&-\sinh\eta\frac{\partial}{\partial\tau}+\frac{1}{\tau}\cosh\eta\frac{\partial}{\partial\eta}.
\end{eqnarray}
In these coordinates, the energy equation \eqref{N29} is a first order differential equation in $\tau$,
\begin{eqnarray}\label{N33}
\frac{\partial\epsilon}{\partial \tau}+\frac{1}{\tau}\left(\epsilon+p_{\|}-BM\right)=0.
\end{eqnarray}
In Sec. \ref{sec3b}, we determine the thermodynamic quantities $p_{\|},p_{\perp}$, and $\epsilon$ as functions of $T$ in an integral form, and show that for massless fermions they coincide with those presented in \eqref{N15}. Plugging the corresponding expressions to these quantities into the energy equation  \eqref{N33} of the ideal relativistic MHD, it becomes a first order differential equation for $T$. In Sec. \ref{sec4}, we numerically solve this equation for a fixed value of $B$, and determine the proper time evolution of the temperature $T$. Plugging this numerical result back into $p_{\|},p_{\perp}$, and $\epsilon$, the proper time evolution of these and a number of other thermodynamic quantities are determined. The numerical results arisen in this approach is then compared with the $1+1$ Bjorken solution \cite{hatsuda-book}
\begin{eqnarray}\label{N34}
T=T_0\left(\frac{\tau_0}{\tau}\right)^{1/\kappa},
\end{eqnarray}
with $T_{0}$ the temperature at the initial time $\tau_0$, and $\kappa=3$ appearing in the  equation of state $\epsilon=\kappa p$ of an ideal gas. The coefficient $\kappa$ is related to the sound velocity $c_s$ by $\kappa=c_s^{-2}$.
Generalizing this result to a $3+1$ dimensional flow, which describes a fluid expanding in both transverse and longitudinal directions, we get
\begin{eqnarray}\label{N35}
T=T_0\left(\frac{\tau_0}{\tau}\right).
\end{eqnarray}
Let us notice that this result is only valid at $r=0$, with $r$ the distance to the center of the collision. In general, it is to be replaced by $T=T_0\left(\frac{\rho_0}{\rho}\right)$, where $\rho\equiv \sqrt{\tau^2-\bs{r}^{2}}$ and $T_{0}$ the temperature at $\rho_0$.
In Sec. \ref{sec4}, we refer to this solution as the Hubble solution at $r=0$.
\par
Another useful comparison is made with the temperature arisen from the $3+1$ dimensional Gubser flow \cite{gubser2010},\footnote{The result originally presented by S. Gubser in \cite{gubser2010} has a different form. In Appendix \ref{appA}, we show that \eqref{N36} is equivalent with $T$ from \cite{gubser2010}.} which describes a fluid expanding not only in the transverse $y$-direction but also in the $x$-$z$ plane,
\begin{eqnarray}\label{N36}
T=T_0\left(\frac{\tau_0}{\tau}\right)^{1/\kappa}\mathcal{G}(r,r_;\tau,\tau_0),
\end{eqnarray}
with
\begin{eqnarray}\label{N37}
\mathcal{G}(r,r_;\tau,\tau_0)\equiv \left(\frac{1+q^4(\tau_{0}^2-r_0^2)^{4}+2q^{2}(\tau_{0}^2+r_0^2)}{1+q^4(\tau^2-r^2)^2+2q^2(\tau^2+r^2)}\right). \nonumber\\
\end{eqnarray}
The factor $q$ that characterizes the Gubser temperature is typically given by $q=1/4.3$ fm$^{-1}$. The parameter $r$ is the radial distance from the origin in the $x$-$z$ plane, and $T_0$ is the initial temperature at $\tau_0$ and $r_0$. Moreover, in \eqref{N37}, ${\cal{G}}(r_0,\tau_0;r_0,\tau_0)=1$.
\subsection{Proper time evolution of the magnetic field in an expanding relativistic plasma}\label{sec2d}
\begin{figure*}[hbt]
\includegraphics[width=8cm,height=6cm]{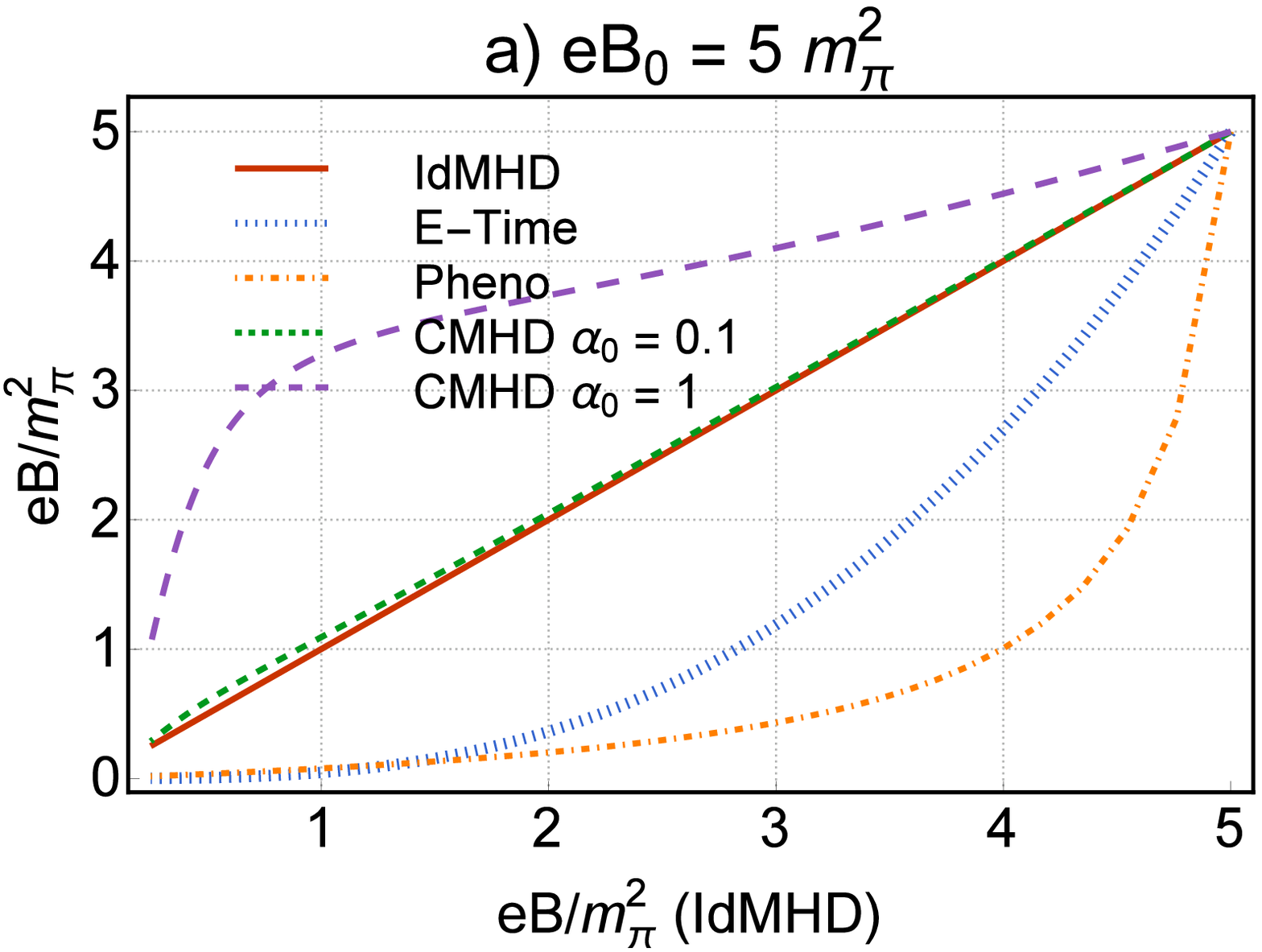}
\includegraphics[width=8cm,height=6cm]{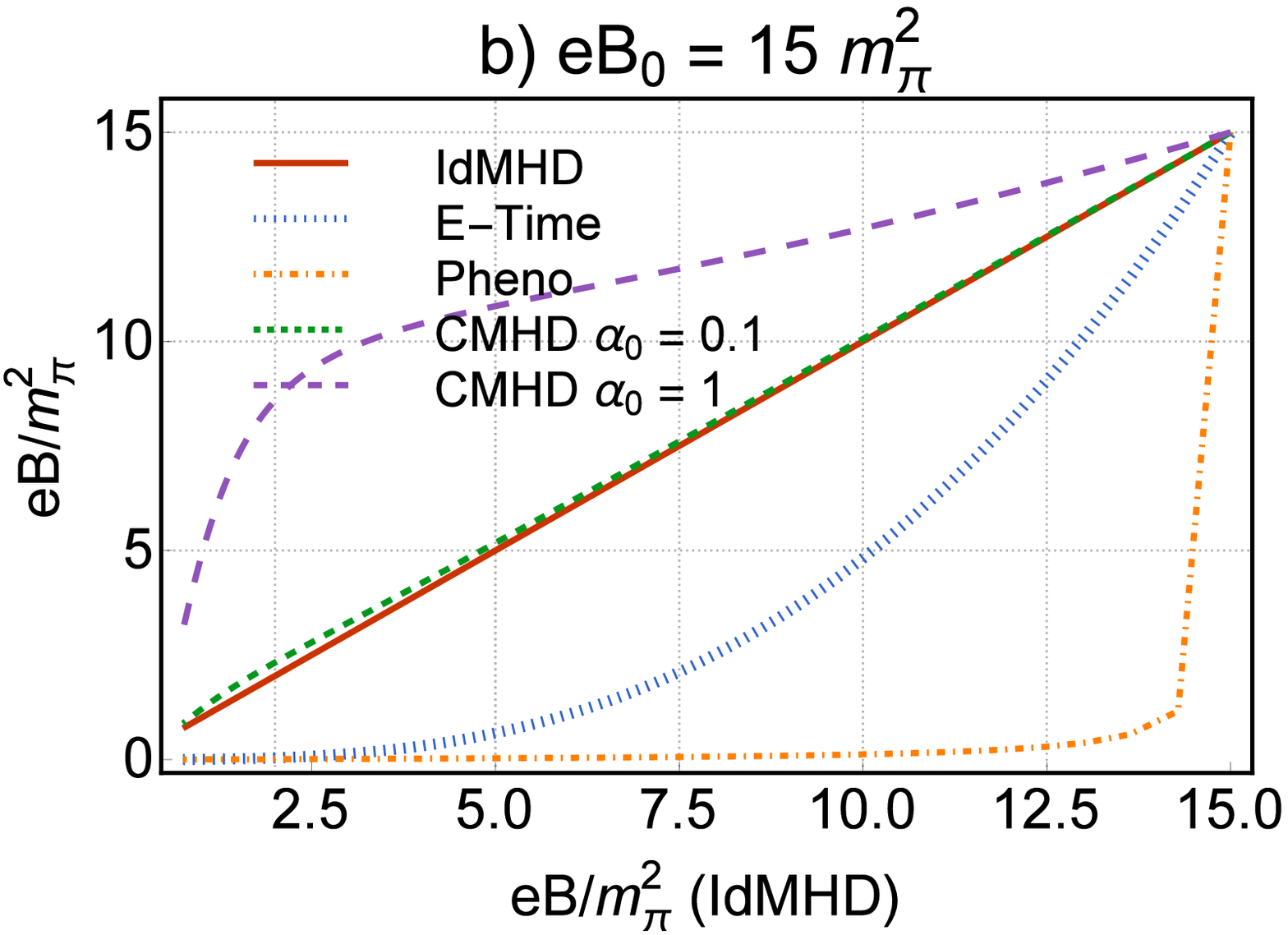}
\caption{(color online). Comparison of the $\tau$ dependence of $eB/m_{\pi}^{2}$ arising from E-time (dashed blue), Pheno (dotted-dashed orange), and CMHD for $\alpha_0=0.1$ (solid red) and $\alpha_0=1$ (dashed magenta) with the $\tau$ dependence of $eB/m_{\pi}^{2}$ arising from IdMHD for the initial magnetic field a) $eB_0=5m_\pi^{2}$ and b) $eB_0=15m_{\pi}^{2}$. As it is shown, the magnetic field arising from CMHD for $\alpha_{0}=0.1$ is very similar to the Bjorken expansion. Later, we will show that despite the difference between these curves, the temperature $T^{\star}$ arising from these magnetic fields are similar (see Sec. \ref{sec4}).}\label{fig-1}
\end{figure*}
In this paper, for the sake of simplicity, we consider a plasma consisting of only one flavor with electric charge $q=+1$. It is not difficult to generalize our results to the more realistic case of an expanding plasma of quarks and gluons consisting of various quark flavors. This is the plasma that is created in the early stages of heavy-ion collisions. Very strong and short-living magnetic fields are also believed to be generated in the same stage of these collisions. As it turns out, after a few fm/c, they become too weak to be able to affect the dynamics of the QGP. The proper time dependence of the electromagnetic field created in HICs was indeed the subject of intensive theoretical and experimental studies in the last few years (see e.g. \cite{warringa2007, skokov2009, huang2015, shokri-MHD, inghirami2019}). The question about the effect of spacetime dependent magnetic fields on the evolution of thermodynamic quantities is still open. One of the main purposes of this paper is to answer this question. The Ritus eigenfunction method, described in Sec. \ref{sec2a} is however only valid for constant magnetic fields. Consequently, the proper time dependence of the temperature and other thermodynamic quantities are only valid for constant magnetic fields.
\par
In Sec. \ref{sec4}, we show that it is possible to generalize the results arisen from our numerical approach for the case of a constant magnetic field to the case of a spacetime dependent magnetic field in an expanding plasma. To do this, we divide the time interval under study into a large number of infinitesimal time slices.
We then assume the magnetic field to be constant in each infinitesimal time slice. Solving the aforementioned differential equation in this time slice, and then varying the magnetic field according to a given phenomenological or theoretical ansatz, we repeat the same computation for the next time slice. Repeating this procedure for all infinitesimal time interval, we arrive at the proper time dependence of the temperature for a point to point varying magnetic field (see Sec. \ref{sec4} for more details). In what follows, we present a number of solutions for the spacetime dependence of the magnetic field.
\par
As it is argued in \cite{huang2015}, the early time dynamics of the magnetic field is described by
\begin{eqnarray}\label{N38}
eB(\tau)=\frac{eB(0)}{\left(1+\tau^{2}/t_{B}^{2}\right)^{3/2}},
\end{eqnarray}
where $eB(0)$ is some constant initial value for the magnetic field at $\tau=0$ fm/c and $t_{B}\sim 0.065$ fm/c at RHIC top energies. In Sec. \ref{sec4}, we refer to this solution as the early time (E-time) solution. Another phenomenological expression, that describes the evolution of the magnetic field in near-central collisions is
\begin{eqnarray}\label{N39}
eB(r,\tau)=\frac{m_{\pi}^{2}}{a+b \tau}\exp\left(-\frac{r^{2}}{\sigma_{r}^{2}}\right).
\end{eqnarray}
Here, $r$ is the distance to the center of the collision, $m_\pi\sim 140$ MeV is the pion mass, $a$ and $b$ are two phenomenological parameters given by $a=78.2658$ and $b=79.5457$ fm$^{-1}$, and $\sigma_r=3.5$ fm for a zero impact parameter \cite{pheno-B}. In Sec.\ref{sec4}, we refer to this solution as the phenomenological Gaussian (Pheno) solution.
\par
Apart from these phenomenological expressions, the evolution of the electromagnetic fields is derived from relativistic MHD. These kinds of derivation are based on a number of assumptions, including those listed in Sec. \ref{sec2c}. In \cite{rischke-MHD, shokri-MHD}, it is also assumed that the magnetic field is perpendicular to its velocity, $\bs{v}$. As it turns out, the expansion of the QGP plays a crucial role in the proper time evolution of the magnetic field. In the simplest case of ideal MHD described by $1+1$ dimensional transverse MHD, and based on the Bjorken flow \eqref{N31}, the evolution of $B$ is given by
\begin{eqnarray}\label{N40}
B=B_{0}\left(\frac{\tau_0}{\tau}\right),
\end{eqnarray}
where $B_0$ is the magnetic field strength at the initial time $\tau_0$. In \cite{shokri2018}, we extended this ideal MHD result to a $3+1$ dimensional Gubser flow \cite{gubser2010}. To do this we solve the MHD equation in a conformally flat dS$^3\times$ E$^{1}$ spacetime. Performing an appropriate Weyl transformation back into the Minkowski spacetime, we arrive at
 \begin{eqnarray}\label{N41}
B(r,\tau)=B_0{\mathcal{H}}(r,r_0;\tau,\tau_0),
\end{eqnarray}
with
\begin{eqnarray}\label{N42}
{\mathcal{H}}(r,r_0;\tau,\tau_0)\equiv \left(\frac{
\frac{1}{r^2 \tau^2}+\frac{\alpha_0^2}{r_0^2\tau_0^2}\mathcal{G}\left(r,r_0;\tau,\tau_0\right)}{\frac{1+\alpha_0^2}{r_0 \tau_0}}\right)^{1/2}.\nonumber\\
\end{eqnarray}
Here, similar to ${\cal{G}}(r,r_0;\tau,\tau_0)$ from \eqref{N37}, ${\cal{H}}(r_0,\tau_0;r_0,\tau_0)=1$. Moreover, according to \cite{shokri2018}, the factor $\alpha_{0}$ is the ratio of the longitudinal and transverse components of the magnetic field $\alpha\equiv B_y/B_x$\footnote{In the more realistic case of relativistic HIC, $\alpha_{0}$ is the cotangent of the angle between $\bs{B}$ and the beamline.} In Sec. \ref{sec4}, we refer to this solution as the conformal MHD (CMHD) solution. In Fig. \ref{fig-1}, a comparison is made between the $\tau$ dependence of the most simple IdMHD solution \eqref{N40} and the other solutions introduced in this section, including the E-time, Pheno and CMHD solutions for $\alpha_{0}=0.1$ and $\alpha_0=1$ from \eqref{N38}, \eqref{N39}, and \eqref{N41} in the first few femtoseconds after the collision, $\Delta\tau=4$ fm/c. Assuming the background magnetic field to be $\tau$ dependent, it is possible to determine the $\tau$ dependence of the temperature following the procedure described before.  It turns out that despite obvious difference between the $\tau$ dependence of various solutions for $B(\tau)$, the temperature arisen from these solutions are similar (see Sec. \ref{sec4} for more details).
\section{Thermodynamic quantities in an expanding magnetized plasma: The Wigner function approach}\label{sec3}
\setcounter{equation}{0}
In this section, we first derive in Sec. \ref{sec3a1} the Wigner function of a fermionic system consisting of a positive charge ($q=+1$) in the presence of a constant magnetic field using the quantization given in \eqref{N10}.
Using this result, we then derive the corresponding energy-momentum tensor to this system in Sec. \ref{sec3a2}. In Sec. \ref{sec3b}, we present the corresponding expressions to a number of thermodynamic quantities including, among others, $p_{\|}, p_{\perp}$, and $\epsilon$, and compare the resulting expressions, all in integral forms, with the results presented in \eqref{N15}, \eqref{N19} and \eqref{N20}. To evaluate these integrals in high  and low-temperature approximations in Sec. \ref{sec3b}, and present separately the analytical results corresponding to the massive and massless case.
\subsection{The Wigner function, the energy-momentum tensor, and the resulting thermodynamic quantities}\label{sec3a}
\subsubsection{The Wigner function}\label{sec3a1}
In quantum statistical mechanics, the Wigner function $W(r,k)$ is given by the thermal average of the normal ordered Wigner operator $\hat{W}$ as \cite{hatsuda-book},
\begin{eqnarray}\label{E1}
W_{\alpha\beta}=\langle : \hat{W}_{\alpha\beta}:\rangle,
\end{eqnarray}
where
\begin{eqnarray}\label{E2}
\lefteqn{\hat{W}_{\alpha\beta}(r,k)}\nonumber\\
&&=\int d^{4}r' e^{-ik\cdot r'}\bar{\psi}_{\beta}\left(r+\frac{r'}{2}\right)U\left(+,-\right)\psi_{\alpha}\left(r-\frac{r'}{2}\right). \nonumber\\
\end{eqnarray}
Here, $\psi$ and $\bar{\psi}$ are fermionic field operators, and $U(+,-)$, defined by
\begin{eqnarray}\label{E3}
U(+,-)=\exp\left(+ie r^{\prime \mu}\int_{0}^{1}ds A_{\mu}(r-\frac{r'}{2}+sr')\right), \nonumber\\
\end{eqnarray}
is the Wilson line that guarantees the gauge invariance of the product $\bar{\psi}\psi$, once the point-splitting is performed. The latter is characterized by $r^{\prime}$. In \eqref{E2} and \eqref{E3}, $r^{\mu}=(t,x,y,z)$ and $r'^{\mu}=(t',x',y',z')$. In the presence of a constant magnetic field, aligned in the $z$-direction, i.e. for $A^{\mu}=Bx\bs{e}_{y}$, $U(+,-)$ reads
\begin{eqnarray}\label{E4}
U(+,-)=e^{ieBxy'}.
\end{eqnarray}
Plugging, apart from $U(+,-)$ from \eqref{E4}, the quantization \eqref{N10} of the field operators $\psi$ and $\bar{\psi}$ into \eqref{E2}, and using \eqref{N12}
as well as
\begin{eqnarray}\label{E5}
\lefteqn{\hspace{-1cm}\langle a^{\dagger n',s'}_{\bar{\bs{p}}^{\prime}}
a^{n,s}_{\bar{\bs{p}}}\rangle=
\langle b^{\dagger n',s'}_{\bar{\bs{p}}^{\prime}}
b^{n,s}_{\bar{\bs{p}}}\rangle
}\nonumber\\
&=&
(2\pi)^{2}V\delta^{2}(\bar{\boldsymbol{p}}'-\bar{\boldsymbol{p}})\delta_{nn'}\delta_{ss'}f_{\text{F}}(E_{n},p_z),
\end{eqnarray}
with $\bar{\bs{p}}=(0,p_y,p_z)$ and $f_{\text{F}}(E_n,p_z)$ defined in (\ref{N16}), we arrive first at
\begin{eqnarray}\label{E6}
\lefteqn{
W_{\alpha\beta}(r,k)=\sum_{n=0}\int \frac{d^{4}r' d^{2}\bar{p}}{(2\pi)^{2}}\frac{e^{-i(k\cdot r'-eB xy')}}{2E_{n}}f_{\text{F}}(E_{n},p_z)
}\nonumber\\
&&\times\bigg\{
[\mathbb{P}_{n}^{(+)}(\zeta_-)]_{\alpha\sigma}\left(\gamma\cdot \tilde{p}_{+}+m\right)_{\sigma\rho}[\mathbb{P}_{n}^{(+)}(\zeta_+)]_{\rho\beta}\nonumber\\
&&\times e^{i(E_{n}t'-p_{y}y'-p_{z}z')}
\nonumber\\
&&-
[\mathbb{P}_{n}^{(+)}(\bar{\zeta}_-)]_{\alpha\sigma}\left(\gamma\cdot \tilde{p}_{-}-m\right)_{\sigma\rho}[\mathbb{P}_{n}^{(+)}(\bar{\zeta}_+)]_{\rho\beta}\nonumber\\
&&\times e^{-i(E_{n}t'-p_{y}y'-p_{z}z')}
\bigg\}.
\end{eqnarray}
Here, $\zeta_{\pm}\equiv\xi^{+}\pm\frac{x'}{2\ell_b}$ and $\bar{\zeta}_{\pm}\equiv \xi^{-}\pm\frac{x'}{2\ell_b}$ with $\xi^{\kappa}=\xi_{+}^{\kappa}$, $\kappa=\pm$ for positively charged particles, as defined in Sec. \ref{sec2a}.  Moreover, the projector $\mathbb{P}_{n}^{(n)}$ is in general defined by  $\mathbb{P}_{n}^{(+)}=P_{+}f_n^{+}+\Pi_{n}P_{-}f_n^{-}$ from \eqref{N6}. Plugging $f_n^{\pm}$ from \eqref{N8} into \eqref{E6}, performing the integration over $r'^{\mu}=(t'x',y',z')$ by making, in particular, use of
\begin{widetext}
\begin{eqnarray}\label{E7}
\int dx e^{-x^2}H_m(x+y)H_n(x+z)=2^n\sqrt{\pi}m!z^{n-m}L_m^{n-m}(-2yz), \quad n\geq m,
\end{eqnarray}
where $L_{m}^{k}(u)$ is the associated Laguerre polynomial,
and eventually performing the integral over $p_{y}$ and $p_z$, we arrive after a lengthy but straightforward computation at the Wigner function for a hot and magnetized QED plasma,
\begin{eqnarray}\label{E8}
\lefteqn{
W(r,k)=}\nonumber\\
&&=\frac{f_{\text{F}}(E_{0},k_{z})}{E_{0}}e^{-\frac{\boldsymbol{k}_{\perp}^{2}}{eB}}L_{0}\left(\frac{2\boldsymbol{k}_{\perp}^{2}}{eB}\right)\bigg[\delta(k_{0}-E_{0})(\gamma^{\|}\cdot \tilde{p}_{\|,+}^{(0)}(k_{z})+m)-\delta(k_{0}+E_{0})(\gamma^{\|}\cdot \tilde{p}_{\|,-}^{(0)}(-k_{z})-m)\bigg]P_{+}\nonumber\\
&&+e^{-\frac{\boldsymbol{k}_{\perp}^{2}}{eB}}
\sum_{n=1}\frac{f_{\text{F}}(E_{n},k_{z})}{E_{n}}(-1)^{n} L_{n}\left(\frac{2\boldsymbol{k}_{\perp}^{2}}{eB}\right)
\bigg[\delta(k_{0}-E_{n})(\gamma^{\|}\cdot \tilde{p}_{\|,+}^{(n)}(k_{z})+m)-\delta(k_{0}+E_{n})(\gamma^{\|}\cdot \tilde{p}_{\|,-}^{(n)}(-k_{z})-m)\bigg]P_{+}\nonumber\\
&&-e^{-\frac{\boldsymbol{k}_{\perp}^{2}}{eB}}
\sum_{n=1}\frac{f_{\text{F}}(E_{n},k_{z})}{E_{n}}(-1)^{n} L_{n-1}\left(\frac{2\boldsymbol{k}_{\perp}^{2}}{eB}\right)
\bigg[\delta(k_{0}-E_{n})(\gamma^{\|}\cdot \tilde{p}_{\|,+}^{(n)}(k_{z})+m)-\delta(k_{0}+E_{n})(\gamma^{\|}\cdot \tilde{p}_{\|,-}^{(n)}(-k_{z})-m)\bigg]P_{-}\nonumber\\
&&+e^{-\frac{\boldsymbol{k}_{\perp}^{2}}{eB}}
\sum_{n=1}\frac{f_{\text{F}}(E_{n},k_{z})}{E_{n}}(-1)^{n} L_{n-1}^{1}\left(\frac{2\boldsymbol{k}_{\perp}^{2}}{eB}\right)
\nonumber\\
&&\times
\sqrt{\frac{2}{n}}
\left\{\frac{(k_{y}+ik_{x})}{\sqrt{eB}}\bigg[\delta(k_{0}-E_{n})(\gamma^{\perp}\cdot \tilde{p}_{\perp,+}^{(n)}(k_{z}))-\delta(k_{0}+E_{n})(\gamma^{\perp}\cdot \tilde{p}_{\perp,-}^{(n)}(-k_{z}))\bigg]P_{-}\right.\nonumber\\
&&\hspace{1cm}\left.+\frac{(k_{y}-ik_{x})}{\sqrt{eB}}\bigg[\delta(k_{0}-E_{n})(\gamma^{\perp}\cdot \tilde{p}_{\perp,+}^{(n)}(k_{z}))-\delta(k_{0}+E_{n})(\gamma^{\perp}\cdot \tilde{p}_{\perp,-}^{(n)}(-k_{z}))\bigg]P_{+}
\right\}.
\end{eqnarray}
\end{widetext}
Here, $\bs{k}_{\perp}\equiv (0,k_x,k_y,0)$, $\gamma^{\|}\equiv (\gamma^{0},0,0,\gamma^{3})$, $\gamma^{\perp}\equiv (0,\gamma^{1},\gamma^{2},0)$, $\tilde{p}_{\pm}^{\mu(n)}(k_z)\equiv (E_n, 0,\mp\sqrt{2neB},k_z)$, $\tilde{p}_{\|,\pm}^{\mu(n)}(k_z)\equiv (E_n, 0,0,k_z)$ with  $E_n=E_n^{(+)}$ defined in \eqref{N5}, $\tilde{p}_{\perp,\pm}^{\mu(n)}(k_z)\equiv (0, 0,\mp\sqrt{2neB},0)$, and   $L_{m}(u)\equiv L_{m}^{0}(u)$. To simplify the combinations $P_{\pm}(\gamma\cdot\tilde{p}_{\pm}^{(n)}\pm m)P_{\pm}$ or $P_{\pm}(\gamma\cdot\tilde{p}_{\pm}^{(n)}\pm m)P_{\mp}$, we used
$
P_{\pm}\gamma^{\mu_\|}=\gamma^{\mu_{\|}}P_{\pm}$, and  $P_{\pm}\gamma^{\mu_{\perp}}=\gamma^{\mu_{\perp}}P_{\mp}$, and
$P_{\pm}^{2}=P_{\pm}$ as well as $P_{\pm}P_{\mp}=0$ \cite{taghinavaz2015}. Let us notice that the expression on the first line of \eqref{E8} including $L_0(u)=1$ is the contribution from the lowest Landau level (LLL) while the remaining terms are those from higher Landau levels (HLL). In what follows, we use the above Wigner function to determine the energy-momentum tensor of this hot and magnetized QED plasma.
\subsubsection{The energy-momentum tensor and thermodynamic quantities $p_{\|}, p_{\perp}$, and $\epsilon$ }\label{sec3a2}
The energy-momentum tensor of a fermionic system is given by \cite{blaschke2016}
\begin{eqnarray}\label{E9}
\mathcal{T}^{\mu\nu}&=&\frac{i}{4}\bar{\psi}\left(\gamma^{\mu}\overrightarrow{D}^{\nu}+\gamma^{\nu}
\overrightarrow{D}^{\mu}-\gamma^{\mu}\overleftarrow{D}^{\nu}-\gamma^{\nu}
\overleftarrow{D}^{\mu}\right)\psi\nonumber\\
&&-g^{\mu\nu}{\cal{L}},
\end{eqnarray}
with the Lagrangian density
\begin{eqnarray}\label{E10}
\hspace{-0.5cm}{\cal{L}}=\frac{1}{2}\bar{\psi}\left(i\gamma^{\mu}\overrightarrow{D}_{\mu}-m\right)\psi-\frac{1}{2}\bar{\psi}\left(i\gamma^{\mu}\overleftarrow{D}_{\mu}-m\right)\psi.\nonumber\\
\end{eqnarray}
Here, $\overrightarrow{D}_{\mu}=\overrightarrow{\partial}_{\mu}-ieA_{\mu}$ and $\overleftarrow{D}_{\mu}=\overleftarrow{\partial}_{\mu}+ieA_{\mu}$ denote the right ($\rightarrow$) and left ($\leftarrow$) covariant derivatives.
As it turns out, the thermal average of this operator leads to
\begin{eqnarray}\label{E11}
\lefteqn{T^{\mu\nu}=\langle:\mathcal{T}^{\mu\nu}:\rangle}\nonumber\\
&=&\bigg[\frac{1}{2}\left(\delta_{~\lambda}^{\mu}\delta^{\nu}_{~\sigma}+\delta^{\nu}_{~\lambda}
\delta^{\mu}_{~\sigma}\right)-g^{\mu\nu}g_{\lambda\sigma}\bigg]\int\frac{d^{4}k}{(2\pi)^{4}}k^{\sigma}\mbox{tr}\left(\gamma^{\lambda}W\right).\nonumber\\
\end{eqnarray}
Plugging the Wigner function \eqref{E8} into \eqref{E11}, and using
\begin{widetext}
\begin{eqnarray}\label{E12}
\lefteqn{\hspace{-2cm}
\mbox{tr}\left(\gamma^{\lambda}[\delta(k_{0}-E_{n})(\gamma^{\|}\cdot \tilde{p}_{\|,+}^{(n)}(k_{z})+m)-\delta(k_{0}+E_{n})(\gamma^{\|}\cdot \tilde{p}_{\|,-}^{(n)}(-k_{z})-m)]P_{\pm}\right)}\nonumber\\
&&=2\bigg[\delta(k_0-E_{n})\left(g^{\lambda 0}E_n- g^{\lambda 3}k_{z}\right)-\delta(k_0+E_n)\left(g^{\lambda 0}E_n+g^{\lambda 3}k_{z}\right)\bigg],\nonumber\\
\lefteqn{\hspace{-2cm}
\mbox{tr}\left(\gamma^{\lambda}[\delta(k_{0}-E_{n})(\gamma^{\perp}\cdot \tilde{p}_{\perp,+}^{(n)}(k_{z}))-\delta(k_{0}+E_{n})(\gamma^{\perp}\cdot \tilde{p}_{\perp,-}^{(n)}(-k_{z}))]P_{\mp}\right)}\nonumber\\
&&=2\bigg[\delta(k_{0}-E_{n})\left(g^{\lambda 2}\mp ig^{\lambda 1}\right)+\delta(k_{0}+E_{n})\left(g^{\lambda 2}\mp ig^{\lambda 1}\right)\bigg]\sqrt{2neB},
\end{eqnarray}
to perform the trace over Dirac matrices, we arrive after some works first at
\begin{eqnarray}\label{E13}
T^{\mu\nu}&=&4\int_{-\infty}^{+\infty}\frac{dk_z}{2\pi}\frac{f_{\text{F}}(E_0,k_z)}{E_{0}}\bigg\{\bigg[\frac{1}{2}\left(g^{\mu 0}\delta^{\nu 0}+\delta^{\mu 0}g^{\nu 0}\right)-g^{\mu\nu}\bigg]E_0^{2}-\bigg[\frac{1}{2}\left(g^{\mu 3}\delta^{\nu 3}+\delta^{\mu 3}g^{\nu 3}\right)-g^{\mu\nu}\bigg]k_{z}^{2}\bigg\}\int _{-\infty}^{+\infty}\frac{d^2 k_{\perp}}{(2\pi)^{2}}e^{-\frac{\bs{k}^{2}_{\perp}}{eB}}
\nonumber\\
&&+4\sum_{n=1}^{\infty}(-1)^{n}\int_{-\infty}^{+\infty}\frac{dk_{z}}{2\pi}\frac{f_{\text{F}}(E_n,k_z)}{E_n}\bigg\{\bigg[\frac{1}{2}\left(g^{\mu 0}\delta^{\nu 0}+\delta^{\mu 0}g^{\nu 0}\right)-g^{\mu\nu}\bigg]E_n^{2}-\bigg[\frac{1}{2}\left(g^{\mu 3}\delta^{\nu 3}+\delta^{\mu 3}g^{\nu 3}\right)-g^{\mu\nu}\bigg]k_{z}^{2}\bigg\}\nonumber\\
&&\quad\times\int_{-\infty}^{+\infty}\frac{d^{2}k_{\perp}}{(2\pi)^{2}}e^{-\frac{\boldsymbol{k}^{2}_{\perp}}{eB}}\bigg[L_{n}\left(\frac{2\boldsymbol{k}^{2}_{\perp}}{eB}\right)-L_{n-1}\left(\frac{2\boldsymbol{k}^{2}_{\perp}}{eB}\right)\bigg]\nonumber\\
&&+8\sum_{n=1}^{\infty}(-1)^{n}\int_{-\infty}^{+\infty}\frac{dk_{z}}{2\pi}\frac{f_{\text{F}}(E_{n},k_z)}{E_{n}}\nonumber\\
&&\times\int_{-\infty}^{+\infty}\frac{d^{2}k_{\perp}}{(2\pi)^{2}}e^{-\frac{\boldsymbol{k}_{\perp}^{2}}{eB}}L_{n-1}^{1}\left(\frac{2\boldsymbol{k}^{2}_{\perp}}{eB}\right)\bigg[
k_{x}^{2}\left(
g^{\mu 1}\delta^{\nu 1}+\delta^{\mu 1}g^{\nu 1}\right)+
k_{y}^{2}\left(
g^{\mu 2}\delta^{\nu 2}+\delta^{\mu 2}g^{\nu 2}
\right)\bigg]\nonumber\\
&&-16g^{\mu\nu}\sum\limits_{n=1}^{\infty}(-1)^{n}\int_{-\infty}^{+\infty}\frac{dk_z}{2\pi}\frac{f_{\text{F}}(E_n,k_z)}{E_n}\int_{-\infty}^{+\infty}\frac{d^{2}k_{\perp}}{(2\pi)^{2}}\boldsymbol{k}_{\perp}^{2}e^{-\frac{\boldsymbol{k}^{2}_{\perp}}{eB}}L_{n-1}^{1}\left(\frac{2\boldsymbol{k}_{\perp}^{2}}{eB}\right).
\end{eqnarray}
\end{widetext}
To perform the integration over $\bs{k}_{\perp}$, we use
\begin{eqnarray}\label{E14}
\int _{-\infty}^{+\infty}\frac{d^2 k_{\perp}}{(2\pi)^{2}}e^{-\frac{\bs{k}^{2}_{\perp}}{eB}}=\frac{eB}{4\pi},
\end{eqnarray}
as well as
\begin{eqnarray}\label{E15}
\int_{-\infty}^{+\infty}\frac{d^{2}k_{\perp}}{(2\pi)^{2}}e^{-\frac{\boldsymbol{k}^{2}_{\perp}}{eB}}L_{n}\left(\frac{2\boldsymbol{k}^{2}_{\perp}}{eB}\right)&=&(-1)^{n}\frac{eB}{4\pi},\nonumber\\
\int_{-\infty}^{+\infty}\frac{d^2 k_{\perp}}{(2\pi)^{2}}e^{-\frac{\boldsymbol{k}^{2}_{\perp}}{eB}}
\boldsymbol{k}^{2}_{\perp} L_{n}^{1}\left(\frac{2\boldsymbol{k}^{2}_{\perp}}{eB}\right)&=&(-1)^{n}
\frac{(eB)^{2}(n+1)}{4\pi}. \nonumber\\
\end{eqnarray}
Moreover, we use symmetry arguments\footnote{Here, $k_i^2, i=x,y$ is first replaced by $\frac{1}{2}\bs{k}_{\perp}^{2}$, and then \eqref{N16} is used. } to perform the following integral for $i=x,y$:
\begin{eqnarray}\label{E16}
\int_{-\infty}^{+\infty}\frac{d^2 k_{\perp}}{(2\pi)^{2}}e^{-\frac{\boldsymbol{k}^{2}_{\perp}}{eB}}
k^{2}_{i} L_{n}^{1}\left(\frac{2\boldsymbol{k}^{2}_{\perp}}{eB}\right)&=&(-1)^{n}
\frac{(eB)^{2}(n+1)}{8\pi}.\nonumber\\
\end{eqnarray}
We finally arrive at
\begin{widetext}
\begin{eqnarray}\label{E17}
T^{\mu\nu}&=&\frac{m^{2}eB}{4\pi^2}\left(g^{\mu 0}\delta^{\nu 0}+\delta^{\mu 0}g^{\nu 0}-2g^{\mu\nu}\right)\sum_{n=0}^{\infty}\alpha_{n}\int_{-\infty}^{+\infty}\frac{dk_z}{E_n}~
f_{\text{F}}(E_n,k_z)\nonumber\\
&&+\frac{eB}{4\pi^{2}}\left(g^{\mu 0}\delta^{\nu 0}+\delta^{\mu 0}g^{\nu 0}-g^{\mu 3}\delta^{\nu 3}-\delta^{\mu 3}g^{\nu 3} \right)\sum_{n=0}^{\infty}\alpha_{n}\int_{-\infty}^{+\infty}\frac{dk_z}{E_n} k_z^2~f_{\text{F}}(E_n,k_z)\nonumber\\
&&+\frac{(eB)^{2}}{4\pi^{2}}\big[2\left(g^{\mu 0}\delta^{\nu 0}+\delta^{\mu 0}g^{\nu 0}\right)-\left(g^{\mu 1}\delta^{\nu 1}+\delta^{\mu 1}g^{\nu 1}+
g^{\mu 2}\delta^{\nu 2}+\delta^{\mu 2}g^{\nu 2}\right)\big]\sum_{n=0}^{\infty}n\alpha_{n}\int_{-\infty}^{+\infty}\frac{dk_z}{E_n} f_{\text{F}}(E_n,k_z).\nonumber\\
\end{eqnarray}
\end{widetext}
Here, $\alpha_{n}=2-\delta_{n, 0}$, as introduced previously in Sec. \ref{sec2b}. The energy-momentum tensor presented in \eqref{E17} is symmetric and diagonal. Its diagonal elements are given by
\begin{eqnarray}\label{E18}
T^{00}&=&\frac{eB}{\pi^{2}}\sum_{n=0}^{\infty}\alpha_n\int_{0}^{\infty}\frac{dk_z}{E_n} \left(E_n^2-m^2\right)f_{\text{F}}(E_n,k_z),\nonumber\\
T^{11}&=&\frac{eB}{\pi^{2}}\sum_{n=0}^{\infty}\alpha_n\int_{0}^{\infty}\frac{dk_z}{E_n}
\left(neB+m^{2}\right)
f_{\text{F}}(E_n,k_z),\nonumber\\
T^{22}&=&\frac{eB}{\pi^{2}}\sum_{n=0}^{\infty}\alpha_n\int_{0}^{\infty}\frac{dk_z}{E_n}
\left(neB+m^{2}\right)
f_{\text{F}}(E_n,k_z),\nonumber\\
T^{33}&=&\frac{eB}{\pi^{2}}\sum_{n=0}^{\infty}\alpha_n\int_{0}^{\infty}\frac{dk_z}{E_n} \left(k_{z}^{2}+m^{2}\right)f_{\text{F}}(E_n,k_z).\nonumber\\
\end{eqnarray}
Here, as expected, the trace of the energy-momentum tensor is proportional to the fermion mass, and reads
\begin{eqnarray}\label{E19}
\hspace{-1.5cm}T^{\mu}_{~\mu}=-\frac{3eB m^{2}}{\pi^{2}}\sum_{n=0}^{\infty}\alpha_{n}\int_{0}^{\infty}\frac{dk_z}{E_n}f_{\text{F}}(E_n,k_z).
\end{eqnarray}
It vanishes for $m=0$, as expected from conformal symmetry for massless fermions.  Identifying, at this stage, $T^{00}$ with the energy density $\epsilon$, $T^{11}=T^{22}$ with the transverse pressure $p_{\perp}$, and $T^{33}$ with the longitudinal pressure $p_{\|}$, it turns out that the results for these quantities in the massless case coincide, as aforementioned, with the matter ($T$ dependent) part of $\epsilon,p_{\|}$, and $p_{\perp}$ from \eqref{N15}, \eqref{N19} and \eqref{N20}, respectively. The quantity $BM$, including the magnetization $M$, is given by $BM=p_{\|}-p_{\perp}$, or
\begin{eqnarray}\label{E20}
T^{33}-T^{11}&=&\frac{eB}{\pi^{2}}\sum_{n=0}^{\infty}\alpha_n\int_{0}^{\infty}\frac{dk_z}{E_n}\left(k_{z}^{2}-neB\right)f_{\text{F}}(E_n,k_z),\nonumber\\
\end{eqnarray}
that is comparable with $BM$ from \eqref{N15}. Assuming the fermions to be massless, the transverse pressure vanishes in the LLL approximation ($n=0$), while, according to \eqref{E18}, we have
\begin{eqnarray}\label{E21}
p_{\|,\text{LLL}}=\epsilon_{\text{LLL}}=\frac{eB T^2}{12}.
\end{eqnarray}
This result arises from
\begin{eqnarray}\label{E22}
\int_{-\infty}^{+\infty}dk_z~\frac{|k_z|}{e^{\beta |k_z|}+1}=\frac{\pi^{2}T^{2}}{6},
\end{eqnarray}
and indicates a sound speed equal to the light speed $c_s=c=1$
\subsection{Thermodynamic quantities in the high  and low-temperature  approximations}\label{sec3b}
To present the analytic results for $p_{\|}, p_{\perp}$, and $\epsilon$ in the high- and low-temperature approximations, let us introduce first  the following integral \cite{kapusta-book}:
\begin{eqnarray}\label{E23}
h_{\ell}(y)=\frac{1}{\Gamma(\ell)}\int_{0}^{\infty} \frac{dx x^{\ell-1}}{\sqrt{x^{2}+y^{2}}}\frac{1}{\exp\left(\sqrt{x^2+y^2}\right)+1},\nonumber\\
\end{eqnarray}
with $x\equiv k_z/T$, $y\equiv M^2$, and $M$ a generic mass. According to our results from previous section, $p_{\|}, p_{\perp}$, and $\epsilon$ are then given by
\begin{eqnarray}\label{E24}
\hspace{-1cm}p_{\|}&=&\frac{eB}{\pi^{2}}\sum_{n=0}^{\infty}\alpha_{n}[2T^{2}h_{3}(y_n)+m^{2}h_{1}(y_n)],\nonumber\\
\hspace{-1cm}p_{\perp}&=&\frac{eB}{\pi^{2}}\sum\limits_{n=0}^{\infty}\alpha_{n}(neB+m^2)h_{1}(y_{n}),\nonumber\\
\hspace{-1cm}\epsilon&=&\frac{eB}{\pi^{2}}\sum\limits_{n=0}^{\infty}\alpha_{n}[2T^{2}h_{3}(y_{n})+2neB h_{1}(y_{n})],
\end{eqnarray}
where $y_n\equiv \beta M_{n}$ with $\beta=T^{-1}$ is defined in terms of the magnetic mass $M_n\equiv \left(m^{2}+2neB\right)^{1/2}$.
In what follows, we present the results for $p_{\|}, p_{\perp}$, and $\epsilon$ in the high- and low-temperature approximations in the  massive and massive cases separately.
\subsubsection{High temperature approximation for the massive and massless fermions}\label{sec3b1}
In the high-temperature expansion, $h_{\ell}$-integrals in \eqref{E23} are evaluated according to \cite{weldon1982},  (see also Appendix \ref{appB} for more details),
\begin{widetext}
\begin{eqnarray}\label{E25}
h_{2 \ell +1}(y) &=& \frac{1}{2 \Gamma(\ell+1)} \sum_{k=0}^{\ell - 1} \left\{ \left( 1 - 2^{1 - 2 \ell - 2 k} \right)  \left( \frac{y}{2} \right)^{ 2 k} \frac{ (-1)^{k}  \Gamma(\ell - k) \zeta(2 \ell - 2 k)}{\Gamma( k +1)}  \right\} \nonumber \\
&& \left. - \frac{(-1)^{\ell}}{2 \Gamma^{2}(\ell+1)}  \left( \frac{y}{2} \right)^{ 2 \ell}  \left( \frac{1}{2} \gamma_{E} -\frac{1}{2} \psi(1+ \ell) +  \ln \left( \frac{y}{\pi} \right) \right)  \right. \nonumber \\
&& +  \frac{1}{2 \Gamma(\ell+1)}  \left( \frac{y}{2} \right)^{ 2 \ell} \sum_{k=1}^{\infty} \left\{    \left( 1 - 2^{1 + 2 k} \right)  \left( \frac{y}{4 \pi} \right)^{ 2 k} \frac{ (-1)^{\ell + k}  \Gamma(2 k + 1) \zeta(2 k + 1)}{\Gamma( k +1) \Gamma(\ell + k + 1)}    \right\}.
\end{eqnarray}
\end{widetext}
This leads in particular to (see also \cite{kapusta-book})
\begin{eqnarray}\label{E26}
h_{1}(y)&\simeq &-\frac{1}{2}\ln\left(\frac{y}{\pi}\right)-\frac{1}{2}\gamma_{E}+{\cal{O}}\left(\left(\frac{y}{2}\right)^2\right).\nonumber\\
h_{3}(y)&\simeq&\frac{\pi^{2}}{24}+{\cal{O}}\left(\left(\frac{y}{2}\right)^2\right).
\end{eqnarray}
In \eqref{E25}, $\Gamma(z)=(z-1)!$,
\begin{eqnarray*}
\zeta(s)=\sum_{n=1}^{\infty}\frac{1}{n^s},\quad \text{and}\quad  \psi(z)=\frac{\Gamma'(z)}{\Gamma(z)},
\end{eqnarray*}
 are the gamma-, Riemann zeta-, and digamma functions, respectively.  Moreover, $\gamma_{E}$ is the Euler-Mascheroni constant,
\begin{eqnarray*}
\gamma_{E}=\lim\limits_{n\to \infty}\left(-\ln n+\sum_{k=1}^{n}\frac{1}{k}\right)\approx 0.577.
\end{eqnarray*}
Assuming the summation over Landau levels $n$ in \eqref{E24} is limited to $N$, the corresponding results to $p_{\|},p_{\perp}$, and $\epsilon$ for nonvanishing fermion mass read
\begin{eqnarray}\label{E27}
p_{\|}&\simeq&\frac{eB(1+2N)T^{2}}{12}-\frac{eB m^{2}}{2\pi^{2}}\sum_{n=0}^{N}\alpha_{n}\bigg[\ln\left(\frac{y_n}{\pi}\right)+\gamma_{E}\bigg], \nonumber\\
p_{\perp}&\simeq&-\frac{eB}{2\pi^{2}}\sum_{n=0}^{N}\alpha_n (m^2+neB)\bigg[\ln\left(\frac{y_{n}}{\pi}\right)+\gamma_{E}\bigg], \nonumber\\
\epsilon&\simeq&\frac{eBT^{2}(1+2N)}{12}-\frac{2(eB)^{2}}{\pi^{2}}\sum_{n=1}^{N}n\bigg[\ln\left(\frac{y_{n}}{\pi}\right)+\gamma_{E}\bigg],\nonumber\\
\end{eqnarray}
with $y_{n}=\beta M_n$, as before. Here, $\sum_{n=0}^{N}\alpha_{n}=1+2N$ is used. In the massless case, for $eB\ll T^{2}$, we arrive, in particular, at
\begin{eqnarray}\label{E28}
p_{\|}&\simeq&\frac{eB\left(1+2N\right)T^{2}}{12},\nonumber\\
p_{\perp}&\simeq&-\frac{(eB)^{2}}{\pi^{2}}\sum_{n=1}^{N}n\left(\ln\left(\frac{z_n}{\pi}\right)+\gamma_{E}\right),\nonumber\\
\epsilon&\simeq&\frac{eB\left(1+2N\right)T^{2}}{12}-\frac{2(eB)^2}{\pi^{2}}\sum_{n=1}^{N}n\left(\ln\left(\frac{z_n}{\pi }\right)+\gamma_{E}\right),\nonumber\\
\end{eqnarray}
with $z_n\equiv\beta \sqrt{2neB}$. Setting, $N=0$ in \eqref{E28}, we obtain $p_{\perp}=0$, and $\epsilon=p_{\|}=\frac{eBT^2}{12}$, as expected from \eqref{E21}.\footnote{We notice that the LLL approximation is only valid in the low-temperature approximation $eB\gg T^2$. } The results presented in \eqref{E28} can be in particular regarded as the high-temperature approximation of the $T$ dependent part of $p_{\|},p_{\perp}$, and $\epsilon$ arising from the effective action \eqref{N13} [see \eqref{N15}, \eqref{N19} and \eqref{N20} for details].
\subsubsection{Low temperature approximation for massless fermions}\label{sec3b2}
In the low-temperature expansion, $h_{\ell}$-integrals in \eqref{E24} are evaluated according to  \cite{weldon1982}, (see also Appendix \ref{appB} for more details)
\begin{widetext}
\begin{eqnarray}\label{E29}
h_{n}(y) = \frac{\Gamma (\frac{n}{2})}{\Gamma(n)} \sum_{k=0}^{\infty} \frac{\Gamma(k+\frac{n}{2})}{\Gamma(\frac{n}{2}-k) k!} \bigg[ \left( \frac{1}{2 y} \right)^{k - \frac{n}{2}+1} \text{Li}_{k + \frac{n}{2}}(e^{- y}) - \frac{1}{2^{2 k}} \left( \frac{1}{ y} \right)^{k - \frac{n}{2}+1} \text{Li}_{k + \frac{n}{2}}(e^{- 2 y}) \bigg],
\end{eqnarray}
\end{widetext}
that leads in particular to
\begin{eqnarray}\label{E30}
\hspace{-0.5cm}h_{1}(y)&\simeq&\sqrt{\frac{\pi}{2y}} \mathcal{U}_{1/2}(y),\nonumber \\
\hspace{-0.5cm}h_{3}(y)&\simeq& \frac{\sqrt{\pi y}}{4} \mathcal{U}_{3/2}(y),
\end{eqnarray}
with
\begin{eqnarray}\label{E31}
\hspace{-1cm}\mathcal{U}_{3/2}(y)&\equiv&\bigg[
\sqrt{2}\mbox{Li}_{3/2}
\left(e^{-y}\right)-
\mbox{Li}_{3/2}
\left(e^{-2y}\right)
\bigg],\nonumber\\
\hspace{-1cm}\mathcal{U}_{1/2}(y)&\equiv&\bigg[
\mbox{Li}_{1/2}
\left(e^{-y}\right)-
\sqrt{2}\mbox{Li}_{1/2}
\left(e^{-2y}\right)\bigg].
\end{eqnarray}
Here, $\text{Li}_{s}(z)$ is the polylogarithm function,
\begin{eqnarray*}
\text{Li}_{s}(z)=\sum_{k=1}^{\infty}\frac{z^{k}}{k^{s}}.
\end{eqnarray*}
Plugging these expressions into \eqref{E24}, and setting $m=0$,  $p_{\|}, p_{\perp}$, and $\epsilon$ in the low-temperature (large magnetic field) approximation in the massless case read
 \begin{eqnarray}\label{E32}
p_{\|}&\simeq&\frac{eBT^{2}}{12}+\frac{2^{1/4}(eB)^{5/4}T^{3/2}}{\pi\sqrt{\pi}}\sum\limits_{n=1}^{N}n^{1/4}\mathcal{U}_{3/2}(z_{n})\nonumber\\
p_{\perp}&\simeq&\frac{2^{1/4}(eB)^{7/4}T^{1/2}}{\pi\sqrt{\pi}}\sum\limits_{n=1}^{N}n^{3/4}\mathcal{U}_{1/2}(z_{n}),\nonumber\\
\epsilon&\simeq&\frac{eBT^{2}}{12}+\frac{2^{1/4}(eB)^{5/4}T^{3/2}}{\pi\sqrt{\pi}}\sum_{n=1}^{N}n^{1/4}\mathcal{U}_{3/2}(z_{n})\nonumber\\
&&+\frac{2^{5/4}(eB)^{7/4}T^{1/2}}{\pi\sqrt{\pi}}\sum_{n=1}^{N}n^{3/4}\mathcal{U}_{1/2}(z_{n}),
\end{eqnarray}
with $\mathcal{U}_{3/2}(z_{n})$ and $\mathcal{U}_{3/2}(z_{n})$ from \eqref{E31}. Here, $z_{n}=\beta\sqrt{2neB}$. Moreover, as in \eqref{E27} and \eqref{E28}, the summation over Landau levels is limited to $N$.
\par
In the next section, we use these quantities to determine the proper time evolution of the temperature. Plugging the corresponding numerical results for $T$ back into  \eqref{E27}, \eqref{E28} and \eqref{E32}, we arrive at the proper time evolution of $p_{\|},p_{\perp}$, and  $\epsilon$ in the high- and low-temperature approximations in the massless and massive cases for a constant magnetic field.
\section{Numerical results}\label{sec4}
\setcounter{equation}{0}
In this section, we present a number of numerical results arising from the analytical ones in the previous sections. In Sec. \ref{sec4a}, we determine the $\tau$ dependence of $T$ by numerically solving the energy equation arising from the relativistic MHD. The necessary inputs are the pressures $p_{\|}$ and $p_{\perp}$ as well as the energy density $\epsilon$, which were determined analytically in the Wigner quantum kinetic theory approach.  We also present a number of comparisons between our results with the $\tau$ dependence of $T$ arising from relativistic hydrodynamics (Bjorken, Hubble, and Gubser flow) for zero magnetic fields. The corresponding results are also compared with the resulting temperatures for a point to point varying magnetic field.
\par
In Sec. \ref{sec4b}, we use our results for the $\tau$ dependence of $T$ to determine the $\tau$ dependence of $p_{\|}, p_{\perp}$, and $\epsilon$. To answer the question of whether a constant magnetic field has a significant impact on the $\tau$ dependence of these quantities is answered by comparing our results with the corresponding results to a $1+1$ dimensional Bjorken flow.  The $\tau$ dependence of other thermodynamic quantities that are particularly defined in terms of $p_{\|}$, $p_{\perp}$, and $\epsilon$ is presented in Sec. \ref{sec4c}. These quantities include $\chi_m,c_{s,\|}$ and $c_{s,\perp}$.  In Sec. \ref{sec4d}, we combine the $\tau$ dependence of these quantities and that of $T$, and determine the $T$ dependence of these quantities in the high- and low-temperature approximations.
\subsection{The $\bs{\tau}$ dependence of $\bs{T}$  in constant and decaying magnetic fields}\label{sec4a}
\begin{figure*}[hbt]
\includegraphics[width=8cm,height=6cm]{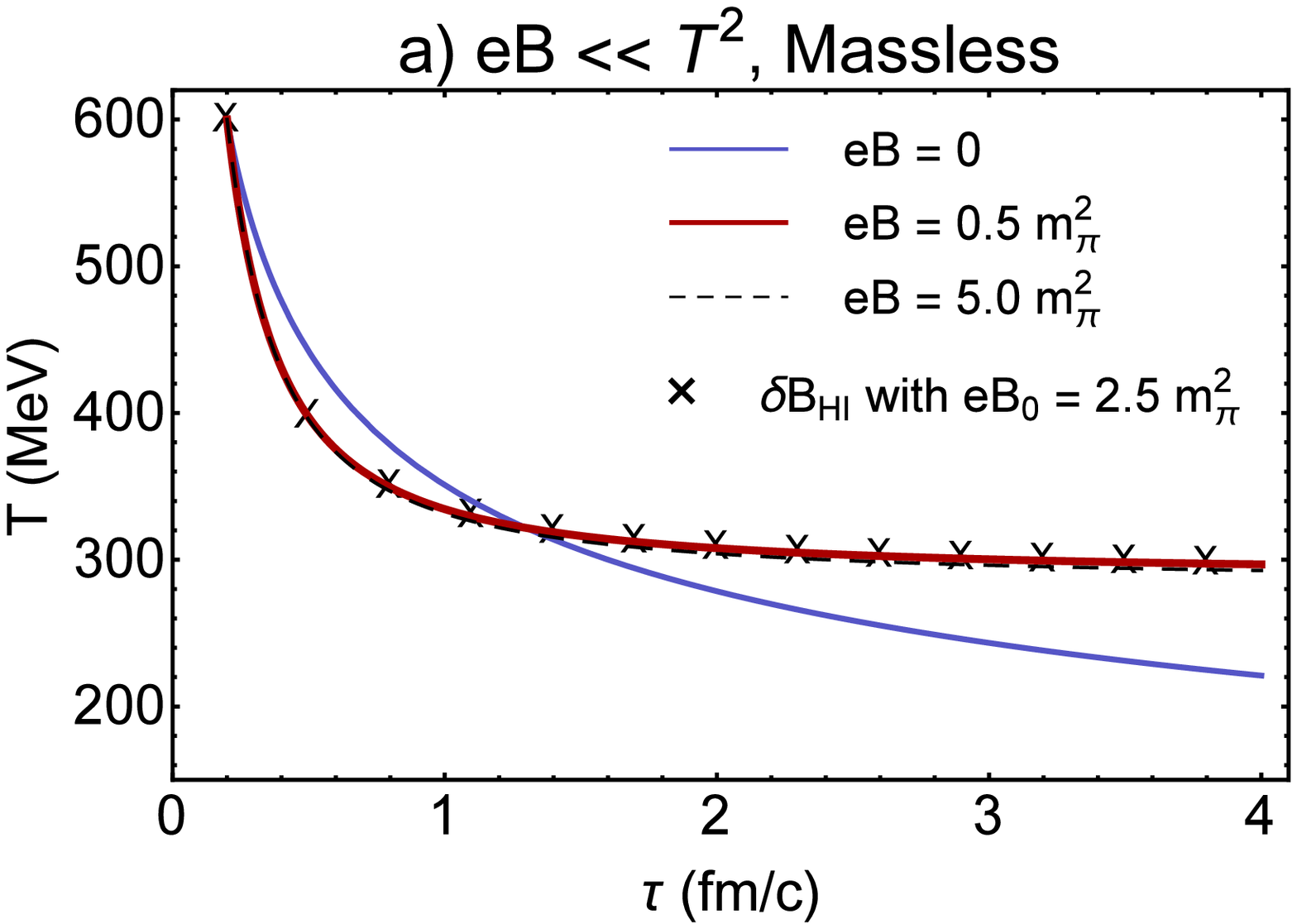}
\includegraphics[width=8cm,height=6cm]{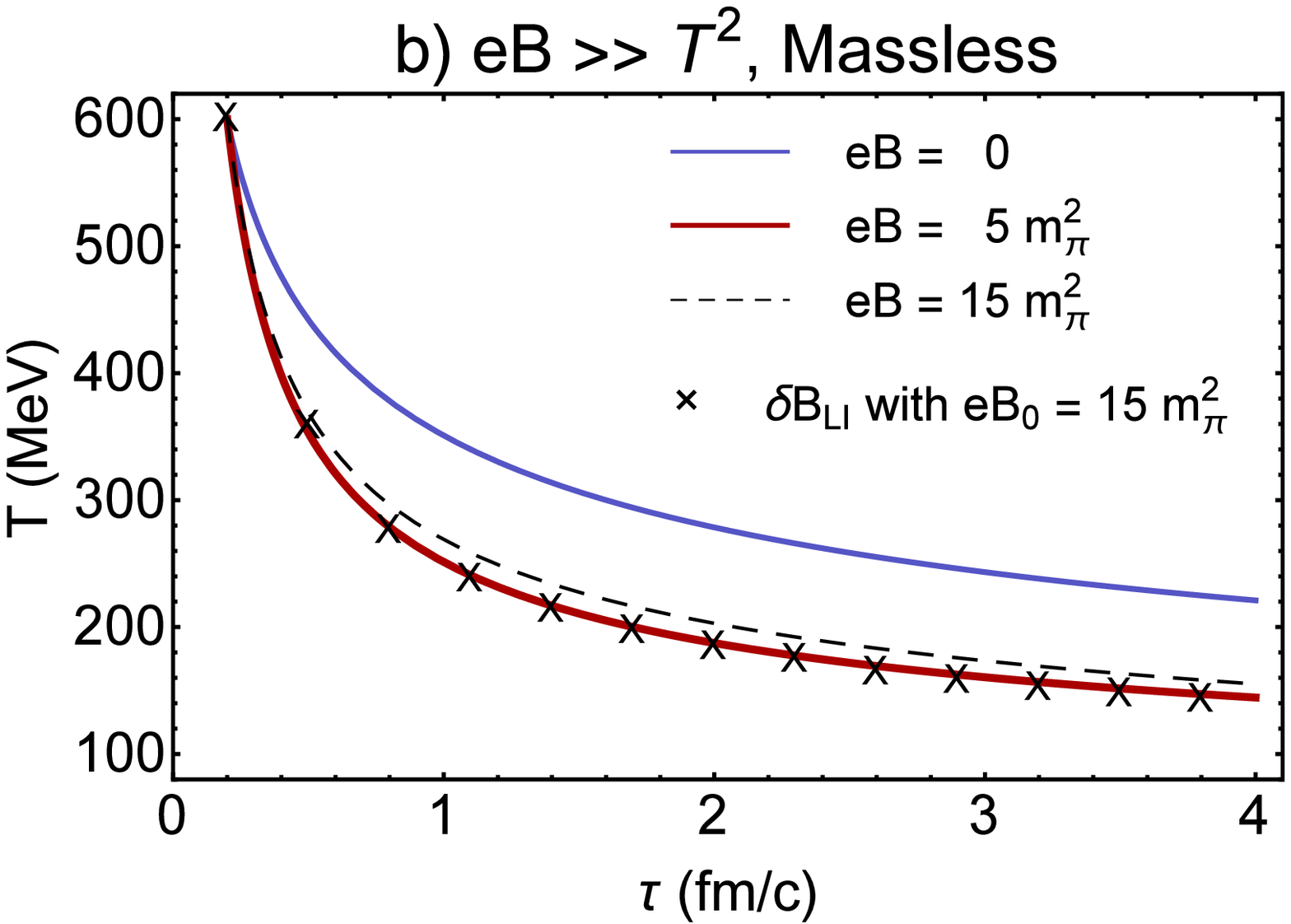}
\caption{(color online). The $\tau$ dependence of $T$ arising from the solution of the energy equation \eqref{N33} is plotted for high-temperature (panel a) and low-temperature (panel b) approximations. The decay of $T$ is then compared with $T(\tau)$ from \eqref{N34} arising from the $1+1$ dimensional Bjorken flow. The latter is demonstrated with the dashed and solid curves. A comparison is also made with the $\tau$ dependence of $T$ arising from point to point varying magnetic fields. Here, we choose $B(\tau)$ from \eqref{N40} (Ideal MHD). The resulting temperature for varying magnetic fields, denoted with black crosses, are computed for the initial magnetic field $eB_0=2.5 m_{\pi}^{2}$ and $eB_0=15 m_{\pi}^{2}$ for the high ($\delta B_{\text{HI}}$) and low ($\delta B_{\text{LI}}$) temperature approximations in panel a and b, respectively. }
\label{fig-2}
\end{figure*}
\begin{figure*}[hbt]
\includegraphics[width=8cm,height=6cm]{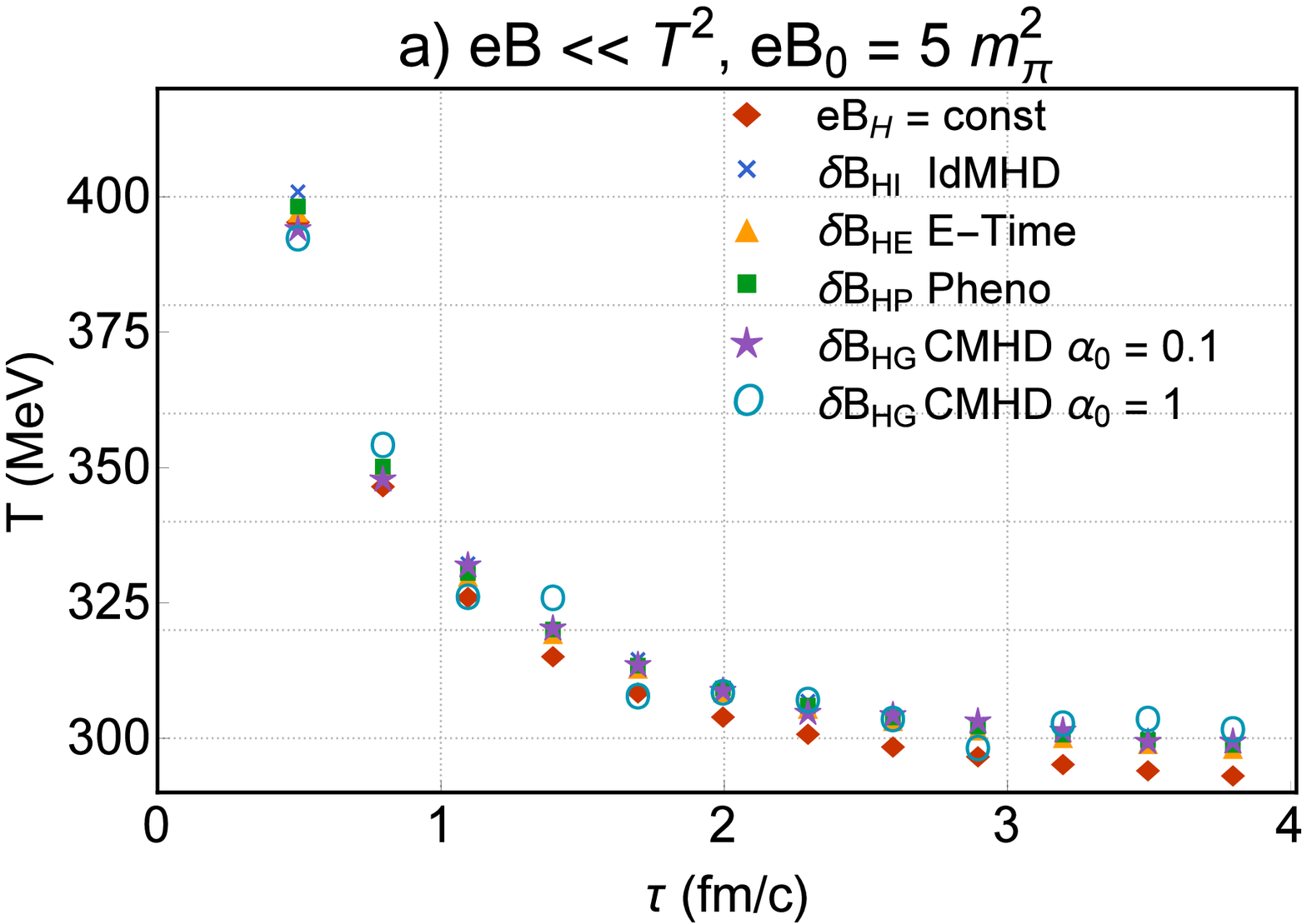}
\includegraphics[width=8cm,height=6cm]{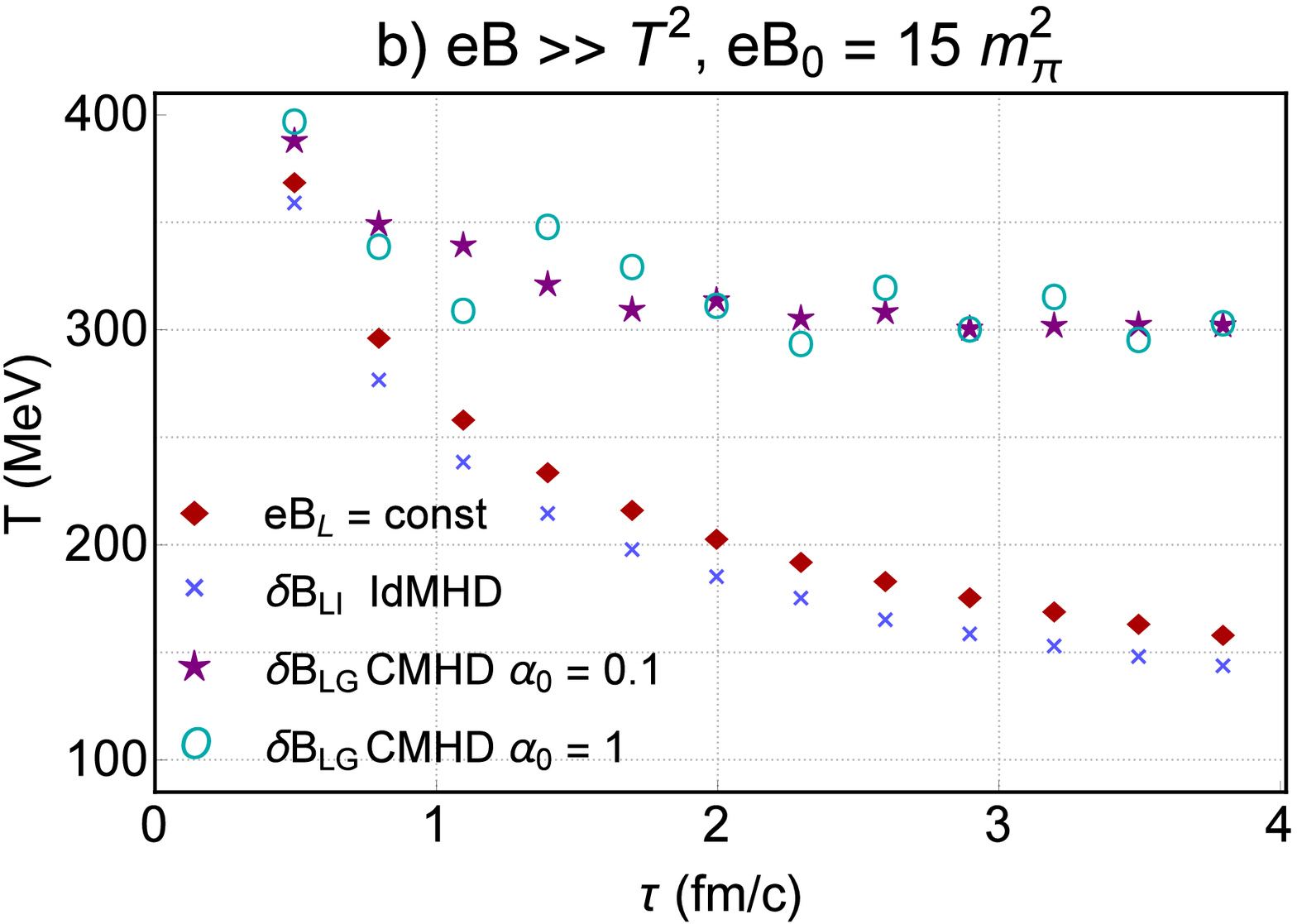}
\caption{(color online). The $\tau$ dependence of $T$ arising from a constant magnetic field $eB=5m_{\pi}^{2}$ (red diamond) is compared with the $\tau$ dependence of $T$ arising from the energy equation with point to point varying magnetic fields. The initial magnetic fields for different solutions are $eB_{0}=5m_{\pi}^{2}$  and $eB_{0}=15m_{\pi}^{2}$ for $eB\ll T^{2}$ and $eB\gg T^{2}$, respectively. According to Fig. \ref{fig-1}, the $B$ field arising from CMHD for $\alpha_{0}=0.1$ almost coincides with the $B$ field arising from IdMHD. Although in $eB\ll T^{2}$ the $\tau$ dependence of $T$ arising from CMHD with $\alpha_{0}=0.1, 1$ almost coincides with the $\tau$ dependence of $T$ arising from IdMHD, this is not the case in $eB\gg T^2$. Small oscillations appearing for the CMHD solution in the low-temperature approximation sustain once smaller intervals are chosen during which the magnetic field is assumed to be constant.}\label{fig-3}
\end{figure*}
\begin{figure}[hbt]
\includegraphics[width=8cm,height=6cm]{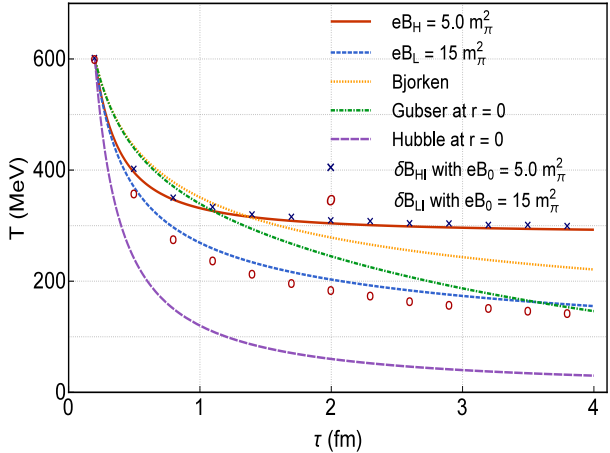}
\caption{The $\tau$ dependence of the temperature for the high- and low-temperature approximations with constant magnetic fields $eB_{\text{H}}=5 m_{\pi}^{2}$ and $eB_{\text{L}}=15 m_{\pi}^{2}$, as well as $\delta B_{HI}$ and $\delta B_{LI}$ results for point to point varying magnetic fields according to IdMHD solution for the $B$ field are compared with the $\tau$ dependence of the $1+1$ dimensional Bjorken, $3+1$ dimensional Hubble, and Gubser flows at $r=0$. According to this result, large magnetic fields slow up the fast decay of the temperature in the early time.  }\label{fig-4}
\end{figure}
 \begin{figure*}[hbt]
\includegraphics[width=8cm,height=6cm]{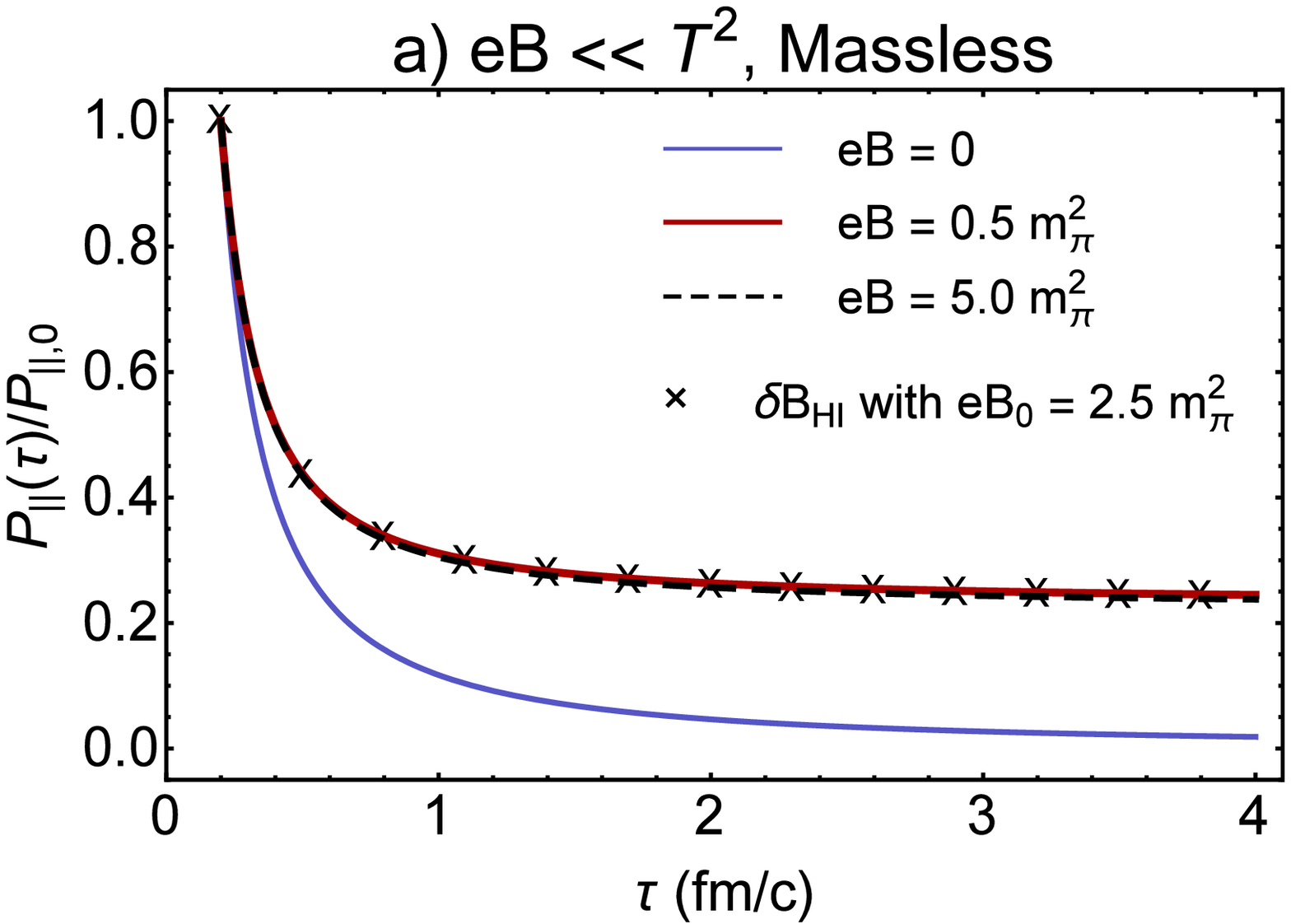}
\includegraphics[width=8cm,height=6cm]{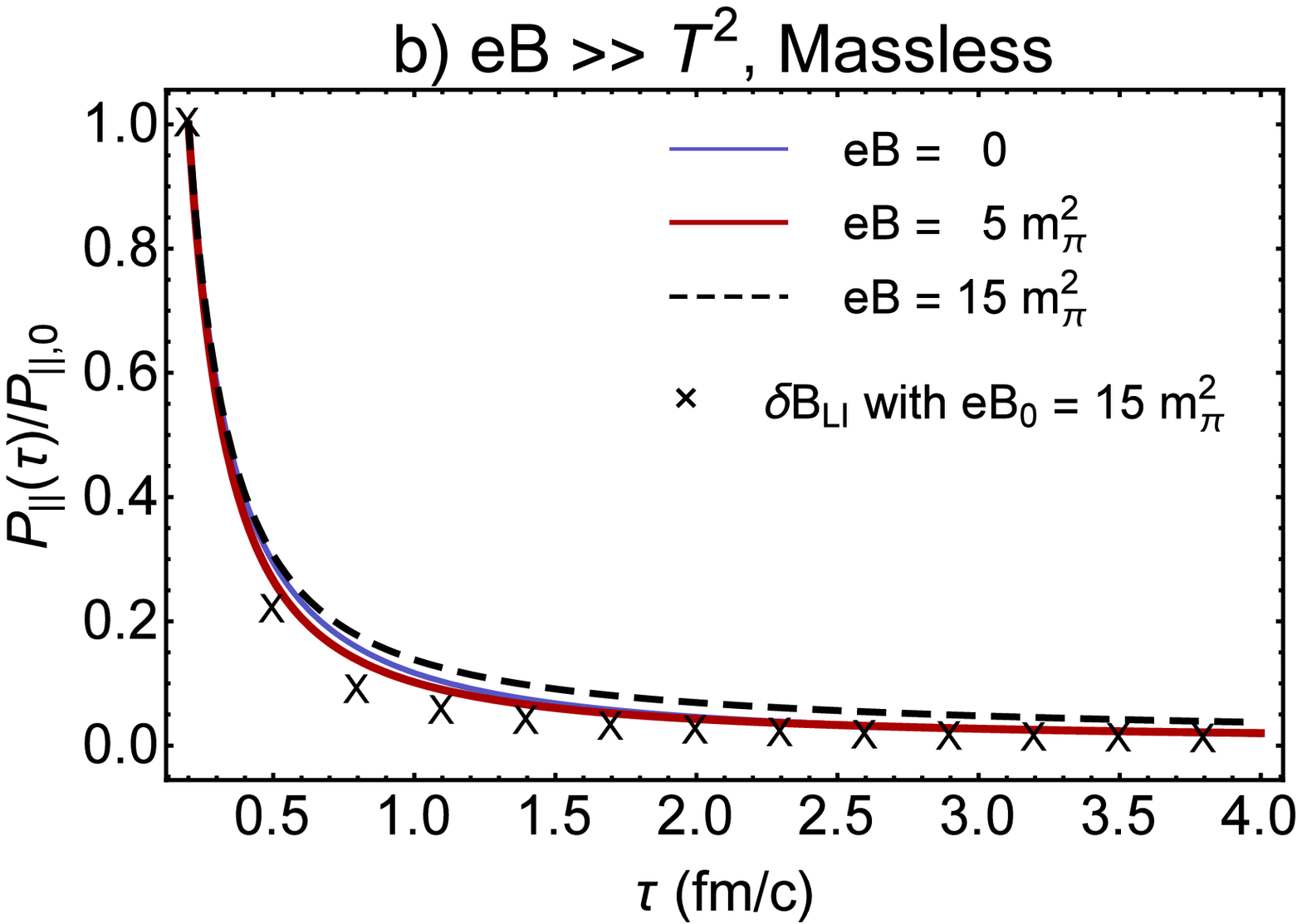}
\includegraphics[width=8cm,height=6cm]{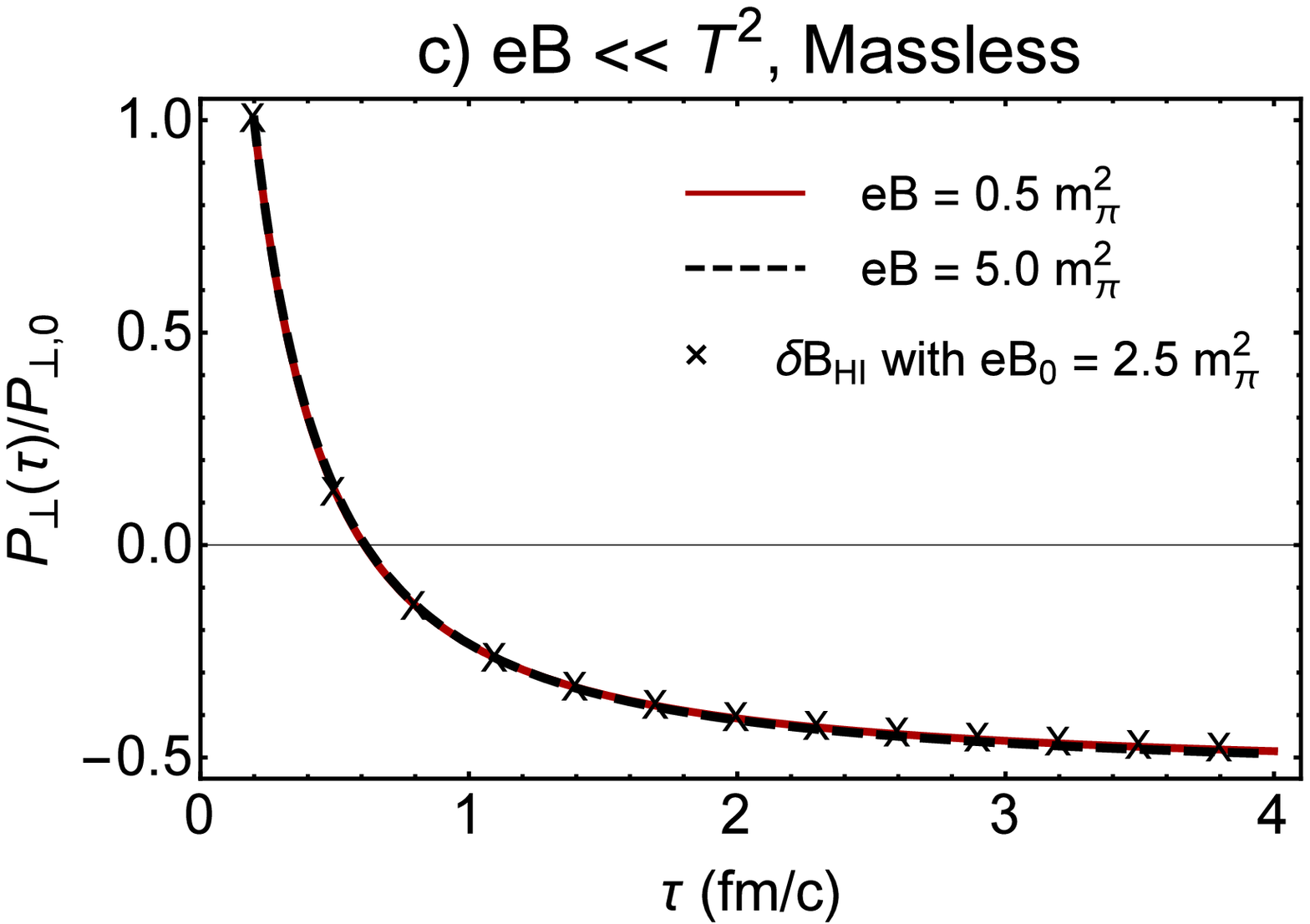}
\includegraphics[width=8cm,height=6cm]{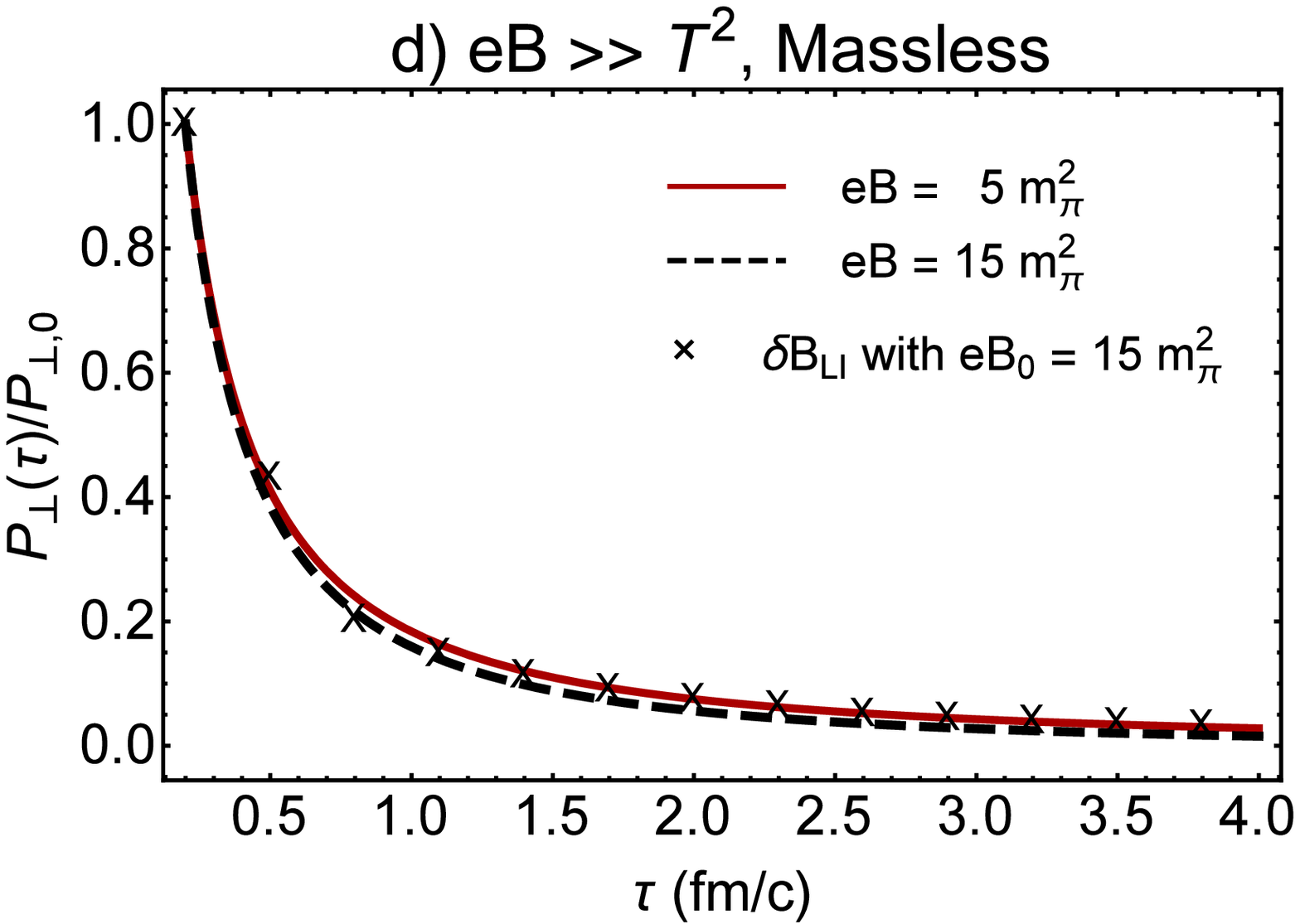}
\par\hspace{0.3cm}
\includegraphics[width=8cm,height=6cm]{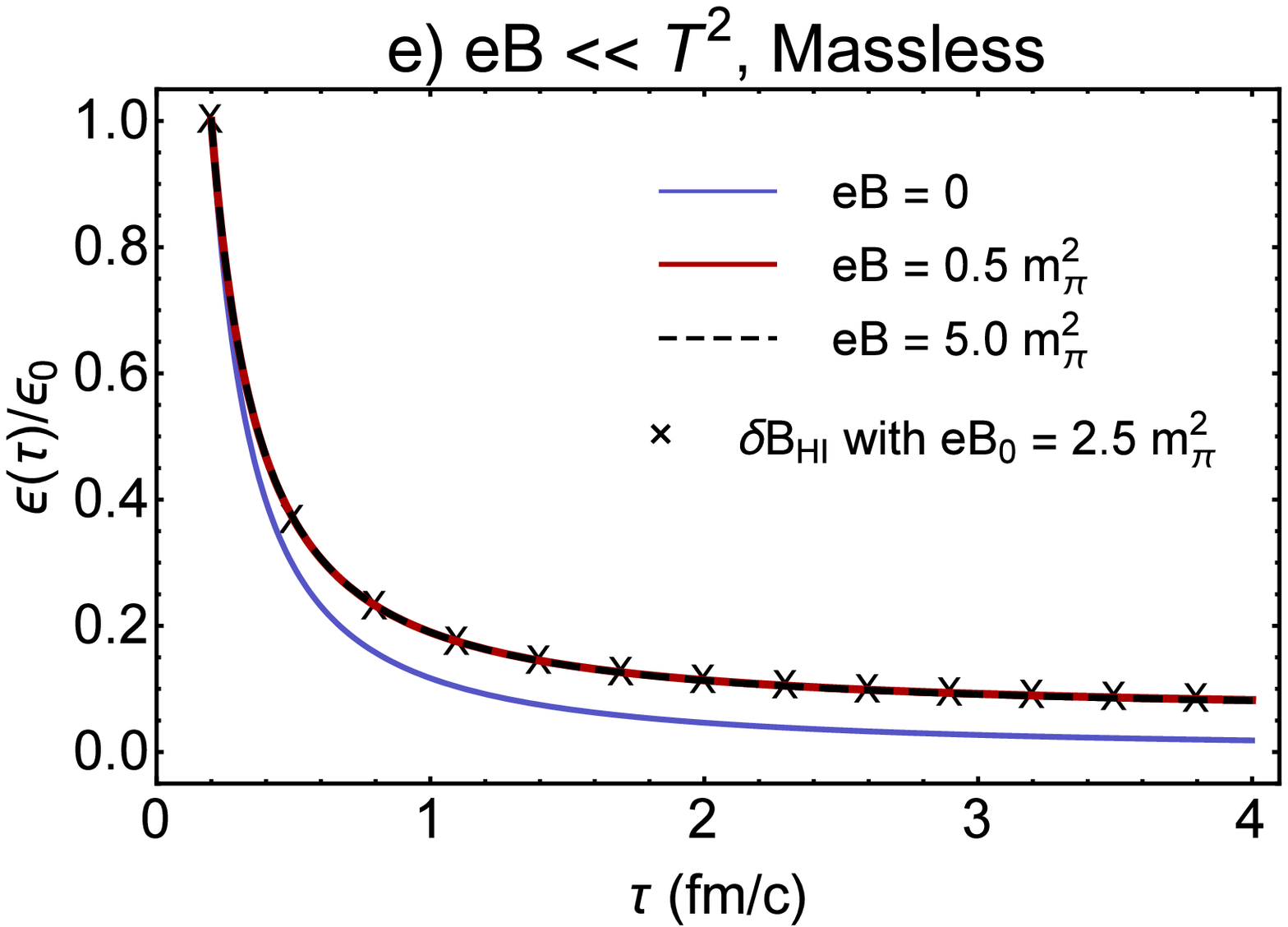}
\includegraphics[width=8cm,height=6cm]{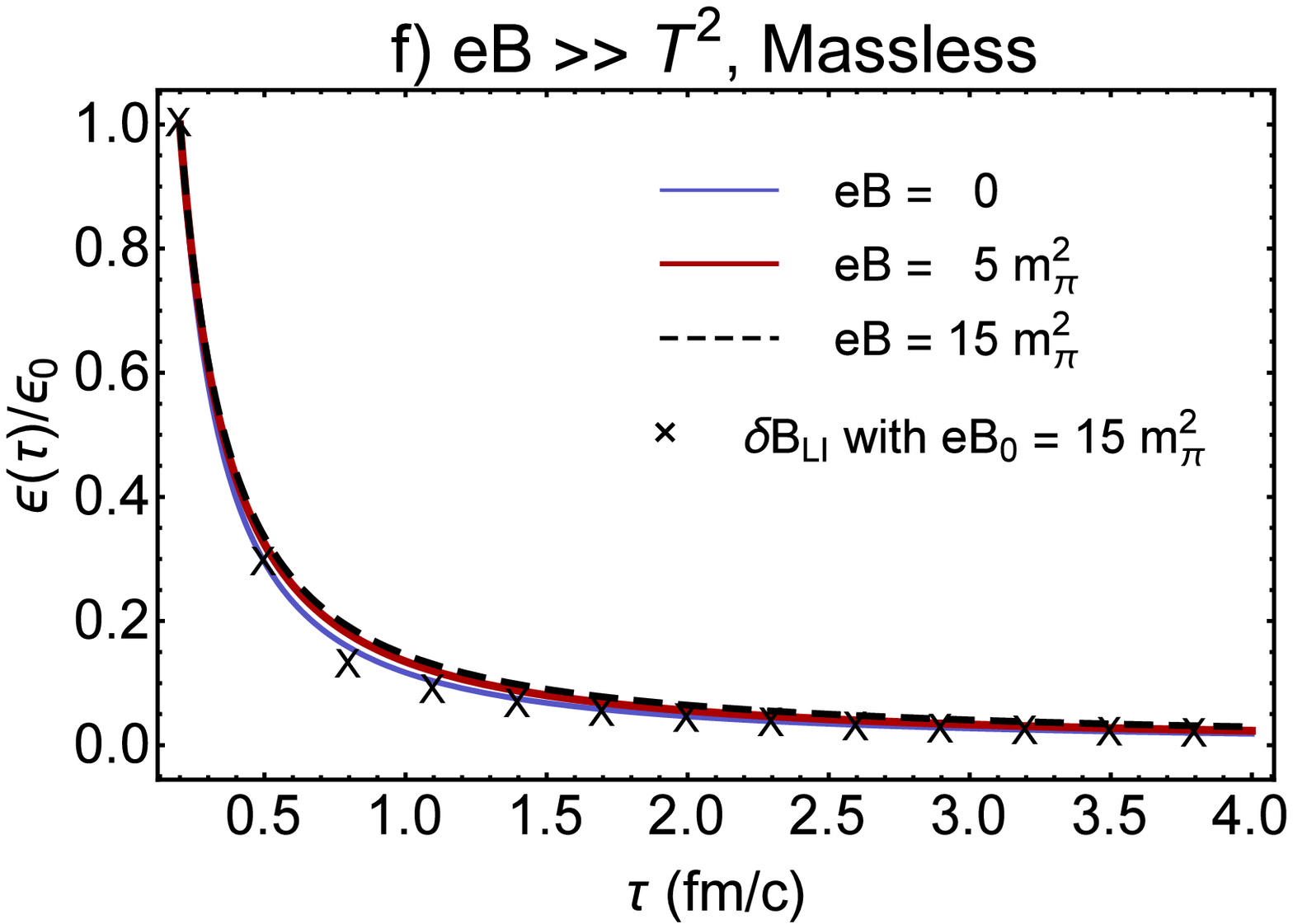}
\caption{(color online). The $\tau$ dependence of $p_{\|}(\tau)/p_{\|,0}$ (panels a and b), $p_{\perp}(\tau)/p_{\perp,0}$ (panels c and d), and $\epsilon(\tau)/\epsilon_{0}$ (panels e and f) from \eqref{E28} and \eqref{E32} with $T$ arising from the numerical solution of the energy equation \eqref{N33} is compared with the corresponding data from point to point decaying magnetic fields $\delta B_{\text{HI}}$ and $\delta B_{\text{LI}}$, and the expressions from \eqref{D2} in a $1+1$ dimensional Bjorken solution of relativistic hydrodynamics. In the low-temperature (large magnetic field) approximation, these results almost coincide. At high-temperature (weak magnetic field), the magnetic field slows up the decay of $\epsilon$ and $p_{\perp}$, once compared with the Bjorken solutions \eqref{D2} (blue solid lines for $eB=0$ in panels a,b,e, and f). }\label{fig-5U}
\end{figure*}
 \begin{figure}[hbt]
\includegraphics[width=8cm,height=6cm]{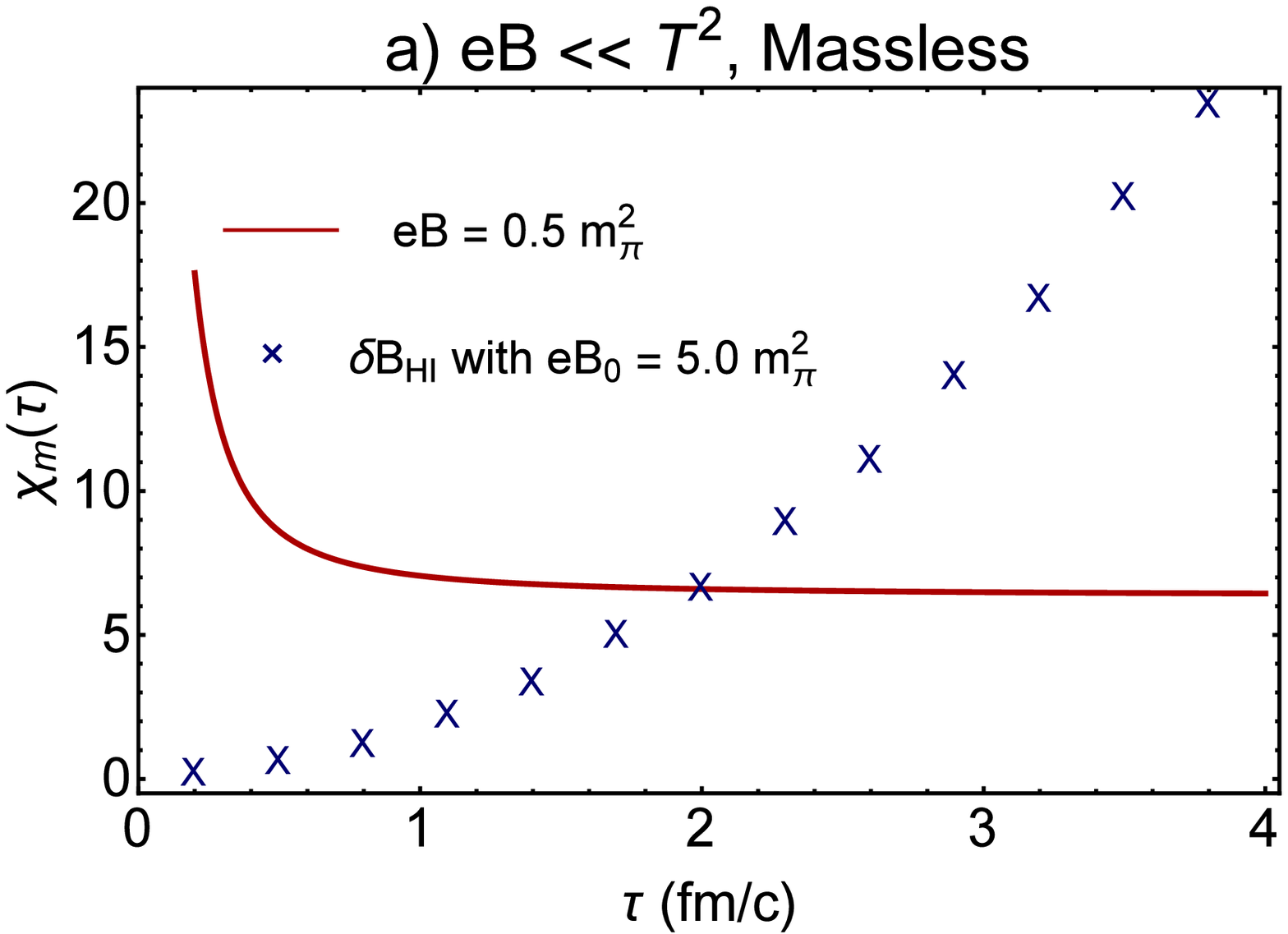}
\includegraphics[width=8cm,height=6cm]{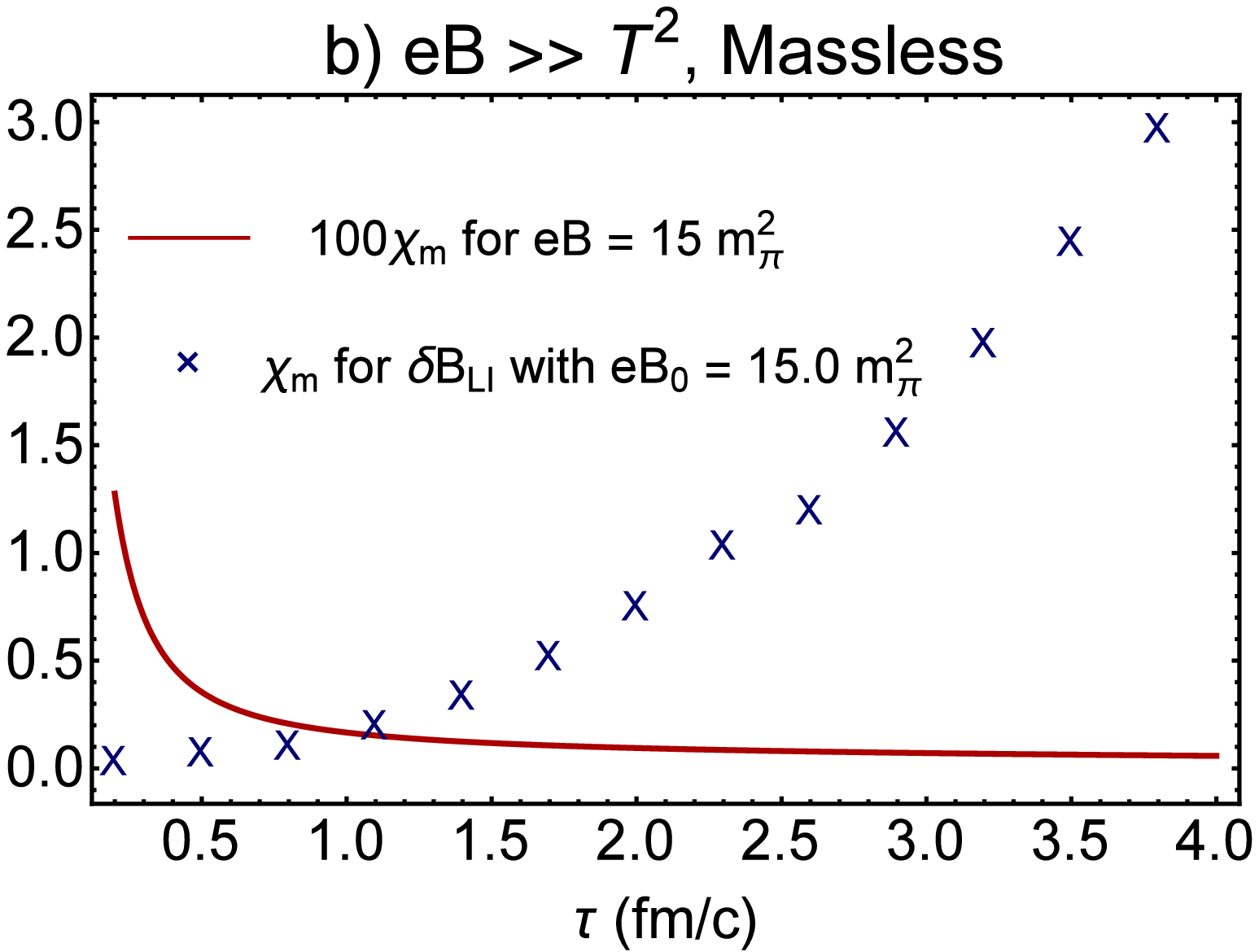}
\caption{(color online) The $\tau$ dependence of $\chi_m$ arising from $\chi_m=M/B$ with $BM=p_{\|}-p_{\perp}$ is plotted in the high (panel a) and low (panel b) temperature approximations. To do this, we use the results of $p_{\|}$, and $p_{\perp}$ from Fig. 5(a)-5(d) for the high- and low-temperature approximations with a constant background magnetic field as well as for a point to point decaying magnetic field. As it turns out, whereas for constant magnetic fields $\chi_m$ decreases with $\tau$ (red curves), it increases with increasing $\tau$ for a decaying magnetic field according to \eqref{N40}. The latter result is expected from \cite{tabatabaee2019}, arising in a classical kinetic theory approach. }\label{fig-6}
\end{figure}
\begin{figure*}[hbt]
\includegraphics[width=8cm,height=6cm]{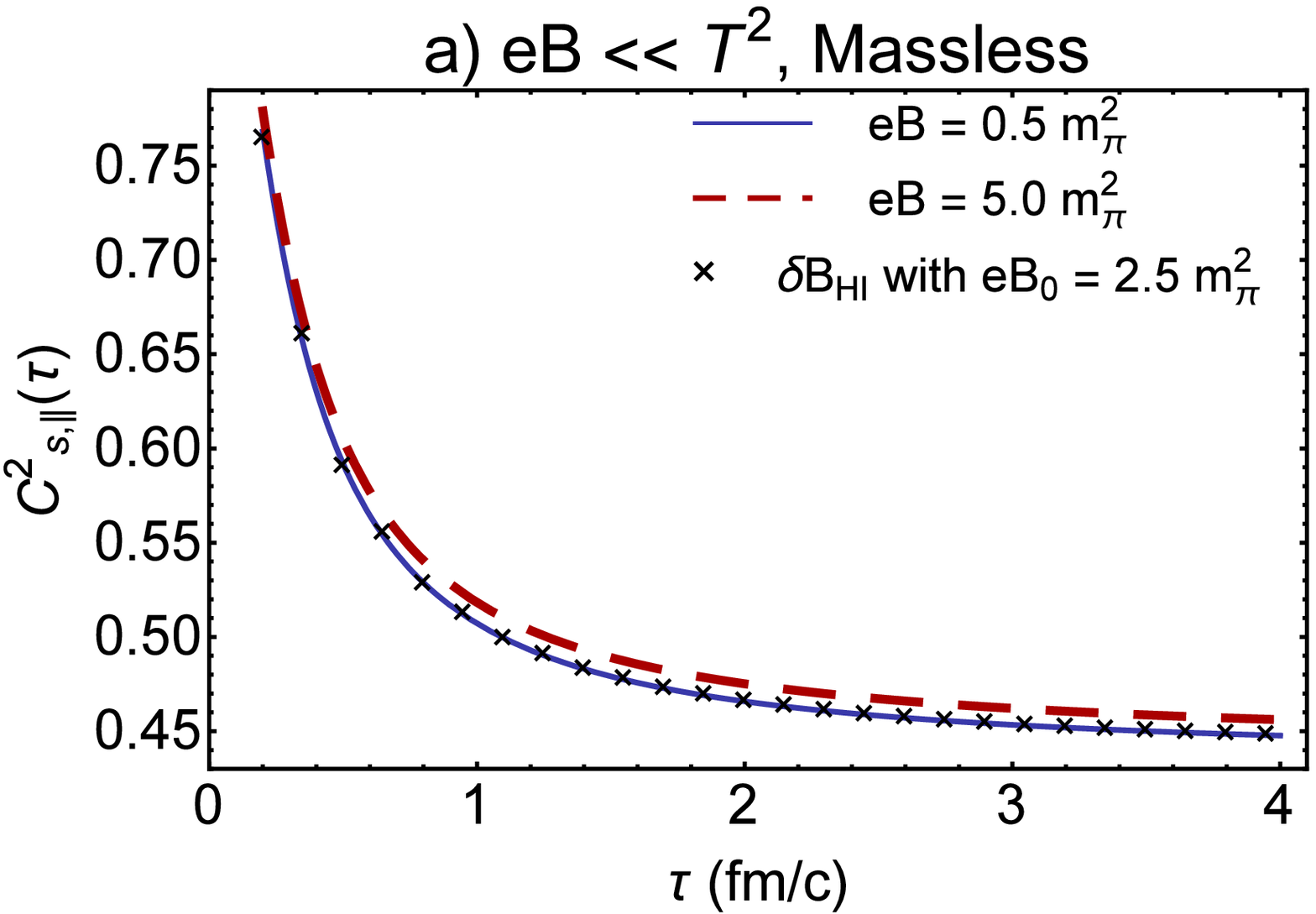}
\includegraphics[width=8cm,height=6cm]{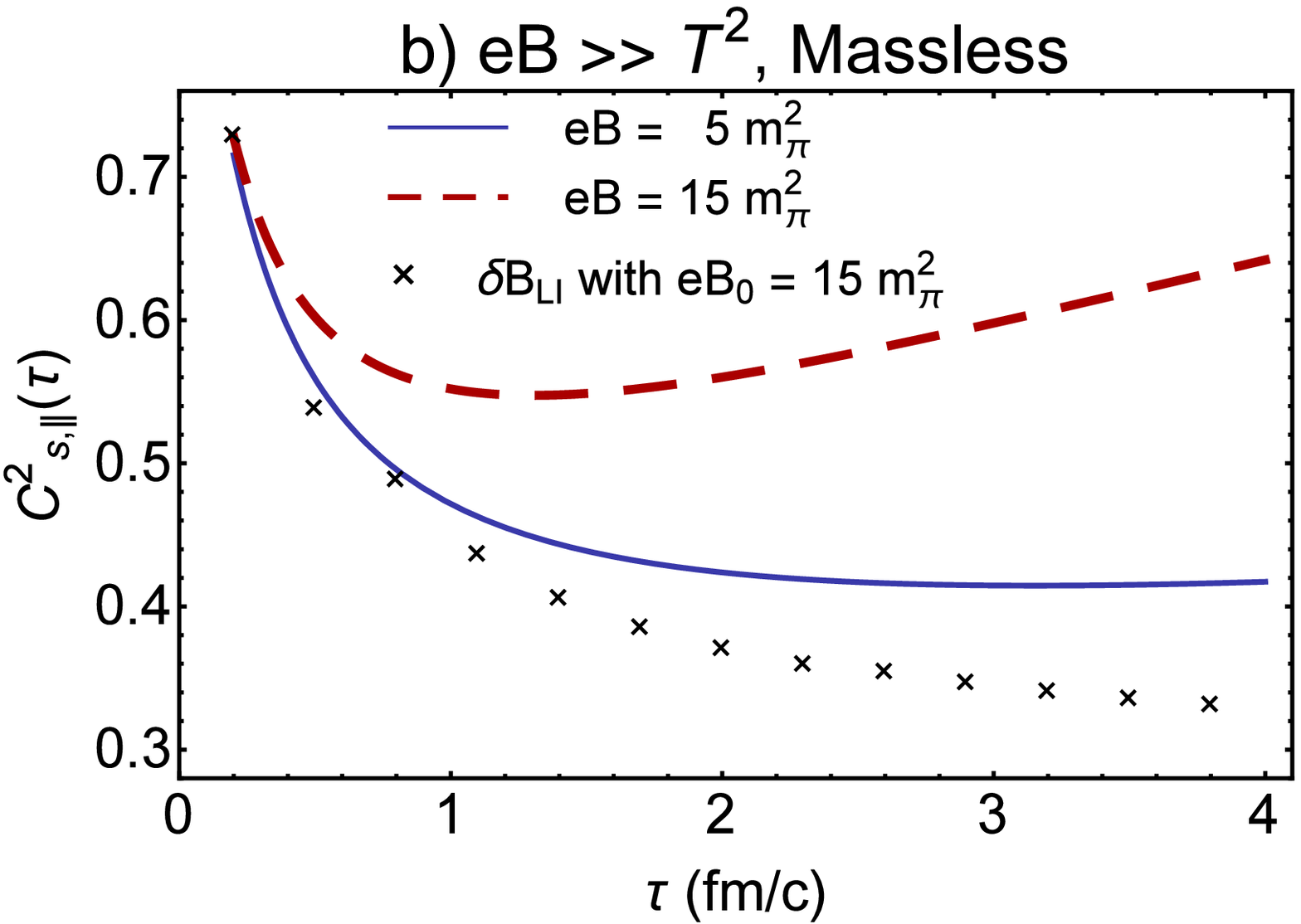}
\includegraphics[width=8cm,height=6cm]{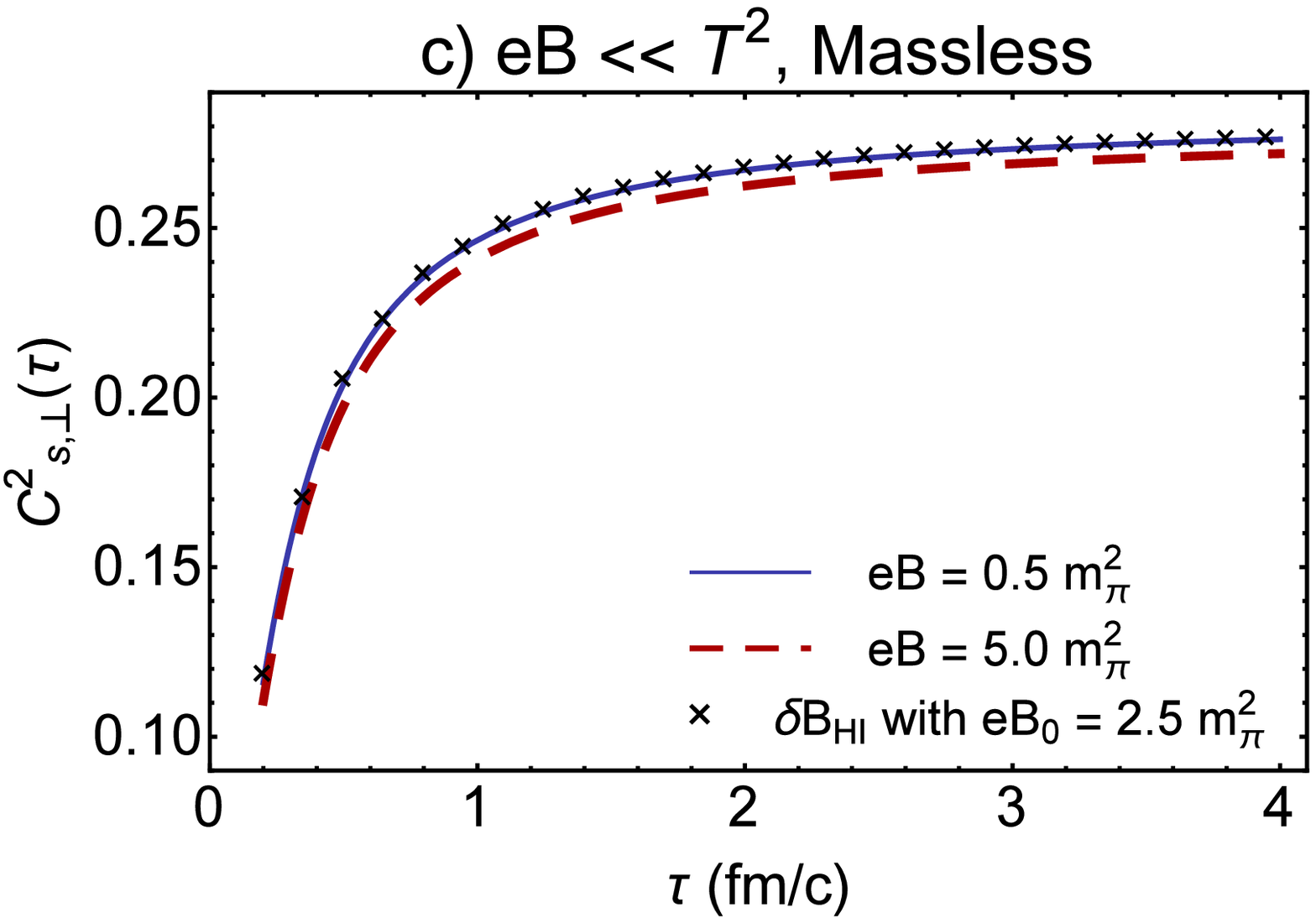}
\includegraphics[width=8cm,height=6cm]{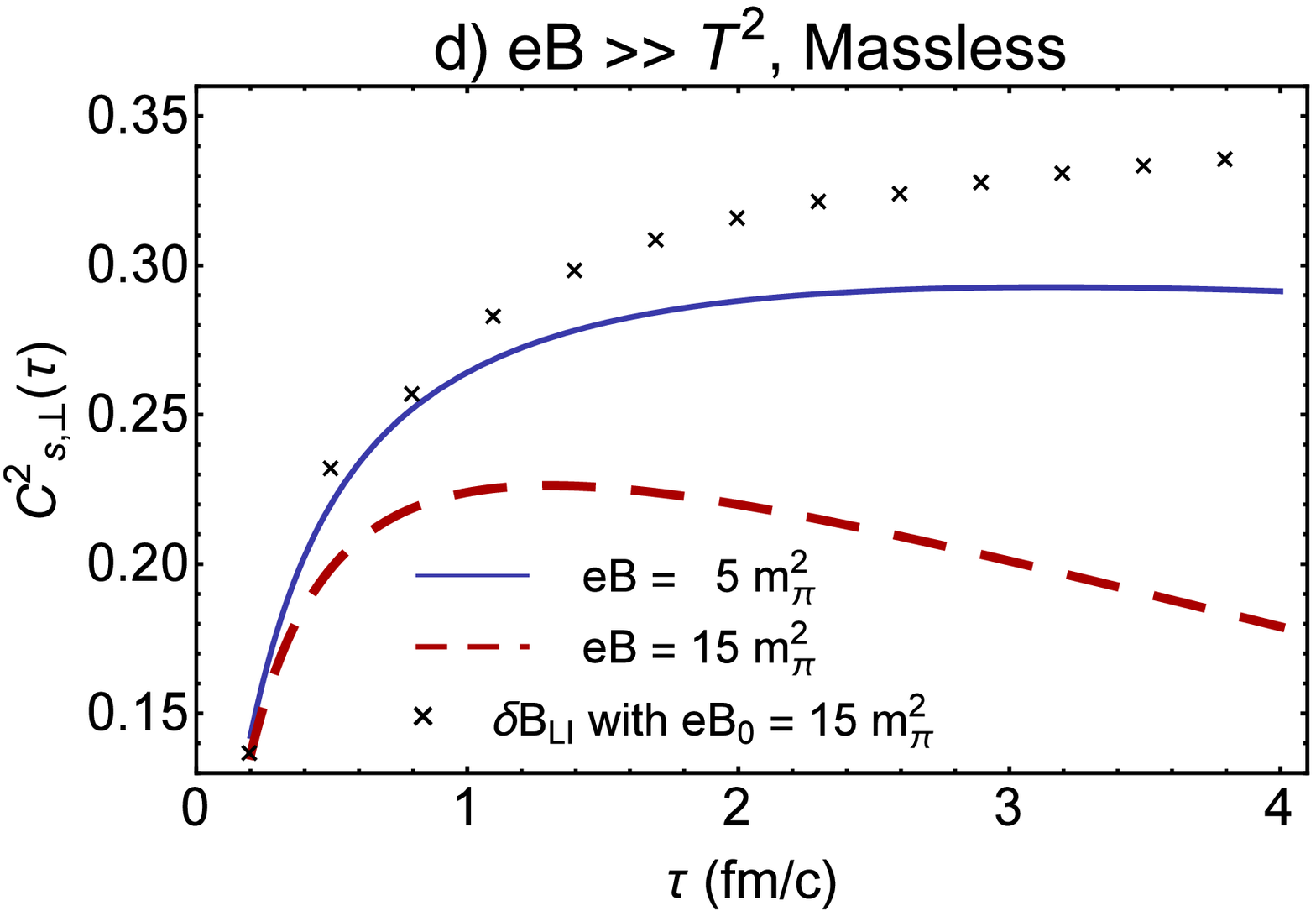}
\caption{(color online) The $\tau$ dependence of the longitudinal and transverse speed of sound is plotted in the high (panels a and  c) and low (panel b and d) temperature approximations. The curves correspond to the results for constant magnetic fields, while the crosses denote the results for a point to point varying magnetic field according to \eqref{N40}. The qualitative difference between $c_{s,\|}^{2}$ and $c_{s,\perp}^{2}$ is similar to that presented in  \cite{tabatabaee2019}, arising in a classical kinetic theory approach. }\label{fig-7}
\end{figure*}
\begin{figure}[hbt]
\includegraphics[width=8cm,height=6cm]{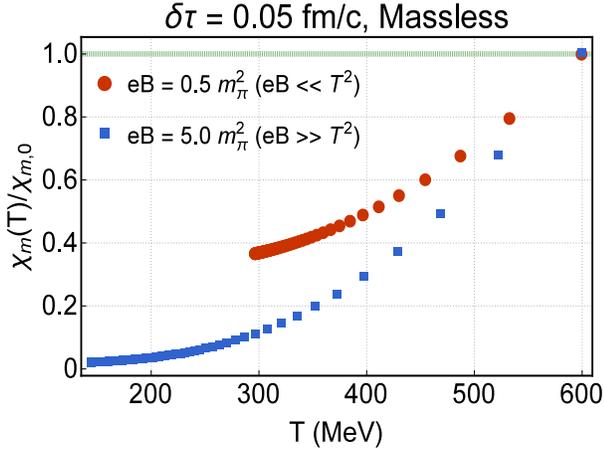}
\caption{(color online). The $T$ dependence of $\chi_{m}(T)/\chi_{m,0}$ is plotted in the high (orange circles) and low (blue squares) temperature approximations. Here, we used the $\tau$ dependence of $T$ and $\chi_m$. The proper time interval between two successive points in each curve is $\delta\tau=0.05$ fm/c.  According to these results, larger magnetic fields have a larger impact on the decay (increase) of $\chi_{m}$ within a fixed $\tau$ ($T$) interval.  }
\label{fig-8}
\end{figure}
\begin{figure*}[hbt]
\includegraphics[width=8cm,height=6cm]{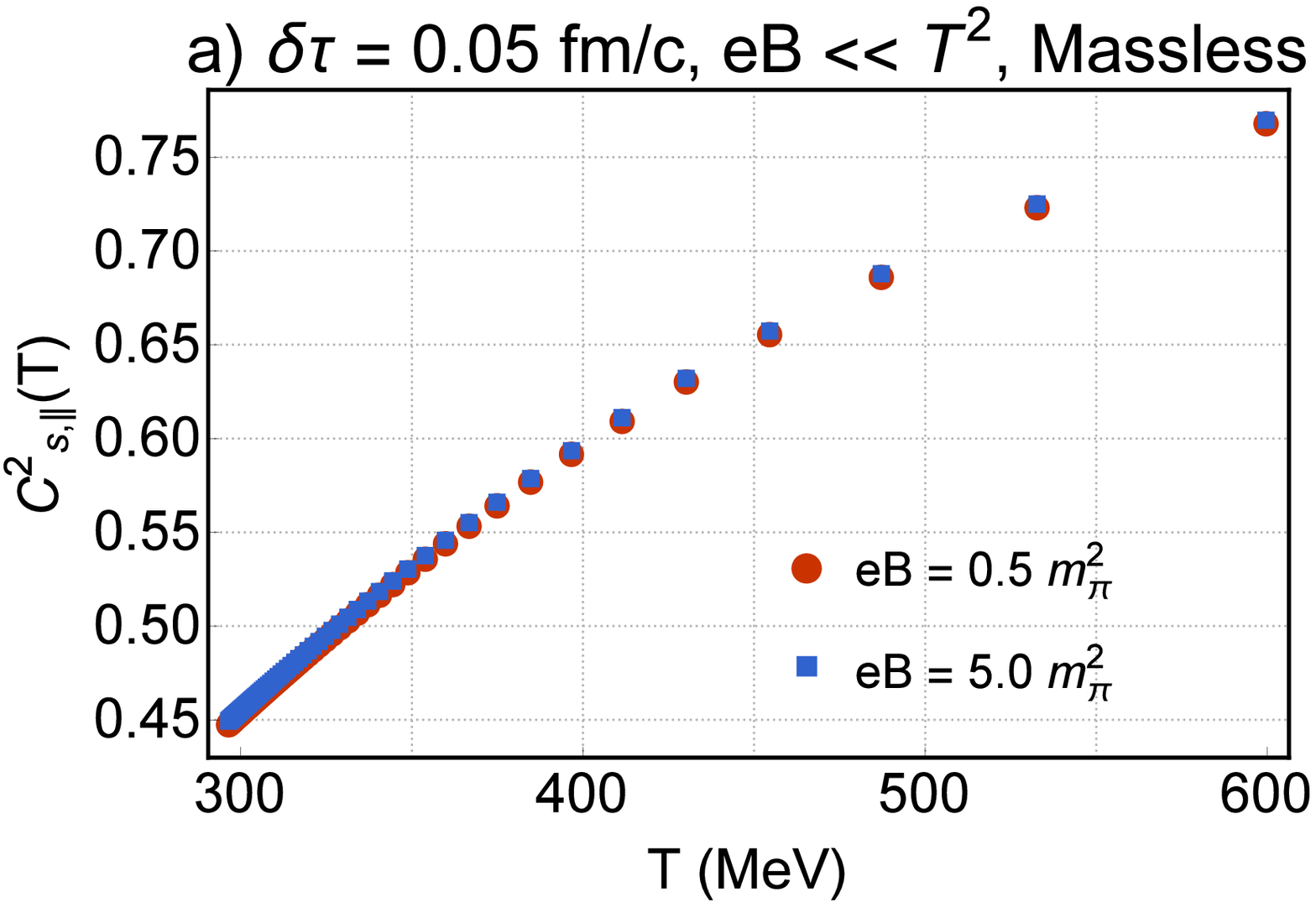}
\includegraphics[width=8cm,height=6cm]{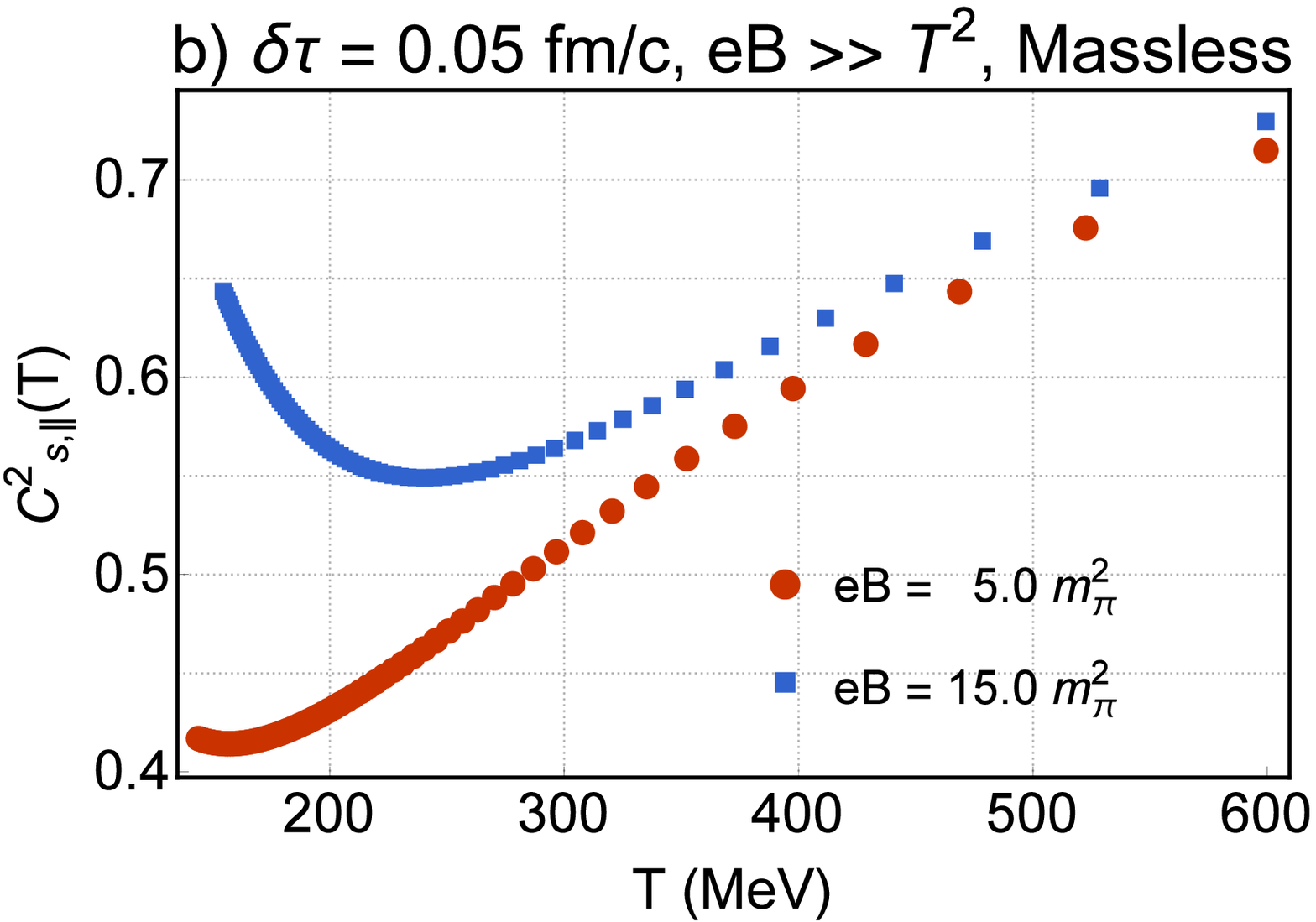}
\par
\includegraphics[width=8cm,height=6cm]{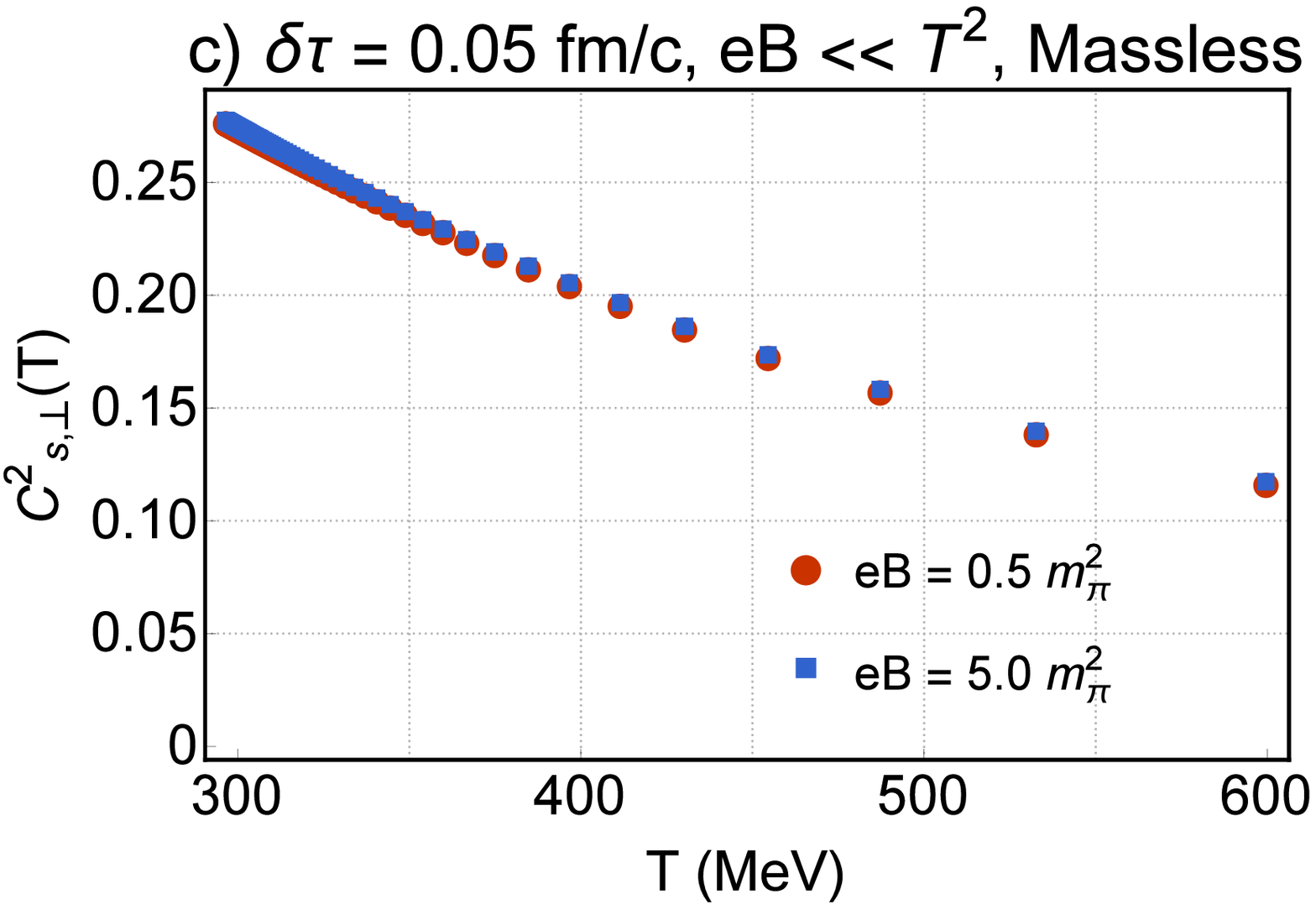}
\includegraphics[width=8cm,height=6cm]{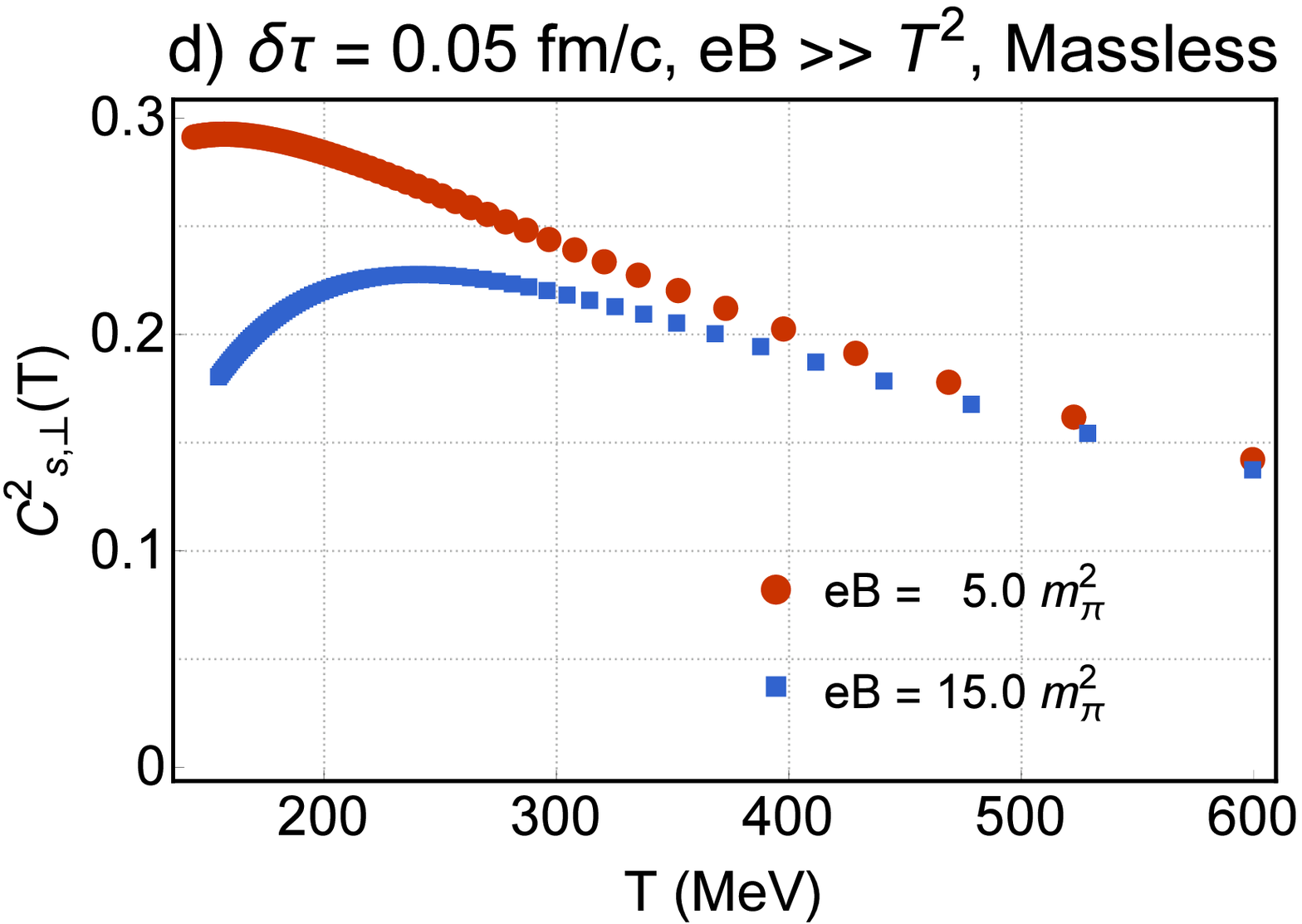}
\caption{(color online). The $T$ dependence of the longitudinal and transverse speed of sound $c_{s,\|}^{2}$ and $c_{s,perp}^{2}$ is plotted in the high (panels a and c) and low (panels b and d) temperature approximations. Increasing the magnetic field up to one order of magnitude does not affect the results in the high-temperature (weak magnetic field) limit. In contrast, in the low-temperature (strong magnetic field) limit the qualitative behavior of $c_{s,\|}^{2}$ and $c_{s,perp}^{2}$ is strongly affected by increasing the magnetic field up to a factor of three, especially in the low-$T$ regime, $T\leq 300$ MeV. }
\label{fig-9}
\end{figure*}
In Sec. \ref{sec3}, we determined the analytical $T$ dependence of $p_{\|},p_{\perp}$, and $\epsilon$ in high- and low-temperature approximations in the massless [\eqref{E28} and \eqref{E32}] and massive cases [\eqref{E27}].  Plugging the corresponding expressions into the energy equation \eqref{N33}, and solving the resulting first order differential equation numerically, we arrive at the $\tau$ dependence of the temperature $T$ in these approximations. The fermion mass $m$ appearing in \eqref{E27}, is chosen to be the thermal mass of the electron \cite{lebellac-book},
\begin{eqnarray}\label{D1}
m(T)=\left(m_0^{2}+\frac{e^{2}T^{2}}{8}\right)^{1/2}.
\end{eqnarray}
Here, $m_{0}=0.5$ MeV is the electron's rest mass, and $e$ its electric charge. It is given by $e\equiv \sqrt{4\pi\alpha}$ with the fine structure constant $\alpha=1/137$. As concerns the initial value of the proper time, $\tau_0$ and $T(\tau_{0})$, we choose $\tau_{0}=0.2$ fm/c and $T_{0}\equiv T(\tau_0)=600$ MeV.  The summation over Landau levels are truncated at $N=\lfloor\frac{T_0^{2}}{eB} \rfloor$ for $T_0=600$ MeV and fixed $eB$.  In Figs. \ref{fig-2}(a) and (b), we compare the $\tau$ dependence of $T$ arising from Bjorken flow \eqref{N34} [blue curves in Figs. \ref{fig-2}(a) and \ref{fig-2}(b)] with the $\tau$ dependence of $T$ arising from the solution of  the aforementioned differential equation in the high- and low-temperature approximations. In Fig. \ref{fig-2}(a), corresponding to $eB\ll T^{2}$, the initial value of the magnetic field is chosen to be $eB_0=0.5,5 m_{\pi}^{2}$ and  in Fig. \ref{fig-2}(b), corresponding to $eB\gg T^{2}$, the initial value of the magnetic field is chosen to be $eB_0=5,15 m_{\pi}^{2}$. The results demonstrated in Fig. \ref{fig-2} are for massless fermions. We repeated the same computation for massive fermions by plugging \eqref{E27} for $eB\ll T^{2}$ into \eqref{N33}, and arrived at the same result as presented in Fig. \ref{fig-2}(a). We thus conclude that once the magnetic field is chosen to be constant, the fermion thermal mass has not a crucial effect on the $\tau$ dependence of $T$.
\par
At this stage, following the procedure described in Sec. \ref{sec2d}, we determined the proper time dependence of $T$ for a point to point decaying magnetic field. For simplicity, we choose $B(\tau)$ from \eqref{N40} arising in the ideal transverse MHD (IdMHD). The corresponding results are denoted by $\delta B_{\text{HI}}$ for the high-temperature expansion with the initial magnetic field $eB_0=2.5 m_{\pi}^{2}$ [see black crosses in Fig. \ref{fig-2}(a)] and $\delta B_{\text{LI}}$ for low-temperature expansion with the initial magnetic field $eB_0=15 m_{\pi}^{2}$ [see black crosses in Fig. \ref{fig-2}(b)]. We repeat the same computation for magnetic fields varying according to \eqref{N38} (E-time), \eqref{N39} (Pheno), and \eqref{N41} (CMHD) with $\alpha_0=0.1$ and $\alpha_0=1$. The results for the $\tau$ dependence of $T$ are plotted in Fig. \ref{fig-3}.
\par
Let us notice that according to the above results for the high-temperature approximation, after an abrupt decay within $1$ fm/c, the temperature remains almost constant around $300$ MeV. Changing the initial magnetic field up to one order of magnitude does not affect this result.  As concerns the low-temperature expansion, there is almost no difference between the $\tau$ dependence of $T$ for constant (red curves) and varying (black crosses) magnetic fields. In the low-temperature approximation, however, the temperature decays fast, the result for $eB=5 m_\pi^2$ and $eB=15 m_\pi^2$ are different, while the results for constant $eB=15 m_{\pi}^{2}$ and a point to point varying magnetic field with the initial magnetic field $eB_0=15 m_{\pi^{2}}$ almost coincide [see Fig. \ref{fig-3}(b)].
\par
In Fig. \ref{fig-4}, the $\tau$ dependence of the temperature $T$ is plotted for different solutions: $eB_{\text{H}}$ and $eB_{\text{L}}$ denote the results arising from the solution of the energy equation for constant magnetic fields $eB=5 m_{\pi}^{2}$ and $eB=15 m_{\pi}^{2}$ in the high- (H) and low- (L) temperature approximations, respectively. The solutions presented in \eqref{N34}, \eqref{N35}, and \eqref{N36} for vanishing magnetic fields are denoted by ``Bjorken'', ``Hubble at $r=0$'' and ``Gubser at $r=0$'', respectively. The results corresponding to the solution of the energy equation with point to point varying magnetic fields are denoted by $\delta B_{\text{HI}}$ and $\delta B_{\text{LI}}$ in the high- and low-temperature approximations.  In this case, the subscript ``I'' indicates the decay of the magnetic field according to  \eqref{N40} in the $1+1$ dimensional ideal MHD. As demonstrated in Fig. \ref{fig-4}, the temperature decays very fast for the $3+1$ dimensional Hubble solution. Weak magnetic fields slow up this decay significantly, and keep the temperature almost constant in a relatively large proper time interval. As concerns the Gubser solution, it interpolates between the $B=0$ Bjorken solution in the early times and the low-temperature (large magnetic field) approximation in the late time, as is also denoted in \cite{shokri2018}.
\subsection{The $\bs{\tau}$ dependence of $\bs{p_{\|}, p_{\perp}}$, and $\bs{\epsilon}$  in constant and decaying magnetic fields}\label{sec4b}
Once the $\tau$ dependence of the temperature is determined by solving the  energy equation either with constant or with a point to point varying magnetic field, it is possible to plug the corresponding results back into the analytical expressions for $p_{\|},p_{\perp}$, and $\epsilon$ from  \eqref{E28} and \eqref{E32} in the high- and low-temperature approximations, $eB\ll T^{2}$ and $eB\gg T^{2}$, for massless fermions, and \eqref{E27} in the $eB\ll T^{2}$ approximation for massive fermions. In Fig. \ref{fig-5U}, we plotted the ratios $p_{\|}(\tau)/p_{\|,0}, p_{\perp}(\tau)/p_{\perp,0}$ and $\epsilon(\tau)/\epsilon_{0}$ arising from this computation. Here, $p_{\|,0},p_{\perp,0}$ and $\epsilon_{0}$ are the corresponding quantities at the initial proper time $\tau_0=0.2$ fm/c. In Figs. \ref{fig-5U}(a), \ref{fig-5U}(b), \ref{fig-5U}(e), and \ref{fig-5U}(f), we have compared the results arising from the numerical computation of $p_{\|}(\tau)/p_{\|,0}$ and $\epsilon(\tau)/\epsilon_{0}$ with the corresponding quantities arising in $1+1$ dimensional relativistic hydrodynamics (Bjorken solutions) in the absence of magnetic fields \cite{hatsuda-book}. They are given by
\begin{eqnarray}\label{D2}
\hspace{-1cm}p(\tau)=p_{0}\left(\frac{\tau_0}{\tau}\right)^{1+1/\kappa}, \qquad \epsilon(\tau)=\epsilon_{0}\left(\frac{\tau_0}{\tau}\right)^{1+1/\kappa}.
\end{eqnarray}
For an ideal gas with $\epsilon=3 p$, $\kappa=3$. According to these results, the data arising in the low-temperature (large magnetic field) approximation, $eB\gg T^{2}$, almost coincide with the corresponding results from \eqref{D2} from Bjorken flow [see Figs. \ref{fig-5U}(b) and \ref{fig-5U}(f)]. In the high-temperature (weak magnetic field) limit $eB\ll T^{2}$, however, magnetic field slows up the decay of $p_{\|}$, and $\epsilon$ [see Figs. \ref{fig-5U}(a) and \ref{fig-5U}(e)]. As concerns the results for the point to point varying magnetic field, once the decay is given by \eqref{N40} for $1+1$ dimensional ideal MHD, the results coincide with the numerical data from high- and low-temperature approximations. In Fig. \ref{fig-5U}, these results are denoted by black crosses. In the $eB\ll T^{2}$ and $eB\gg T^{2}$ cases, the initial magnetic field are chosen to be $eB_0=2.5 m_{\pi}^{2}$ and  $eB_0=15 m_{\pi}^{2}$, respectively.
\subsection{The $\bs{\tau}$ dependence of $\bs{\chi_m, c_{s,\|}}$, and $\bs{c_{s,\perp}}$ in constant and decaying magnetic fields}\label{sec4c}
Using the numerical results for $T, p_{\|},p_{\perp}$, and $\epsilon$ from the previous section, it is possible to compute a number of other thermodynamic quantities. Let us start with the magnetic susceptibility $\chi_m$, that is defined by the ratio $\chi_{m}=M/B$. Here, $M$ is the magnetization of the medium, and is determined by the difference between the longitudinal and transverse pressure, $BM=p_{\|}-p_{\perp}$, as defined also in Sec. \ref{sec2b}. Using the results for $p_{\|}$, and $p_{\perp}$, that are plotted in Figs. \ref{fig-5U}(a)-\ref{fig-5U}(d), we plotted the $\tau$ dependence of $\chi_m=BM/B^2=(p_{\|}-p_{\perp})/B^2$ in Fig. \ref{fig-6}. The red curves in \ref{fig-6}(a) and \ref{fig-6}(b) correspond to $\chi_m$ in a background constant magnetic field $eB=0.5 m_{\pi}^{2}$ in the high-$T$ and $eB=15 m_{\pi}^{2}$ in the low-$T$ approximations. The blue crosses correspond to the results arising from a point to point decaying magnetic field according to \eqref{N40}. Whereas in the constant magnetic field $\chi_m$ decreases with increasing $\tau$, it increases abruptly once the magnetic field decreases from point to point. These results coincide with the results demonstrated in \cite{tabatabaee2019}, where the $\tau$ dependence of $\chi_m$ in a uniformly expanding plasma is studied using the classical kinetic theory methods of anisotropic MHD. Assuming that the magnetic field decreases with $\tau$ according to \eqref{N40}, it is shown that within the first few fm/c $\chi_{m}$ increases. Later on, after reaching a maximum, it decreases in the late time $\tau\simeq 10$ fm/c.  Apart from this, $\chi_m$ turns out to be positive. This indicates that the plasma under investigation is a paramagnet. Comparing the values of $\chi_{m}$ in the high- and low-$T$ cases, it turns out that $\chi_{m}$ for $eB=15m_{\pi}^{2}$ is two order of magnitude smaller than $\chi_{m}$ for $eB=5 m_{\pi}^{2}$.
\par
Let us now turn to the numerical results arising for longitudinal and transverse speed of sound, $c_{s,\|}$ and $c_{s,\perp}$. They are defined by \cite{romatschke2009},\footnote{In \cite{klemm2008} an alternative definition of the equation of state is introduced for a conformal fluid $\epsilon=3p_{\|}-2BM$. Another alternative definition for $c_s^2$ is presented in \cite{rischke-MHD, tabatabaee2019}, and reads $c_{s,i}^2=p_{i}/\epsilon$ with $i=\|,\perp$.   }
\begin{eqnarray}\label{D3}
c_{s,i}^2=\frac{dp}{d\epsilon}=\frac{p'_{i}(T)}{\epsilon'(T)}, \qquad i=\|,\perp.
\end{eqnarray}
Here, the primes denote the derivative with respect to the temperature $T$. In Fig. \ref{fig-7}, the $\tau$ dependence of $c_{s,\|}^{2}$ and $c_{s,\perp}^{2}$ are plotted in the high- and low-temperature approximations [see Figs. \ref{fig-7}(a) and \ref{fig-7}(c) for the high-$T$ and \ref{fig-7}(a) and \ref{fig-7}(c) for the low-$T$ approximations]. They are determined by making use of the corresponding numerical data to $p_{\|}, p_{\perp}$, and $\epsilon$ from Fig. 5. As it is shown in these plots, whereas in the high-temperature (weak magnetic field) approximation, the results corresponding to a point to point decaying magnetic field (almost) coincides with those corresponding to constant magnetic fields, in the low-temperature (large magnetic field) approximation the corresponding results to a point to point decaying magnetic field is quite different from those corresponding to constant magnetic fields. The qualitative difference between the $\tau$ dependence of $c_{s,\|}$, and $c_{s,\perp}$ is also observed in the results presented in \cite{tabatabaee2019}, arising from a classical kinetic theory approach.
\subsection{The $\bs{T}$ dependence of $\bs{\chi_m, c_{s,\|}}$, and $\bs{c_{s,\perp}}$ in constant magnetic fields}\label{sec4d}
The $T$ dependence of thermodynamic quantities is of particular interest in the thermal field theory.  Among others, lattice QCD is one of the most prominent numerical methods to compute this dependence. In order to have a comparison with lattice data, we present, in this section, the $T$ dependence of $\chi_m, c_{s,\|}$, and $c_{s,\perp}$ arisen from our previous numerical results. Using the $\tau$ dependence of $T$ demonstrated in Sec. \ref{sec4a}, and combining the resulting data with the data arisen from the $\tau$ dependence of  $\chi_m, c_{s,\|}$, and $c_{s,\perp}$, the $T$ dependence of these quantities are determined. In Fig. \ref{fig-8}, the $T$ dependence of $\chi_{m}(T)/\chi_{m,0}$ is presented in the high- (orange circles) and low- (blue squares) temperature approximations. Here, $\chi_{m,0}$ is the value of $\chi_m$ at $\tau=\tau_0$. The proper time interval between two successive points in each curve is $\delta\tau=0.05$ fm/c. As it turns out, for a given temperature, the value of $\chi_{m}$ arising from the high-$T$ approximation is $10-30 \%$ larger than the corresponding value arising from the low-$T$ approximation. Moreover, according to these results, $\chi_{m}$ decreases with decreasing temperature. Larger magnetic fields have a larger impact on the decay of $\chi_{m}$ within a fixed proper time interval. Here, this interval is chosen to be $\Delta\tau=4$ fm/c.
\par
As concerns the $T$ dependence of the longitudinal and transverse speed of sound, $c_{s,\|}^{2}$ and $c_{s,\perp}^{2}$, we combined the data corresponding to the $\tau$ dependence of $T$ from Fig. \ref{fig-2} and those from $c_{s,\|}^{2}$ and $c_{s,\perp}^{2}$ from Fig. \ref{fig-7}, and arrived at the plots demonstrated in Fig. \ref{fig-9}. According to these results, in the high-temperature approximation $c_{s,\|}^{2}$ ($c_{s,\perp}^{2}$) decreases (increases) monotonically with decreasing temperature, and increasing the magnetic field up to one order of magnitude (from $0.5 m_{pi}^{2}$  to $5 m_{\pi}^{2}$) does not significantly affect them. In contrast, the $T$ dependence of $c_{s,\|}^{2}$ ($c_{s,\perp}^{2}$) exhibits a nonmonotonic decrease (increase) with decreasing $T$ in the low-temperature (large magnetic field) limit. Increasing the magnetic field up to one order of magnitude affects this behavior significantly, in particular in the late time (low-$T$).
\section{Concluding remarks}\label{sec5}
\setcounter{equation}{0}
In this paper, we combined the Wigner function formalism of relativistic quantum kinetic theory with the energy equations of relativistic MHD to determine the proper time evolution of the temperature in an expanding hot and magnetized QED plasma. This novel approach takes, in particular, important quantum corrections, including the contribution of Landau levels in the distribution function of this fermionic system, into account. Using the corresponding numerical results for the proper time dependence of the temperature, we also determined the evolution of a number of other thermodynamic quantities in this plasma.
\par
We used the solution of the Dirac equation to explicitly determine the Wigner function of this plasma. This leads to the corresponding energy-momentum, which, for its part, gives rise to the longitudinal and transverse pressures with respect to the external magnetic fields, $p_{\|}$ and $p_{\perp}$, as well as the energy density of this plasma, $\epsilon$. For further use, we determined these quantities in a high and low-temperature approximation. For massless fermions, these approximations are characterized by $eB\ll T^2$ and $eB\gg T^{2}$. They can also be interpreted as weak and strong magnetic field approximations, respectively. Plugging these expressions in the energy equation of relativistic MHD, and solving the resulting first order differential equation numerically, we arrived at the proper time evolution of the temperature in these high- and low-temperature approximations. Plugging these numerical results for $T$ in the analytical expressions for $p_{\|},p_{\perp}$ and $\epsilon$, the proper time evolution of these quantities were determined.
We compared our results with the proper time evolution of $T,p_{\|},p_{\perp}$, and $\epsilon$ arising from the well-known Bjorken solution of relativistic hydrodynamics. As concerns the $\tau$ dependence of the temperature, we showed that weak (strong) magnetic field decreases (increases) the slope of the decay of $T$ [see Figs. \ref{fig-2}(a) and (b)]. A point to point decaying magnetic field does not change this picture significantly. In the case of pressures and energy density, weak magnetic fields (large $T$) decreases the slope of the decay of $p_{\|}$ and $\epsilon$, while strong magnetic field does not affect the decay of these quantities too much.
\par
We notice that the Bjorken solution is in particular based on the equation of state of a noninteracting ideal gas, $\epsilon=c_s^2 p$, where $c_{s}=1/\sqrt{3}$ is the speed of sound. To analogously determine the relation between the pressure and energy density in this model, we plotted the proper time evolution of the longitudinal and transverse speed of sound with our own data for $p_{\|},p_{\perp}$, and $\epsilon$. The results demonstrated in Fig. \ref{fig-7} show that the proper time evolution of $c_{s,i}^{2}, i=\|,\perp$ is strongly affected by strong magnetic fields $eB\gg T^{2}$. The effect of point to point decaying magnetic fields on the evolution of these quantities is also noticeable only in the case of strong magnetic fields.
\par
We finally eliminated the parameter $\tau$ in the data corresponding to $T(\tau)$ and  $\chi_{m}(\tau)$ as well as $c_{s,i}^{2}(\tau), i=\|,\perp$, and arrive at the $T$ dependence of these quantities. The results are plotted in Figs. \ref{fig-8} and \ref{fig-9}. They indicate that $\chi_{m}$ decreases with decreasing $T$, and, as concerns the $T$ dependence of $c_{s,i}^{2}, i=\|, \perp$, that strong magnetic fields significantly affect the $T$ dependence of these quantities. We emphasize that the $T$ dependence of $c_{s,\|}^{2}$ and $c_{s,\perp}^{2}$ are totally different: Whereas $c_{s,\|}^{2}$ increases, $c_{s,\perp}^{2}$ decreases with increasing $T$. It would be interesting to confirm these results using lattice QCD at finite $T$ and in the presence of constant magnetic fields. As we have noticed before, in \cite{tabatabaee2019} we used the classical kinetic theory in combination with anisotropic MHD, and studied the proper time as well as the temperature dependence of the same thermodynamic quantities that are studied in the present paper. A comparison of the $\tau$ and $T$ dependence of these quantities here with those presented in \cite{tabatabaee2019} shows the effect of the quantum corrections, included in the Wigner function, on the evolution of these quantities.
It would be interesting to combine these two approaches to find a method where the Fermi-Dirac distribution function appearing in the nonvanishing components of $T^{\mu\nu}$ in \eqref{E18} is replaced with the anisotropic distribution function appearing in \cite{tabatabaee2019}, and to take, in this way, the effects of plasma viscosities \textit{and} quantum corrections on the thermodynamic properties of the plasma into account. To do this, it is necessary to appropriately revise \eqref{E5}, in order to incorporate the anisotropy parameter appearing in \cite{tabatabaee2019} in the partition function on the right-hand side of this relation. It is worthwhile to emphasize that the method presented in \cite{tabatabaee2019} is a phenomenological method. It is, in particular, engineered according to the recipe introduced in \cite{florkowski2010,strickland2010} to consider, apart from the effect of plasma viscosities \cite{florkowski2010,strickland2010}, the effect of its magnetization (net spin) on 1) its isotropization 2) its thermodynamic properties. In contrast, the approach introduced in the present paper considers the interplay between the spin and the magnetic field from first principles, as the Wigner function is built from quantized fermions that satisfy the Dirac equation in which the spin of the fermions is inherently included.
\par
Recently, the Wigner function approach is used in many areas of the QGP physics, including, among others in the spin hydrodynamics \cite{florkowski2018}. It would be interesting to study the coupling of the magnetic field and spin in this context. An attempt in this direction is made in \cite{rischke2019}. The analytical result for the Wigner function in the presence of a constant magnetic field, which is presented in this paper may be used in this framework. In particular, it would be interesting to study the effect of the spin-magnetic field coupling on the proper time evolution of the temperature and other thermodynamic quantities, using the method presented in this paper.
\par
Another way to extend the results of this paper is to consider a rotating plasma in a constant magnetic field. As it is shown in \cite{rotation-new}, the Dirac equation, in this case, has an analytical solution, that can be used to determine the corresponding Wigner function. The latter can then be used to determine the proper time dependence of the temperature and other thermodynamic quantities by carrying out the method presented in this paper.
\section{Acknowledgments}
This work is supported by Sharif University of Technology's Office of Vice President for Research under Grant No: G960212/Sadooghi.
\begin{appendix}
\section{The Gubser temperature \eqref{N36}}\label{appA}
\setcounter{equation}{0}
The original expression for the temperature by Gubser is given by \cite{gubser2010}
 \begin{eqnarray}\label{appA1}
\hspace{-0.5cm}
T=\frac{\hat{T}_0}{\tau^{1/3}}\frac{(2q)^{2/3}}{[1+q^4(\tau^{2}-r^{2})^{2}+2q^2(\tau^{2}+r^{2})]^{1/3}}.
\end{eqnarray}
In what follows, we show how \eqref{N36} with $\kappa=3$ is equivalent with \eqref{appA1}.
\par
Let us first notice that $\hat{T}_{0}$ in \eqref{appA1} is a dimensionless constant which is fixed by choosing an appropriate initial value for the temperature at some initial point $(\tau_0,r_0)$. Moreover, $q$ is a dimensionful quantity in fm$^{-1}$, and the proper time $\tau$ is in fm. Hence $T$, as given in \eqref{appA1}, is in fm$^{-1}$.  In order to convert it into MeV, we multiply it with a factor $\alpha_0=200$, arising from the fact that 1 fm$^{-1}$ is equivalent to $200$ MeV.\footnote{This arises from $\hbar c\sim 200\text{MeVfm}$. Setting $\hbar=c=1$, we have fm$^{-1}$=$200 \text{MeV}$.} We thus get
\begin{eqnarray}\label{appA2}
\hspace{-0.5cm}
T=\frac{\alpha_{0}\hat{T}_0}{\tau^{1/3}}\frac{(2q)^{2/3}}{[1+q^4(\tau^{2}-r^{2})^{2}+2q^2(\tau^{2}+r^{2})]^{1/3}},
\end{eqnarray}
that is in MeV. To reformulate \eqref{appA2} in terms of an initial temperature $T_{0}$, given by $T_{0}\equiv T(\tau_{0})$ and similar to $T_{0}$ appearing in the Bjorken temperature \eqref{N34}, we define
\begin{eqnarray}\label{appA3}
\hspace{-0.5cm}
\tilde{T}_{0}\equiv \frac{\hat{T}_{0}}{[1+q^4\left( \tau_0^{2}-r_{0}^{2}\right)^{2}+2q^{2}\left(\tau_{0}^{2}+r_{0}^{2}\right)]^{1/3}
\tau_{0}^{1/3}}.
\end{eqnarray}
The above quantity is in fm$^{-1/3}$. Plugging this expression into (\ref{appA2}), we get the temperature $T$ in MeV,
\begin{eqnarray}\label{appA4}
T=\alpha_{0}\tilde{T}_{0}(2q)^{2/3}\left(\frac{\tau_{0}}{\tau}\right)^{1/3}[\mathcal{G}(r,\tau;r_0,\tau_0)]^{1/3},
\end{eqnarray}
with $\mathcal{G}(r,\tau;r_0,\tau_0)$ defined in \eqref{N37}. Here, $\alpha_{0}\tilde{T}_{0}(2q)^{2/3}$ is in MeV. At this stage, we use the fact that in the limit $q\to 0$ and for fixed $\tilde{T}_{0}(2q)^{2/3}$,\footnote{See the paragraph below Eq. (22) in the first reference of \cite{gubser2010} for a similar argument for $\epsilon$.}  \eqref{appA4} must be equal to the Bjorken temperature (\ref{N34}). At $\tau=\tau_{0}$, we thus require
\begin{eqnarray}\label{appA5}
\lim_{q\to 0}\alpha_{0}\tilde{T}_{0}(2q)^{2/3}\left(\frac{\tau_{0}}{\tau}\right)^{1/3}\bigg|_{\tau=\tau_0}=T_{0}.
\end{eqnarray}
Here, $\lim\limits_{q\to 0}\mathcal{G}(r,\tau;r_0,\tau_0)=1$ is used. This leads to
\begin{eqnarray}\label{appA6}
\tilde{T}_{0}=\frac{T_{0}}{\alpha_{0}(2q)^{2/3}}\qquad \mbox{in fm$^{-1/3}$}.
\end{eqnarray}
Plugging eventually this expression into (\ref{appA4}) yields \eqref{N36} with $\kappa=3$, as expected.
\section{High  and low-temperature expansions}\label{appB}
\setcounter{equation}{0}
In this appendix, we show that the high- and low-temperature expansions of the $h_\ell$-integrals \eqref{E23} are given by \eqref{E25} and \eqref{E29}, respectively. To this purpose, we generalize the method introduced in \cite{weldon1982} to the fermionic case.
\par
Let us first consider \eqref{E23}. Plugging
\begin{eqnarray} \label{appB1}
\frac{1}{e^{a}+1}=\sum_{\ell=1}^{\infty} e^{- \ell a} - 2 \sum_{\ell=1}^{\infty} e^{- 2 \ell a},
\end{eqnarray}
into \eqref{E23}, and using
\begin{eqnarray}\label{appB2}
\int_{0}^{\infty} dx \frac{x^{2\ell} e^{- a \sqrt{x^{2} + y^{2} }}}{\sqrt{x^{2} + y^{2} }} &=&\frac{\Gamma\left(\ell+\frac{1}{2}\right)}{ \sqrt{\pi} } \left( \frac{2 y}{a} \right)^{\ell} K_{\ell} (ay),\nonumber\\
\end{eqnarray}
with $\text{Re}(a)>0 ,  \text{Re}(b)>0$,  as well as
\begin{eqnarray}\label{appB3}
\Gamma^{-1} \left(\frac{\ell+1}{2}\right)  = \frac{ 2^{\ell-1} \Gamma (\frac{\ell}{2})}{\sqrt{\pi} \Gamma(\ell)},
\end{eqnarray}
we arrive after some computations at
\begin{eqnarray}\label{appB4}
h_{\ell}(y)&=&\frac{1}{\Gamma \left(\frac{\ell+1}{2}\right)} \left(\frac{y}{2} \right)^{\frac{\ell-1}{2}} \sum_{p= 1}^{\infty} \bigg[p^{(1-\ell)/2} K_{(\ell-1)/2}(py)\nonumber\\
&& - 2^{\frac{3-\ell}{2}}p^{(1-\ell)/2} K_{(\ell-1)/2}(2 py) \bigg] .
\end{eqnarray}
In the above relations $K_{\ell}(z)$ is the modified Bessel function. To perform the summation over $p$, we use the Mellin transformation technique \cite{davies-book,weldon1982}. This technique is mainly used to perform the summation \cite{davies-book}
\begin{eqnarray}\label{appB5}
I=\sum_{n=1}^{\infty}f(n).
\end{eqnarray}
Using the Mellin transform of $f(n)$
\begin{eqnarray}\label{appB6}
f(n) = \frac{1}{2 \pi i} \int_{C-i \infty}^{C + i \infty} dp n^{-p} F(p) ,
\end{eqnarray}
and the definition of the $\zeta$-function, $\zeta(p)\equiv \sum_{n=1}^{\infty} n^{-p}$, we obtain
\begin{eqnarray}\label{appB7}
I = \frac{1}{2 \pi i} \int_{C-i \infty}^{C + i \infty} dp  F(p) \zeta(p) .
\end{eqnarray}
According to this method, the Mellin transform of \eqref{appB4} is given by
\begin{eqnarray}\label{appB8}
h_{\ell}(y)&=&\frac{1}{4\Gamma\left(\frac{\ell+1}{2}\right)}\left(\frac{y}{2}\right)^{\ell-1}
\frac{1}{2\pi i}\int_{C-i\infty}^{C+i\infty}ds  \left(1-2^{1-s}\right)\nonumber\\
&&\times\left(\frac{y}{2}\right)^{-s}\Gamma\left(\frac{s-\ell+1}{2}\right)\Gamma\left(\frac{s}{2}\right)\zeta(s).
 \end{eqnarray}
To arrive at \eqref{appB8}, we also use \cite{gradshtein-book}
\begin{eqnarray}\label{appB9}
\lefteqn{
\hspace{-0.5cm}\int_{0}^{\infty}dp p^{\mu}K_{\nu}(pz)}\nonumber\\
&&\hspace{-0.5cm}=\frac{1}{4}\left(\frac{2}{z}\right)^{\mu+1}\Gamma\left(\frac{\mu-\nu+1}{2}\right)\Gamma\left(\frac{\mu+\nu+1}{2}\right),
\end{eqnarray}
where $\text{Re}(z)>0$ and $\text{Re}(\mu+1\pm \nu)>0$ is assumed. For our purpose, it is enough to concentrate on $h_{2\ell+1}$. To performing the contour integral appearing in \eqref{appB9}, we shall have in mind that the integrand has double poles in $s=0,-2,-4,\cdots,-2n$ and single poles in $s=1$ and $s=2,4,6,\cdots, 2n$. Hence, using
\begin{eqnarray}\label{appB10}
\lim\limits_{s\to s_0}\mbox{Res}[f(s)]=\lim\limits_{s\to s_{0}}\frac{1}{(m-1)!}\frac{d^{m-1}}{ds^{m-1}}\bigg[
f(s)(s-s_{0})^{m}
\bigg],\nonumber\\
\end{eqnarray}
we obtain first
\begin{eqnarray}\label{appB11}
\lim_{s\to 1} f(s)&=&0\nonumber \\
\lim_{s \to 2k} f(s)&=&\left( 1 - 2^{1- 2 k} \right)\left( \frac{y}{2} \right)^{- 2 k} \Gamma(k) \zeta(2 k)\nonumber\\
&&\times \frac{2 (-1)^{\ell-k}}{\Gamma(\ell - k +1)} \nonumber \\
\lim_{ s \to 0} f(s) &  =&  - \frac{(-1)^{\ell}}{\Gamma(\ell+1)} \bigg[\gamma_{E} - \psi(1+ \ell) + 2 \ln \left( \frac{y}{\pi} \right) \bigg] \nonumber \\
\lim_{ s \to - 2 k } f(s) &  =&  \frac{4 (-1)^{\ell} }{\Gamma(k+1) \Gamma(\ell + k+1)} \left( 1 - 2^{1+ 2 k} \right)\nonumber\\
&&\times \left( \frac{y}{2} \right)^{ 2 k} \zeta'(- 2 k).
\end{eqnarray}
Here, $\gamma_{E}$ and the digamma function $\psi(z)$ are defined in Sec. \ref{sec3} [see the paragraph below \eqref{E26}].  Using then
\begin{eqnarray}\label{appB12}
\zeta'(- 2 k) = \frac{1}{2} (-1)^{k}  (2 \pi)^{- 2 k} \Gamma(2 k + 1) \zeta (2 k + 1) ,\nonumber\\
\end{eqnarray}
we finally arrive at \eqref{E25}, as claimed. This result is used to determine $h_{1}$ and $h_{3}$ in the high-temperature approximation ($eB\ll T^{2}$ for massless fermions).
\par
As concerns the low-temperature approximation ($eB\gg T^{2}$ for massless fermions), let us consider \eqref{E23} again. Using
\begin{eqnarray}\label{appB13}
\frac{1}{e^{a}+1}=\frac{1}{e^{a}-1}-\frac{2}{e^{2a}-1},
\end{eqnarray}
as well as $w\equiv\exp \left( y -  \sqrt{x^{2} + y^{2}} \right)$, and $v \equiv\exp \left( 2 y - 2 \sqrt{x^{2} + y^{2}} \right)$, we arrive first at
\begin{eqnarray} \label{appB14}
h_{\ell} (y)&=&  \frac{1}{\Gamma(\ell)} \left(  \frac{1}{2 y}  \right)^{1 - \frac{\ell}{2}}
\nonumber\\
&&\times \left\{  \int_{0}^{1} dw \frac{ \left( 1 - \frac{\ln w}{2 y}\right)^{\frac{\ell}{2}-1} \left( - \ln w \right)^{\frac{\ell}{2}-1} }{e^{y} - w}\right.\nonumber\\
 &&\left.-   \int_{0}^{1} dv \frac{ \left( 1 - \frac{\ln v}{4 y}\right)^{\frac{\ell}{2}-1} \left( - \ln v \right)^{\frac{\ell}{2}-1} }{e^{2 y} - v}  \right\}.\nonumber \\
\end{eqnarray}
Assuming then $|\frac{\ln w}{2y}|\ll 1$ as well as $|\frac{\ln v}{4y}|\ll 1$, and using
\begin{eqnarray}\label{appB15}
\left( 1 -x \right)^{a-1} = \Gamma\left(a\right) \sum_{k=0}^{\infty} \frac{(- x)^{k}}{\Gamma\left(a-k\right) \Gamma(k+1)},
\end{eqnarray}
as well as the integral representation of the polylogarithm function $\text{Li}_{n}(x)$ for $n>0$ \cite{weldon1982},
\begin{eqnarray}\label{appB16}
\text{Li}_{n}(x) = - \frac{1}{\Gamma(n)} \int_{0}^{1} dt \frac{\left( - \ln t \right)^{n-1}}{t- x^{-1}},
\end{eqnarray}
we arrive at \eqref{E29}, which is used for the low-temperature approximation in Sec. \ref{sec3}.

\end{appendix}

\end{document}